\documentclass[hyper,11pt,paper]{article}
\usepackage{jheppub}
\usepackage{graphicx,amsmath,amssymb,multirow,array,bm,mathrsfs}
\usepackage{epsf,amsfonts}
\usepackage[numbers,sort&compress]{natbib}
\usepackage{hyperref}
\newcommand{\N}{\mathcal{N}}

\newcommand{\beq}{\begin{equation}}
\newcommand{\eeq}{\end{equation}}

\newcommand{\R}{{\mathcal{R}}}

\newcommand{\bea}{\begin{eqnarray}}
\newcommand{\eea}{\end{eqnarray}}
\newcommand{\bdm}{\begin{displaymath}}
\newcommand{\edm}{\end{displaymath}}
\newcommand{\nn}{\nonumber}

\newcommand{\p}{\partial}
\newcommand{\vol}{\text{vol}}

\newcommand{\coe}{Y}

\def\We{\wp}
\def\g{\gamma}

\newcommand{\be}{\begin{equation}}
\newcommand{\ee}{\end{equation}}
\newcommand{\ba}{\begin{eqnarray}}
\newcommand{\ea}{\end{eqnarray}}
\newcommand{\lp}{\left(}
\newcommand{\rp}{\right)}
\newcommand{\ls}{\left[}
\newcommand{\rs}{\right]}

\renewcommand{\ell}{R}

\newcommand{\M}{{\mathcal{M}}}

\title{On Holographic Defect Entropy}

\author[a]{John Estes,}
\author[b,c]{Kristan Jensen,}
\author[d]{Andy O'Bannon,}
\author[]{Efstratios Tsatis}
\author[e]{and Timm Wrase}

\affiliation[a]{Blackett Laboratory, Imperial College, London SW7 2AZ, United Kingdom}
\affiliation[b]{Department of Physics and Astronomy, University of Victoria, Victoria, BC V8W 3P6, Canada}
\affiliation[c]{C.N. Yang Institute for Theoretical Physics, SUNY Stony Brook, Stony Brook, NY 11794-3840 United States}
\affiliation[d]{Rudolf Peierls Centre for Theoretical Physics, University of Oxford, \\ \hspace{.3cm} 1 Keble Road, Oxford OX1 3NP, United Kingdom}
\affiliation[e]{Stanford Institute for Theoretical Physics, Stanford University\\
\hspace{0.3cm} Stanford, CA 94305, United States}

\emailAdd{johnaldonestes@gmail.com}
\emailAdd{kristanj@insti.physics.sunysb.edu}
\emailAdd{obannon@physics.ox.ac.uk}
\emailAdd{efstratiostsatis@hotmail.com}
\emailAdd{timm.wrase@stanford.edu}

\abstract{
We study a number of $(3+1)$- and $(2+1)$-dimensional defect and boundary conformal field theories holographically dual to supergravity theories. In all cases the defects or boundaries are planar, and the defects are codimension-one. Using holography, we compute the entanglement entropy of a (hemi-)spherical region centered on the defect (boundary). We define defect and boundary entropies from the entanglement entropy by an appropriate background subtraction. For some $(3+1)$-dimensional theories we find evidence that the defect/boundary entropy changes monotonically under certain renormalization group flows triggered by operators localized at the defect or boundary. This provides evidence that the $g$-theorem of $(1+1)$-dimensional field theories generalizes to higher dimensions.
}

\preprint{Imperial/TP/2014/JE/01, YITP-SB-14-08,\\\null  \hfill OUTP-14-03p, SU/ITP-14/05}

\begin{document}
\maketitle

%%%%%%%%%%%%%%%%%%%%%%%%%%%%%%%%%%%%%%%%%%%%%%%%%%%%%%%%%%%%%%%%%%%%%%%%%%%%%%%%%%%%%%%%%%%%%%%%%%%%%%%%%%%%%%%%%%
\section{Introduction and Summary}
\label{S:intro}
%%%%%%%%%%%%%%%%%%%%%%%%%%%%%%%%%%%%%%%%%%%%%%%%%%%%%%%%%%%%%%%%%%%%%%%%%%%%%%%%%%%%%%%%%%%%%%%%%%%%%%%%%%%%%%%%%%

%%%%%%%%%%%%%%%%%%%%%%%%%%%%%%%%%%%%%%%%%%%%%%%%%%%%%%%%%%%%%%%%%%%%%%%%%%%%%%%%%%%%%%%%%%%%%%%%%%%%%%%%%%%%%%%%%%
\subsection{Motivation and Review}
%%%%%%%%%%%%%%%%%%%%%%%%%%%%%%%%%%%%%%%%%%%%%%%%%%%%%%%%%%%%%%%%%%%%%%%%%%%%%%%%%%%%%%%%%%%%%%%%%%%%%%%%%%%%%%%%%%

Consider a quantum system in an ensemble described by a density matrix $\rho$, and suppose that the Hilbert space may be decomposed into a product of two subspaces $\mathcal{H}_A$ and $\mathcal{H}_B$. One measure of the quantum entanglement between the subsystems $A$ and $B$ is the entanglement entropy (EE), $S$, defined as the von Neumann entropy of the reduced density matrix $\rho_A$ of the subsystem $A$ obtained by tracing $\rho$ over the states in $\mathcal{H}_B$,
\begin{equation*}
S = - \text{tr}(\rho_{A}\ln\rho_{A})\,.
\end{equation*}
EE  has many possible uses, for example in detecting topological order~\cite{Kitaev:2005dm,Levin:2006zz}.

In this paper we consider EE in the vacuum of local quantum field theories (QFTs) in Minkowski space. In particular, for a fixed time slice we pick $A$ and $B$ to be a spatial region $\mathcal{R}$ and its complement $\overline{\mathcal{R}}$, respectively. We will refer to the surface separating $\mathcal{R}$ and $\overline{\mathcal{R}}$ as the ``entangling surface,'' $\mathcal{M}$. Since the vacuum of a local QFT is a pure state, the EE obtained by first tracing over states in $\overline{\mathcal{R}}$ is the same as that obtained by first tracing over states in $\mathcal{R}$. In this sense, the position-space EE is a non-local observable which depends on $\mathcal{M}$ rather than on $\mathcal{R}$ or $\overline{\mathcal{R}}$. In a continuum QFT, position-space EE diverges due to correlations among highly-entangled short-distance modes near $\M$. To obtain a finite result for the EE, we introduce a short-distance cutoff $\varepsilon$.

Remarkably, position-space EE can be related to certain monotonicity theorems, as follows. In the vacuum state of a $d$-dimensional conformal field theory (CFT), consider the EE for a spherical $\mathcal{M}$ of radius $R$, \textit{i.e.}\ $\M=\mathbb{S}^{d-2}$. For $d=2$, $\mathcal{M}$ consists of the two endpoints of an interval of length $2R$. In $d=2,3$, and $4$, this EE takes the form~\cite{Bombelli:1986rw,Srednicki:1993im,Holzhey:1994we,Calabrese:2004eu}
\begin{align}
\begin{split}
\label{E:sphericalEE}
	S = \left\{
		\begin{array}{lc}
			 C_{log}^{(d=2)}\ln \left( \frac{2R}{\varepsilon}\right) + \tilde{C}_0^{(d=2)}\,, & d=2,
			\\
			C_1\frac{R}{\varepsilon}+C^{(d=3)}_0\,, & d=3,
			\\
			C_2 \frac{R^2}{\varepsilon^2}+C^{(d=4)}_{log}\ln\left( \frac{2R}{\varepsilon}\right) +\tilde{C}_0^{(d=4)}, & d=4,
\end{array}\right.
\end{split}
\end{align}
where the various $C$'s and $\tilde{C}$'s are constants that are independent of $R$ and $\varepsilon$ but depend on the details of the CFT. In eq.~\eqref{E:sphericalEE} we have neglected terms that vanish as $\varepsilon\to 0$, as we will continue to do in all that follows. The quantities $C_1$, $C_2$, and the $\tilde{C}_0$'s depend on the choice of regularization, while the $C_{log}$'s and $C_0^{(d=3)}$ are ``universal'' in that they are invariant under rescalings of $\varepsilon$. Such universal constants are in principle physically observable. In particular, the $C_{log}$'s are proportional to Weyl anomaly coefficients, and $C_0^{(d=3)}$ is minus the free energy of the Euclidean CFT on $\mathbb{S}^3$ of radius $R$~\cite{Holzhey:1994we,Calabrese:2004eu,Myers:2010xs,Myers:2010tj,Casini:2011kv}:\footnote{We choose conventions such that the $d=2,4$ Weyl anomalies are
\begin{align*}
d=2: \quad \langle T_{\mu}^{~\mu}\rangle &= - \frac{c}{24\pi}R,
\\
d=4: \quad \langle T_{\mu}^{~\mu}\rangle & = c W_{\mu\nu\rho\sigma}W^{\mu\nu\rho\sigma}- a E_4,
\end{align*}
with $R$ the intrinsic Ricci scalar of the background metric, $W_{\mu\nu\rho\sigma}$ the Weyl tensor, and $E_4 = R_{\mu\nu\rho\sigma}R^{\mu\nu\rho\sigma}-4R_{\mu\nu}R^{\mu\nu}+R^2$ the four-dimensional Euler density. In $d=3$, if $Z_{\mathbb{S}^3}$ is the partition function of the Euclidean CFT on $\mathbb{S}^3$, then the free energy is $F_{\mathbb{S}^3}\equiv- \ln Z_{\mathbb{S}^3}$. Typically, $F_{\mathbb{S}^3}$ is divergent, so we can extract physical information from $F_{\mathbb{S}^3}$ only after renormalization.}
\beq
\label{E:cToEE}
C_{log}^{(d=2)} = \frac{c}{3}\,, \qquad C_0^{(d=3)} = - F_{\mathbb{S}^3}\,, \qquad C_{log}^{(d=4)} = - 64\pi^2 a\,.
\eeq
Each of these objects obeys a monotonicity theorem, and in that sense counts degrees of freedom. In $d=2$, Zamolodchikov's $c$-theorem~\cite{Zamolodchikov:1986gt} states that $c$ decreases between the endpoints of an RG flow: $c_{UV}\geq c_{IR}$. Similarly, in $d=3$ the $F$-theorem (conjectured in ref.~\cite{Jafferis:2011zi} and proven in ref.~\cite{Casini:2012ei}) states that $F_{\mathbb{S}^3}$ decreases between the endpoints of an RG flow. In $d=4$, the $a$-theorem (conjectured in ref.~\cite{Cardy:1988cwa} and proven in ref.~\cite{Komargodski:2011vj}) states that $a$ decreases between the endpoints of an RG flow.

In $d=2$ another monotonicity theorem exists, for ``defect CFTs'' (DCFTs). A DCFT consists of two CFTs each on a half-line connected at their mutual endpoint by a conformally-invariant defect. For $\R$ an interval of length $2R$ centered on the defect, the EE is~\cite{Calabrese:2004eu,Azeyanagi:2007qj}
\beq
\label{E:EEfor2dDCFT}
S = \frac{S_++S_-}{2}+\ln(g),
\eeq
where $S_{\pm}$ are the EE's for intervals of length $R$ in the CFTs on the two sides of the defect. The quantity $\ln(g)$ is called the defect entropy. The folding trick maps a $d=2$ DCFT to a $d=2$ CFT with a conformal boundary, called a ``boundary CFT'' (BCFT). Denoting the CFTs on the two sides of the defect as $\textrm{CFT}_{\pm}$, the BCFT is the tensor product $\textrm{CFT}_+\otimes \textrm{CFT}_-$ equipped with conformally-invariant boundary conditions characterized by a boundary state~\cite{Cardy:1989ir}. If $\R$ is an interval of length $R$ ending on the boundary, then the EE for the BCFT is equal to that of the DCFT eq.~\eqref{E:EEfor2dDCFT}: in terms of the central charge of the BCFT, the EE is
\beq
\label{eq:bcftboundent}
S= \frac{c}{6} \ln\left( \frac{2R}{\varepsilon} \right) + \tilde{C}_0 + \ln(g).
\eeq
In this context, $\ln(g)$ is called the boundary entropy. In eq.~\eqref{eq:bcftboundent}, $\ln(g)$ looks like a contribution to a non-universal constant. We can prove that in fact $\ln(g)$ is universal via a background subtraction, as follows. If we compute both of $S_{\pm}$ using the same regulator $\varepsilon$, and then compute $S$ in the associated DCFT or BCFT also using the same $\varepsilon$, then in $S - \frac{S_+ + S_-}{2}$ all divergent and non-universal terms (the $\ln\left( \frac{2R}{\varepsilon} \right)$ and $\tilde{C}_0$ terms in eq.~\eqref{eq:bcftboundent}) will cancel, leaving behind a universal contribution, $\ln(g)$.

The $g$-theorem (conjectured in ref.~\cite{Affleck:1991tk} and proven in ref.~\cite{Friedan:2003yc}) states that $g$ decreases along an RG flow between two BCFTs triggered by an operator localized to the boundary, called a ``boundary RG flow.'' For an RG flow triggered by an operator in the ambient CFT, no such monotonicity theorem exists: in such cases, $g$ may either increase or decrease~\cite{Green:2007wr}. Thanks to the folding trick, the $g$-theorem also holds for DCFTs.

Some important open questions are: for BCFTs and DCFTs in $d>2$, can we extract a boundary or defect entropy from EE? If so, can we prove whether it is monotonic along RG flows triggered by defect/boundary-localized operators? Can we prove whether it is monotonic along RG flows triggered by deformations of the ambient CFT? In short, does the $g$-theorem generalize to higher dimensions? These questions are difficult to answer, partly because EE is difficult to compute even in free theories.

%%%%%%%%%%%%%%%%%%%%%%%%%%%%%%%%%%%%%%%%%%%%%%%%%%%%%%%%%%%%%%%%%%%%%%%%%%%%%%%%%%%%%%%%%%%%%%%%%%%%%%%%%%%%%%%%%%
\subsection{The Systems We Study}
\label{S:systems}
%%%%%%%%%%%%%%%%%%%%%%%%%%%%%%%%%%%%%%%%%%%%%%%%%%%%%%%%%%%%%%%%%%%%%%%%%%%%%%%%%%%%%%%%%%%%%%%%%%%%%%%%%%%%%%%%%%

To address these questions, we turn to holography, or more precisely the Anti-de Sitter/CFT (AdS/CFT) correspondence~\cite{Maldacena:1997re}. This correspondence relates certain $d$-dimensional CFTs to string theories on backgrounds that in general consist of a warped product of a $(d+1)$-dimensional AdS factor, $AdS_{d+1}$, and an internal space. In the best-understood examples, the CFTs are non-Abelian gauge theories in the 't Hooft large-$N$ limit with large 't Hooft coupling, $\lambda \gg 1$, and the holographic duals are semiclassical supergravities (SUGRAs).

We use holography simply because it is the easiest way to compute EE for interacting CFTs in $d>2$. For CFTs dual to SUGRA, the prescription to compute EE in a time-independent state, conjectured in refs.~\cite{Ryu:2006bv,Ryu:2006ef} and proven in ref.~\cite{Lewkowycz:2013nqa}, is the following. On a fixed time slice in the bulk, we determine the codimension-one surface of minimal (Einstein-frame) area $\mathcal{A}_{min}$ that approaches $\M$ at the $AdS_{d+1}$ boundary. The EE is then, with $G_N$ the bulk Newton's constant,
\beq
\label{rt}
S = \frac{\mathcal{A}_{min}}{4 G_N}.
\eeq

In principle, we would like to study holographic duals of RG flows between BCFTs and DCFTs. Many gravity solutions exist that describe RG flows between DCFTs, usually involving ``probe'' defects, meaning the defect's contributions to observables (including EE) are suppressed by factors of $N$ relative to the ambient CFT~\cite{Yamaguchi:2002pa}. Few solutions exist describing conformal defects outside of the probe limit~\cite{Gutperle:2012hy,Dias:2013bwa,Korovin:2013gha}. Some \textit{ad hoc} solutions for the holographic duals of BCFTs, and RG flows between BCFTs, appear in refs.~\cite{Takayanagi:2011zk,Fujita:2011fp,Nozaki:2012qd,Gutperle:2012hy}.\footnote{Despite the title of ref.~\cite{Gutperle:2012hy}, the solutions there actually describe fixed points, not RG flows.} In some cases these are genuine solutions of SUGRA theories~\cite{Fujita:2011fp}, and hence we have good reason to believe a pathology-free dual BCFT actually exists.  In general, however, that is not guaranteed.  Moreover, without a specific dual field theory, a comparison between results calculated on the two sides of the correspondence, gravity and field theory, is impossible.\footnote{The bottom-up models for BCFTs of refs.~\cite{Takayanagi:2011zk,Fujita:2011fp,Nozaki:2012qd} also deviate in an essential way from almost all holographic BCFTs that arise in string theory: they are locally AdS. More precisely, in the bottom-up models of refs.~\cite{Takayanagi:2011zk,Fujita:2011fp,Nozaki:2012qd}, the dual spacetime ends on a codimension-one brane which may support some localized matter content. The geometry is locally AdS everywhere away from the ``brane'' , and the shape of the brane is determined by the Israel junction condition involving the brane stress-energy tensor. Currently, the one and only example of such a holographic BCFT in string theory appears in ref.~\cite{Fujita:2011fp}, where the dual spacetime ends on two separated O8$^{-}$ planes, together with two stacks of D8-branes. We do not expect such features to be generic in string theory. In particular, in all other known examples of holographic BCFTs in string theory, the dual spacetime caps off smoothly, rather than ending on a ``brane,'' and the metric only asymptotically approaches AdS. These examples include the $d=4$ BCFT arising from D3-branes ending on D5-branes~\cite{Aharony:2011yc}, the $d=3$ BCFT arising from M2-branes ending on M5-branes~\cite{Bachas:2013vza}, and the $d=2$ BCFTs of refs.~\cite{Chiodaroli:2011fn,Chiodaroli:2012vc}.}

Our goal is a more general analysis, relying as little as possible on special limits such as the probe limit, and using genuine solutions of SUGRA, so that we have good reason to believe dual BCFTs and DCFTs exist. To our knowledge, no SUGRA solutions exist describing RG flows between BCFTs or DCFTs outside of the probe limit. We thus turn to known SUGRA solutions that describe fixed points rather than RG flows. We will be able to extract boundary and defect entropies from our holographic results for EE, but our arguments for their behavior along RG flows will be indirect. Such is the price we pay for working outside the probe limit and demanding that dual field theories exist.

We focus exclusively on two CFTs that we will deform to obtain DCFTs and BCFTs. The first CFT is $(3+1)$-dimensional $\N=4$ supersymmetric (SUSY) $SU(N)$ Yang-Mills (SYM) theory. The second CFT is the $(2+1)$-dimensional $\N=6$ SUSY $U(N)_k\times U(N)_{-k}$ Chern-Simons-matter theory of Aharony, Bergman, Jafferis, and Maldacena (ABJM)~\cite{Aharony:2008ug}. In each theory we work in the 't Hooft large-$N$ limit, with large 't Hooft coupling, in which case the holographic dual is SUGRA on a background with an AdS factor.

We choose these two CFTs for two reasons. First, in the dual SUGRA theories, many solutions are known that describe conformal boundaries and codimension-one defects~\cite{Bak:2003jk,Clark:2005te,D'Hoker:2006uu,D'Hoker:2007xy,D'Hoker:2007xz,D'Hoker:2009gg,Aharony:2011yc,Suh:2011xc,Berdichevsky:2013ija,Bobev:2013yra}. All of these solutions describe a  boundary or defect that is planar. Second, not only are we confident that the dual DCFTs and BCFTs actually exist, in contrast to many bottom-up models, but also in many cases explicit Lagrangians are known for the dual DCFTs and BCFTs~\cite{DeWolfe:2001pq,Erdmenger:2002ex,Clark:2004sb,D'Hoker:2006uv,Kim:2008dj,Gaiotto:2008sa,Honma:2008un,Gaiotto:2008ak,D'Hoker:2009gg}. We will perform a general calculation, applicable to essentially all of the solutions of refs.~\cite{Bak:2003jk,Clark:2005te,D'Hoker:2006uu,D'Hoker:2007xy,D'Hoker:2007xz,D'Hoker:2009gg,Aharony:2011yc,Suh:2011xc,Berdichevsky:2013ija,Bobev:2013yra}, however, we will present explicit results only for a representative sample of the SUGRA solutions in refs.~\cite{Bak:2003jk,D'Hoker:2006uu,D'Hoker:2007xy,D'Hoker:2007xz,D'Hoker:2009gg,Aharony:2011yc}, as we discuss below.

\begin{figure}
  \begin{center}
    \includegraphics[width=.8\textwidth]{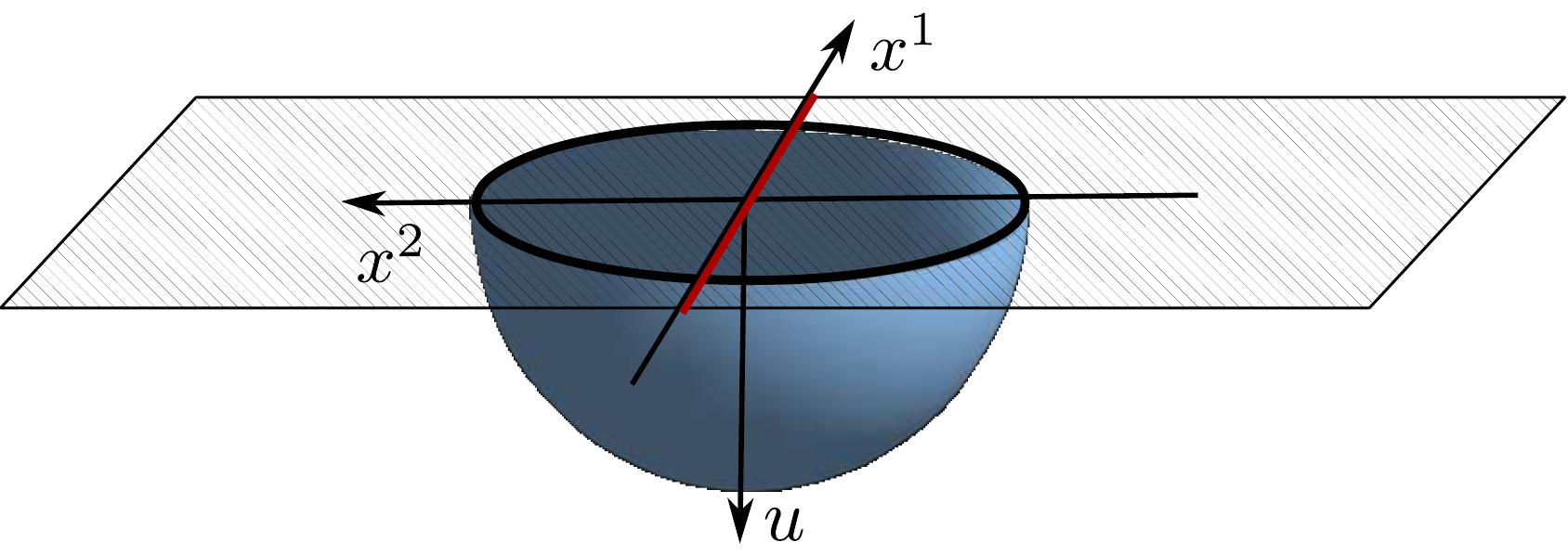}
  \end{center}
  \caption{\label{fig:setup} A cartoon of a $(2+1)$-dimensional DCFT and its holographic dual. The DCFT ``lives'' at the boundary of the holographic dual, depicted as the shaded plane. The two horizontal directions $x^1$ and $x^2$ are the DCFT's spatial directions, while $u$ is the holographic direction. The planar defect is extended along $x^1$, as depicted by the solid red line. We compute the EE of a spherical region centered on the defect. The entangling surface $\M = \mathbb{S}^1$ is depicted as the solid black circle. The blue hemisphere is the minimal-area surface in the bulk which ends on $\mathcal{M}$, and whose area determines the EE via eq.~\eqref{rt}.}
\end{figure}

For the entangling surface $\mathcal{M}$, for DCFTs we choose a sphere centered on the defect, as depicted in fig.~\ref{fig:setup}. We do so for two reasons. First, for a special class of DCFTs we know the spherical EE provides a defect entropy monotonic under a defect RG flow, namely DCFTs in which the defect is a CFT in its own right, completely decoupled from the ambient CFT. In these cases, the spherical EE decomposes into a sum of two spherical EE's, one for the ambient CFT, $S_{\textrm{CFT}}$, and one for the defect CFT, $S_{\textrm{defect}}$, that is, $S=S_{\textrm{CFT}}+S_{\textrm{defect}}$. For a defect of spacetime dimension $2$, $3$, or $4$, and for RG flows triggered by defect-localized operators built out of defect fields, the $c$-, $F$-, and $a$-theorems, combined with eq.~\eqref{E:sphericalEE}, tell us that a certain term in the EE will change monotonically. For instance, if the defect has spacetime dimension $2$, then the defect entropy $S-S_{\textrm{CFT}}$ will include a logarithmic term whose coefficient always decreases under defect RG flows. Analogous statements apply for BCFTs, where we choose $\mathcal{M}$ to be a hemi-sphere centered on the boundary.

Our second reason for studying (hemi-)spherical $\mathcal{M}$ is practical: for any DCFT or BCFT with a holographic dual, an exact solution is known for the minimal area surface that approaches the (hemi-)spherical $\mathcal{M}$ at the boundary~\cite{Jensen:2013lxa}. Using that solution, for many holographic DCFTs and BCFTs we are able to compute defect and boundary entropies exactly, without approximation (beyond the SUGRA approximation to string theory) and without numerics.

To be precise, we define defect and boundary entropies following the $d=2$ example: we regulate the EE in the DCFT or BCFT in the same way as the original CFT, and then define defect entropy, $S_{\textrm{defect}}$, or boundary entropy, $S_{\partial}$, via a background subtraction,
\beq
\label{E:defDefBdyS}
S_{\textrm{defect}}  \equiv S - \frac{S_++S_-}{2},
\qquad
S_{\partial}  \equiv S - \frac{S_{CFT}}{2},
\eeq
where $S_{\pm}$ are the spherical EE's for the ambient CFTs on either side of a defect, and $S_{CFT}$ is the spherical EE for the ambient CFT far away from the boundary. We emphasize that in the holographic calculation, matching the cutoff $\varepsilon$ of the DCFT or BCFT to the original CFT is non-trivial (and sometimes difficult), because the DCFT or BCFT is dual to SUGRA in a very different spacetime from that of the original CFT.

%%%%%%%%%%%%%%%%%%%%%%%%%%%%%%%%%%%%%%%%%%%%%%%%%%%%%%%%%%%%%%%%%%%%%%%%%%%%%%%%%%%%%%%%%%%%%%%%%%%%%%%%%%%%%%%%%%
\subsection{Summary of Results}
%%%%%%%%%%%%%%%%%%%%%%%%%%%%%%%%%%%%%%%%%%%%%%%%%%%%%%%%%%%%%%%%%%%%%%%%%%%%%%%%%%%%%%%%%%%%%%%%%%%%%%%%%%%%%%%%%%

In section~\ref{S:holo} and in the appendix we perform a general calculation of $S_{\textrm{defect}}$ and $S_{\partial}$ applicable to essentially all of the solutions of refs.~\cite{Bak:2003jk,Clark:2005te,D'Hoker:2006uu,D'Hoker:2007xy,D'Hoker:2007xz,D'Hoker:2009gg,Aharony:2011yc,Suh:2011xc,Berdichevsky:2013ija,Bobev:2013yra}, namely DCFTs and BCFTs in $d=3,4$ holographically dual to ten- or eleven-dimensional SUGRA. For our holographic DCFTs, we find that generically $S_{\textrm{defect}}$ takes the form
\beq
\label{E:Sdefect}
S_{\textrm{defect}} = \left\{ \begin{array}{lc} D_{log}^{(d=3)} \ln\left( \frac{2R}{\varepsilon}\right) + \tilde{D}_0^{(d=3)}, & \quad d=3, \\ D_1^{(d=4)} \frac{R}{\varepsilon} + D_0^{(d=4)}, & \quad d=4,\end{array}\right.
\eeq
where the various $D$'s and $\tilde{D}_0^{(d=3)}$ are constants that are independent of $\varepsilon$ and $R$ but that depend on the details of the DCFT. Note that $S_{\textrm{defect}}$ takes the same form as the spherical EE for a CFT, eq.~\eqref{E:sphericalEE}, of the same spacetime dimension as the defect, $d-1$. Our holographic calculation makes clear that $\tilde{D}_0^{(d=3)}$ and $D_1^{(d=4)}$ are non-universal while $D_{log}^{(d=3)}$ and $D_0^{(d=4)}$ are universal. For our holographic BCFTs, we find that generically $S_{\partial}$ takes the form
\beq
\label{E:Sbdy}
S_{\partial} =  \left\{ \begin{array}{lc} B_{log}^{(d=3)} \ln\left( \frac{2R}{\varepsilon}\right) + \tilde{B}_0^{(d=3)}, &\quad d=3, \\ B_1^{(d=4)} \frac{R}{\varepsilon} + B_0^{(d=4)}, & \quad d=4, \end{array}\right.
\eeq
where the $B$'s and $\tilde{B}_0^{(d=3)}$ are constants that are independent of $\varepsilon$ and $R$ but that depend on the details of the BCFT. Our holographic calculation makes clear that $\tilde{B}_0^{(d=3)}$ and $B_1^{(d=4)}$ are non-universal while  $B_{log}^{(d=3)}$ and $B_0^{(d=4)}$ are universal. Eq.~\eqref{E:Sbdy} is not surprising: in free $d=4$ BCFTs, with hemi-spherical $\M$ centered on the boundary, $S_{\partial}$ takes the same form~\cite{Fursaev:2013mxa}.

For a CFT in $d=3$, the $F$-theorem, by way of eqs.~\eqref{E:sphericalEE} and~\eqref{E:cToEE}, tells us that minus the constant piece of the spherical EE monotonically decreases under RG flow. By analogy, for DCFTs and BCFTs in $d=4$, we propose that $-D_0^{(d=4)}$ and $-B_0^{(d=4)}$ decrease under RG flows triggered by defect- or boundary-localized operators. On similar grounds, for DCFTs and BCFTs in $d=3$ we propose that the coefficients of the logarithmic terms, $D_{log}^{(d=3)}$ and $B_{log}^{(d=3)}$, also decrease under such RG flows. This latter conjecture was made, and proven for the bottom-up holographic BCFTs of refs.~\cite{Takayanagi:2011zk,Fujita:2011fp}, already in ref.~\cite{Fujita:2011fp}.

Our holographic calculation makes clear that $D_{log}^{(d=3)}$, $B_{log}^{(d=3)}$, $D_0^{(d=4)}$, and $B_0^{(d=4)}$ depend on the entire geometry of the holographic dual, and not just a region near the defect or boundary. Moreover, we expect that $D_{log}^{(d=3)}$ and $B_{log}^{(d=3)}$ are related to Weyl anomalies supported on the defect or boundary. As a result, these coefficients could potentially decrease even under flows in which the ambient CFT is deformed.

In section~\ref{S:examples} we compute $D_{log}^{(d=3)}$, $D_0^{(d=4)}$, and $B_0^{(d=4)}$ explicitly in various examples. (Our examples do not include a holographic BCFT in $d=3$, so we present no examples of $B_{log}^{(d=3)}$.) Our examples involving $\N=4$ SYM are: the DCFT obtained from the $(2+1)$-dimensional intersection of D3- and D5-branes~\cite{Karch:2000gx,DeWolfe:2001pq,Erdmenger:2002ex,Gomis:2006cu,D'Hoker:2007xy,D'Hoker:2007xz}, the BCFT obtained from D3-branes ending on D5-branes~\cite{Aharony:2011yc}, the DCFT obtained by coupling the so-called $T[SU(N)]$ theory (a CFT in $d=3$)~\cite{Gaiotto:2008ak} to $\N=4$ SYM~\cite{Assel:2011xz}, and certain so-called Janus deformations of $\N=4$ SYM~\cite{Bak:2003jk,Clark:2004sb,D'Hoker:2006uu,D'Hoker:2006uv,D'Hoker:2007xy,Gaiotto:2008sd}, in which the YM coupling takes different values on two halves of space. Our example involving ABJM theory also involves a Janus-like defect~\cite{D'Hoker:2009gg,Estes:2012vm,Bobev:2013yra}.

Although we cannot compute EE holographically along RG flows between DCFTs or BCFTs, in principle we can compare the (hemi-)spherical EE between fixed points connected by RG flows. Fortunately, thanks to SUSY we can identify fixed points connected by RG flows within the class of examples that we study. Our prime example is the D3/D5 DCFT~\cite{Karch:2000gx,DeWolfe:2001pq,Erdmenger:2002ex,Gomis:2006cu,D'Hoker:2007xy,D'Hoker:2007xz}, $\N=4$ SYM with gauge groups $SU(N_{3}^{\pm})$ on the two sides of a codimension-one defect that supports a number $N_5$ of $(2+1)$-dimensional hypermultiplets preserving eight real supercharges. We can trigger a defect RG flow by introducing a hypermultiplet mass. A mass deformation preserving eight real supercharges exists, allowing us to identify the IR fixed point unambiguously: it is the D3/D5 theory again, but with a reduced value of $N_5$. We can trigger a bulk RG flow by moving onto the Higgs branch of the SUSY moduli space. In that case, SUSY again allows us to identify the IR fixed point: it is the D3/D5 theory with reduced values of $N_3^{\pm}$. Analogous statements apply for the D3/D5 BCFT.

Our holographic calculation reveals that $-D_0^{(d=4)}$ or $-B_0^{(d=4)}$ always decreases under the defect or boundary RG flow in which $N_5$ decreases, and may either increase or decrease under the bulk RG flow in which $N_3^{\pm}$ decreases. Such behavior is highly reminiscent of the original $g$-theorem in $d=2$, and provides non-trivial evidence supporting our conjecture for a $g$-theorem in $d=4$.

Our other examples provide additional circumstantial evidence for our conjectures, and raise additional questions. For example, for a $T[SU(N)]$ defect in $\N=4$ SYM, in the limit where $N$ is much greater than the rank of the $\N=4$ SYM gauge group, our holographic calculation reveals that $-D_0^{(d=4)} = - F_{\mathbb{S}^3}$, where here $F_{\mathbb{S}^3}$ is the free energy of the Euclidean $T[SU(N)]$ theory on $\mathbb{S}^3$. In our ABJM Janus example, we find $D_{log}^{(d=3)}=0$, the meaning of which remains mysterious to us. (The same thing happened in a bottom-up holographic model for a $d=3$ DCFT in ref.~\cite{Korovin:2013gha}.) We leave further details of our examples to section~\ref{S:examples}.

%%%%%%%%%%%%%%%%%%%%%%%%%%%%%%%%%%%%%%%%%%%%%%%%%%%%%%%%%%%%%%%%%%%%%%%%%%%%%%%%%%%%%%%%%%%%%%%%%%%%%%%%%%%%%%%%%%
\subsection{Outlook}
%%%%%%%%%%%%%%%%%%%%%%%%%%%%%%%%%%%%%%%%%%%%%%%%%%%%%%%%%%%%%%%%%%%%%%%%%%%%%%%%%%%%%%%%%%%%%%%%%%%%%%%%%%%%%%%%%%

Our work is just the tip of the iceberg of higher-dimensional defect and boundary entropies. What follows is our own somewhat idiosyncratic list of promising directions for future research.

In $d=2$, Zamolodchikov's $c$-function~\cite{Zamolodchikov:1986gt}, built from the two-point function of the stress-energy tensor, decreases monotonically along RG flows and coincides with the central charge $c$ at the fixed points. Similarly, a $g$-function exists~\cite{Friedan:2003yc}, defined in terms of the one-point function of the stress-energy tensor, which decreases monotonically along a defect or boundary RG flow, and coincides with $g$ at the fixed points.

As proven in refs.~\cite{Casini:2004bw,Casini:2012ei}, in $d=2$ the ``renormalized EE''~\cite{Liu:2012eea} of an interval, $R \frac{dS}{dR}$, also acts as a $c$-function, although the relation to Zamolodchikov's $c$-function remains mysterious~\cite{Casini:2004bw,Casini:2012ei}. The method of proof in refs.~\cite{Casini:2004bw,Casini:2012ei} relied on Lorentz boost symmetries that are broken when we introduce a defect or boundary, hence such methods cannot immediately provide a $g$-function. To date, a $g$-function defined in terms of (renormalized) EE has not been found.\footnote{The proof of refs.~\cite{Casini:2004bw,Casini:2012ei} also does not immediately generalize to higher $d$. In $d=3$ the renormalized EE of a circle, $(R \frac{d}{dR}-1)S$, provides an $F$-function~\cite{Casini:2012ei}, albeit one that may not be stationary at fixed points~\cite{Klebanov:2012va}. Moreover, in $d=4$ holography provides evidence that renormalized EE of a sphere, $\frac{1}{2} R \frac{d}{dR}\left(R \frac{d}{dR}-2\right)S$, does not always change monotonically under RG flows, and hence may not provide an $a$-function~\cite{Liu:2012eea}.} Clearly, some important open questions are: in $d=2$, can we define a $g$-function from EE? In $d>2$, can we obtain $g$-functions, using EE or otherwise?

If we wish to address these questions using holography, then we necessarily need gravity solutions describing RG flows between DCFTs or BCFTs, rather than just fixed points. Generically, holographic $c$-theorems invoke the null energy condition in the bulk~\cite{Freedman:1999gp} to guarantee monotonicity of certain terms in the EE~\cite{Myers:2010xs,Myers:2010tj}: at fixed points these terms coincide with either an $a$-type central charge (for even $d$) or $(-1)^{(d-1)/2}$ times the free energy of the Euclidean theory on $\mathbb{S}^d$ (for odd $d$). Holographic $g$-functions have been proposed which invoke a null energy condition for the stress-energy tensor of a brane on the gravity side, either a probe brane dual to a defect~\cite{Yamaguchi:2002pa} or the ``brane'' on which spacetime ends in the bottom-up holographic models of BCFTs of refs.~\cite{Takayanagi:2011zk,Fujita:2011fp,Nozaki:2012qd}. What physical observables these $g$-functions are dual to in the field theory is not always clear. A natural question is whether they are dual to some contribution to an EE. Probe branes may provide the simplest way to address this question, since several techniques exist to calculate a probe brane's contribution to EE~\cite{Chang:2013mca,Jensen:2013lxa,Karch:2014ufa}.

In a $d=4$ CFT the coefficient of the $\ln\left(2R/\varepsilon\right)$ term in the EE is determined completely by $\mathcal{M}$ and the central charges $a$ and $c$. For spherical $\mathcal{M}$, the coefficient is $\propto a$, as in eq.~\eqref{E:sphericalEE}, while for cylindrical $\mathcal{M}$ it is $\propto c$~\cite{Casini:2011kv}. The central charge $c$ obeys no monotonicity theorem: explicit examples show that $c$ can either increase or decrease under RG flows (see for example refs.~\cite{Anselmi:1997am,Anselmi:1997ys}). In this paper we focus on (hemi-)spherical $\mathcal{M}$, but what about other $\mathcal{M}$? Can we characterize the $\ln\left(2R/\varepsilon\right)$ terms in defect/boundary entropy by $\mathcal{M}$ and a finite set of ``central charges''? The results of ref.~\cite{Fursaev:2013mxa} for $d=4$ BCFTs, for $\mathcal{M}$ that intersects the boundary, suggest that this may be the case. What about $\mathcal{M}$ that do not intersect the defect/boundary? Studying different $\mathcal{M}$ could be useful for identifying and studying candidates for defect/boundary entropies, for example by eliminating some candidates (like $c$ in $d=4$).

There are proposals to use EE to detect topological order in $d=3$~\cite{Kitaev:2005dm,Levin:2006zz} and $d\geq 4$~\cite{Grover:2011fa}. Holography can describe many topologically non-trivial phases,  and so can help to test these proposals. For example, two kinds of holographic models exist for time-reversal invariant fractional topological insulators in $d=4$. The first kind uses probe branes~\cite{Maciejko:2010tx,HoyosBadajoz:2010ac}, for which EE could be computed using the methods of refs.~\cite{Chang:2013mca,Jensen:2013lxa,Karch:2014ufa}. The second kind uses Janus solutions of SUGRA~\cite{Estes:2012nx}, including some of the examples we study in sections~\ref{ssec:NSJanus} and~\ref{ssec:Janus}. (The two kinds of models may be closely related~\cite{Estes:2012nx}.) A natural questions is: to what extent do our results in sections~\ref{ssec:NSJanus} and~\ref{ssec:Janus} characterize the topological order of these states?

Lastly, SUSY localization has been used to compute a SUSY version of R\'enyi entropy for certain Chern-Simons-matter theories in $d=3$~\cite{Nishioka:2013haa}. The EE may be extracted from this SUSY R\'enyi entropy~\cite{Nishioka:2013haa}. Moreover, SUSY localization has also been used to compute the partition functions of SUSY theories on manifolds with boundaries~\cite{Sugishita:2013jca,Honda:2013uca,Hori:2013ika}. Presumably these two things can be combined: for SUSY theories on manifolds with boundaries, SUSY localization could be used to compute EE. Such calculations could provide exact results for boundary entropies, which could be very useful for testing higher-dimensional $g$-theorems.

This paper is organized as follows. In section~\ref{S:holo}, we discuss the calculation of spherical EE for general holographic DCFTs and BCFTs. We pay special attention to the regularization of the EE, so that we can perform the background subtractions in our definitions of $S_{\textrm{defect}}$ and $S_{\partial}$ in eq.~\eqref{E:defDefBdyS}. Section~\ref{S:examples} is a case-by-case study of spherical EE in our various examples of DCFTs and BCFTs. The appendix contains the technical details of our general analysis of spherical EE in $d=3$,$4$, including in particular the derivation of eqs.~\eqref{E:Sdefect} and~\eqref{E:Sbdy}.

%%%%%%%%%%%%%%%%%%%%%%%%%%%%%%%%%%%%%%%%%%%%%%%%%%%%%%%%%%%%%%%%%%%%%%%%%%%%%%%%%%%%%%%%%%%%%%%%%%%%%%%%%%%%%%%%%%
\section{Holographic Calculation}
\label{S:holo}
%%%%%%%%%%%%%%%%%%%%%%%%%%%%%%%%%%%%%%%%%%%%%%%%%%%%%%%%%%%%%%%%%%%%%%%%%%%%%%%%%%%%%%%%%%%%%%%%%%%%%%%%%%%%%%%%%%

%%%%%%%%%%%%%%%%%%%%%%%%%%%%%%%%%%%%%%%%%%%%%%%%%%%%%%%%%%%%%%%%%%%%%%%%%%%%%%%%%%%%%%%%%%%%%%%%%%%%%%%%%%%%%%%%%%
\subsection{Review: No Defect or Boundary}
\label{sec:2.1}
%%%%%%%%%%%%%%%%%%%%%%%%%%%%%%%%%%%%%%%%%%%%%%%%%%%%%%%%%%%%%%%%%%%%%%%%%%%%%%%%%%%%%%%%%%%%%%%%%%%%%%%%%%%%%%%%%%

We start with the simple case of $AdS_{d+1}$ and a spherical $\M$. In this case, the first holographic calculation of EE was in refs.~\cite{Ryu:2006bv,Ryu:2006ef}. We will give an alternative derivation of the same result, highlighting several points that will be useful to us later. In particular, the duals of DCFTs and BCFTs will have $SO(d-1,2)$ isometry, so from the beginning we will make manifest an $SO(d-1,2)$ subgroup of the $SO(d,2)$ isometry of $AdS_{d+1}$.

The metric of $AdS_{d+1}$ with radius $L$ in Poincar\'e slicing is
\beq
g = \frac{L^2}{z^2} \left(dz^2 - dt^2 + d\vec{x}^2 \right),
\eeq
with $\vec{x} = (x^1,\ldots,x^{d-1})$ and with the $AdS_{d+1}$ boundary at $z \to 0$. To make manifest the $SO(d-1,2)$ subgroup of the $SO(d,2)$ isometry, we change coordinates as
\beq
\label{AdSFG}
z = \frac{u}{\cosh x}\,, \qquad x^{d-1} = u \tanh x,
\eeq
which puts the $AdS_{d+1}$ metric into $AdS_d$ slicing,
\beq
\label{eq:AdSslicing}
g = L^2 \left(dx^2 + \cosh^2(x) g_{AdS_{d}}\right).
\eeq
with $g_{AdS_d}$ the metric of a unit-radius $AdS_d$ in Poincar\'e slicing,
\begin{align}
\label{E:gAdSd}
g_{AdS_d} = \frac{1}{u^2} (du^2 -dt^2 + dr^2 + r^2 g_{\mathbb{S}^{d-3}}),
\end{align}
where $g_{\mathbb{S}^{d-3}}$ is the metric of a unit-radius $(d-3)$-sphere, $\mathbb{S}^{d-3}$. For $d=3$, we use the $r \to -r$ symmetry to choose the convention that $r\in \mathbb{R}^+$ with $\text{vol}(\mathbb{S}^0)=2$. The $SO(d-1,2)$ subgroup of the $SO(d,2)$ isometry acts as the isometry of the $AdS_d$ slice. The $AdS_d$ slicing splits $AdS_{d+1}$ into two regions, $x>0$ and $x<0$. In particular, from eq.~\eqref{AdSFG} we see that the $AdS_{d+1}$ boundary $z \to 0$ splits into two pieces at $x = \pm \infty$. These two pieces are glued together at the boundary of the $AdS_{d}$ slice, $u \to 0$, or equivalently at $x^{d-1} = 0$. In the dual CFT, we can think of $x^{d-1}=0$ as the location of a fictitious codimension-one planar defect.

We now consider a spherical $\M$ of radius $R$ centered on the fictitious defect, or more precisely centered at the origin $\vec{x}=\vec{0}$. Following Ryu and Takayanagi~\cite{Ryu:2006bv,Ryu:2006ef}, to compute this EE holographically we must compute the area of the minimal surface which lives on a fixed time-slice and approaches $\M$ as $z \to 0$. That minimal area surface wraps the $\mathbb{S}^{d-3}$ and so is described by a hypersurface in the $(x,u,r)$-space. If we describe that surface as $r(x,u)$, then the area functional becomes
\beq
\label{eq:AdSslicingarea}
\mathcal{A} = \vol(\mathbb{S}^{d-3})L^{d-1} \int du \, dx \, r^{d-3} \, \frac{\cosh^{d-2}(x)}{u^{d-2}} \sqrt{1+ (\p_u r)^2 + \frac{\cosh^2(x)}{u^2} (\p_x r)^2}\,.
\eeq
We will discuss the endpoints of the $u$ and $x$ integrations in eq.~\eqref{eq:AdSslicingarea} momentarily. The Euler-Lagrange equation arising from eq.~\eqref{eq:AdSslicingarea} is a complicated partial differential equation for $r(x,u)$. However, the minimal area surface that we want has a simple description in Poincar\'e slicing~\cite{Ryu:2006bv,Ryu:2006ef}: $z^2 + (x^{d-1})^2 + r^2 = R^2$, which at the $AdS_{d+1}$ boundary $z \to 0$ clearly describes a sphere of radius $R$ centered at the origin. Switching to $AdS_d$ slicing via eq.~\eqref{AdSFG}, the solution for the minimal area surface becomes
\beq
\label{eq:minsurface}
u^2 + r^2 = R^2.
\eeq
A straightforward exercise shows that the solution for $r(x,u)$ given by eq.~\eqref{eq:minsurface} indeed solves the Euler-Lagrange equation arising from eq.~\eqref{eq:AdSslicingarea}. Notice that the solution for $r(x,u)$ given by eq.~\eqref{eq:minsurface} depends on $u$ but not on $x$.

Let us now compute the value of the minimal area. To do so, we insert the solution in eq.~\eqref{eq:minsurface} into the area functional eq.~\eqref{eq:AdSslicingarea} and then integrate in $x \in (-\infty,\infty)$ and $u \in [0,R]$. The integrand in eq.~\eqref{eq:AdSslicingarea} diverges exponentially in the asymptotically $AdS_{d+1}$ regions at large $|x|$, and hence $\mathcal{A}$ is divergent. From the CFT perspective, these are the expected short-distance divergences from highly-entangled modes near $\mathcal{M}$. Again following Ryu and Takayanagi~\cite{Ryu:2006bv,Ryu:2006ef}, we regulate the divergence by introducing a Fefferman-Graham (FG) cutoff: in the Poincar\'e-sliced coordinates we introduce a cutoff surface $z = \varepsilon$. In the $AdS_d$ slicing, the FG cutoff becomes a surface in the $(x,u)$-space. Via eq.~\eqref{AdSFG}, that surface is described as the union of two surfaces $\chi_{\pm}(\frac{\varepsilon}{u})$ given by
\beq
\chi_{\pm}\left(\frac{\varepsilon}{u}\right) \equiv \pm \text{arccosh}\left(\frac{u}{\varepsilon}\right)= \pm \ln \left( \frac{2u}{\varepsilon} \right) \pm \ln \left[ \frac{1}{2} + \frac{1}{2} \sqrt{1- \frac{\varepsilon^2}{u^2}} \right]\,,
\eeq
where $u\in [\varepsilon,R]$. Note that the cutoff surface is real and continuous for this choice of lower bound on $u$. Using these cutoffs and the solution for $r(x,u)$ in eq.~\eqref{eq:minsurface}, the integral for the minimal area becomes
\beq
\label{eq:sphereee1}
\mathcal{A}_{min} = \vol(\mathbb{S}^{d-3}) \, L^{d-1} \, R \int^R_{\varepsilon} du \, \frac{(R^2 - u^2)^\frac{d-4}{2}}{u^{d-2}} \int_{\chi_-(\frac{\varepsilon}{u})}^{\chi_+(\frac{\varepsilon}{u})} dx \, \cosh^{d-2}(x)\,.
\eeq
We are interested in the cases $d=3,4$, for which
\beq
\label{eq:minarearesult}
\mathcal{A}_{min} =  \left\{ \begin{array}{ll}  2 \pi L^2 \left( \frac{R}{\varepsilon} - 1 \right)\,, & \quad d =3\,, \\ & \\ 2\pi \, L^{3} \left( \frac{R^2}{\varepsilon^2} - \ln\left( \frac{2R}{\varepsilon} \right) - \frac{1}{2} \right)\,, & \quad d=4\,.
\end{array}\right.
\eeq
Following eq.~\eqref{rt}, we multiply eq.~\eqref{eq:minarearesult} by $1/(4G_N)$ to obtain the EE, which reproduces the results of refs.~\cite{Ryu:2006bv,Ryu:2006ef}, as advertised.

In this work, we study DCFTs and BCFTs where the ambient CFT is either the ABJM theory or $\mathcal{N}=4$ SYM. The holographic dual of $U(N)_k\times U(N)_{-k}$ ABJM theory is eleven-dimensional M-theory on $AdS_4 \times \mathbb{S}^7/\mathbb{Z}_k$, where the $AdS_4$ has radius $L$ and the $\mathbb{S}^7/\mathbb{Z}_k$ has radius $2L$. In the $N \gg k$ and $N\gg k^5$ limits, the M-theory is well-approximated by eleven-dimensional SUGRA. The minimal-area surface wraps the internal space $\mathbb{S}^7/\mathbb{Z}_k$, so the result for $\mathcal{A}_{min}$ is the $d=3$ result in eq.~\eqref{eq:minarearesult} times the volume of $\mathbb{S}^7/\mathbb{Z}_k$, $\frac{\pi^4}{3k}(2L)^7$. In eleven dimensions the gravitational constant is given by $4 G_N = 2^6 \pi^7 l_p^9$ and the AdS radius is related to field theory quantities as $L^6 = \frac{\pi^2}{2} N k \, l_p^6$, where $l_p$ is the Planck length~\cite{Aharony:2008ug}. The spherical EE then follows from eq.~\eqref{rt},
\beq
\label{E:EEforABJM}
S = \frac{\pi \sqrt{2}}{3} \, k^\frac{1}{2} \, N^\frac{3}{2} \left[ \frac{R}{\varepsilon} - 1 \right]\,. \qquad \textrm{(ABJM theory)}
\eeq

The holographic dual of $\N=4$ SYM theory is type IIB string theory on $AdS_5 \times \mathbb{S}^5$, where both the $AdS_5$ and $\mathbb{S}^5$ have radius $L$. In the $N,\lambda \gg 1 $ limits (with $\lambda\equiv g_{YM}^2N \ll N$), the string theory is well-approximated by type IIB SUGRA. The minimal-area surface wraps the internal space $\mathbb{S}^5$, so the result for $\mathcal{A}_{min}$ is the $d=4$ result in eq.~\eqref{eq:minarearesult} times the volume of the $\mathbb{S}^5$, $\pi^3 L^5$. In ten dimensions and in Einstein frame, the gravitational constant is given by $4 G_N = 2^5 \pi^6 (\alpha')^4$ and the AdS radius is given in terms of SYM quantities as $L^4 = 4 \pi N (\alpha')^2$, where $\alpha'$ is the string length squared. The spherical EE then follows from eq.~\eqref{rt},
\beq
\label{E:EEforSYM}
S = N^2 \left[ \frac{R^2}{\varepsilon^2} - \ln\left( \frac{2R}{\varepsilon} \right) - \frac{1}{2} \right]\,. \qquad \textrm{($\N=4$ SYM theory)}
\eeq 

%%%%%%%%%%%%%%%%%%%%%%%%%%%%%%%%%%%%%%%%%%%%%%%%%%%%%%%%%%%%%%%%%%%%%%%%%%%%%%%%%%%%%%%%%%%%%%%%%%%%%%%%%%%%%%%%%%
\subsection{General Defect or Boundary}
%%%%%%%%%%%%%%%%%%%%%%%%%%%%%%%%%%%%%%%%%%%%%%%%%%%%%%%%%%%%%%%%%%%%%%%%%%%%%%%%%%%%%%%%%%%%%%%%%%%%%%%%%%%%%%%%%%

\label{S:generalDefect}

We now turn our attention to the calculation of $S$ for a general DCFT or BCFT in $d=4$ or $d=3$ holographically dual to type IIB string theory or M-theory. For now we will discuss DCFTs with holographic duals, saving BCFTs for the end of this subsection. We consider only codimension-one planar defects, so the $(d+1)$-dimensional DCFT will have $SO(d-1,2)$ conformal symmetry, and the dual ten- or eleven-dimensional geometry will include an $AdS_d$ factor. The (Einstein-frame) metrics that we study all have the form
\beq
\label{E:generalMetric}
g = \left[ f(x,y^a)^2 g_{AdS_d} + \rho(x,y^a)^2 dx^2 + G_{bc}(x,y^a)dy^b dy^c\right],
\eeq
where we will use the $AdS_d$ metric of eq.~\eqref{E:gAdSd},
\begin{equation*}
g_{AdS_d} = \frac{1}{u^2}\left( du^2-dt^2+dr^2+r^2g_{\mathbb{S}^{d-3}}\right),
\end{equation*}
and where $y^a$ are the coordinates of a compact internal space with metric $G_{bc}(x,y^a)$.\footnote{The most general metric with $SO(d-1,2)$ isometry is of the form in eq.~\eqref{E:generalMetric} plus mixed $dx\,dy^a$ terms. We can always choose $\{x,y^a\}$ to remove those mixed terms locally, but whether we can always remove such terms globally in such a way as to preserve the asymptotic AdS regions as $|x| \to \pm \infty$ is not clear. In all of the examples we consider in this paper, however, such a global choice always exists, hence we restrict our analysis to metrics of the form in eq.~\eqref{E:generalMetric}.} The backgrounds dual to DCFTs possess two asymptotic $AdS_{d+1}$ regions. We will choose $x$ so that these regions are located at $x\to \pm \infty$. In the DCFT, the ambient CFTs on the two sides of the defect need not be the same, so in the holographic dual the $AdS_{d+1}$ radii of curvature in the two regions, $L_{\pm}$, need not be the same. More generally, the warp factors $f(x,y)^2$ and $\rho(x,y)^2$ and the metric $G_{bc}(x,y^a)$ may approach distinct values in the $x \to \pm \infty$ limits. In the $x \to \pm \infty$ limits, the metric functions admit the following expansions in $e^{\pm x}$:
\begin{align}
\begin{split}\label{eq:expansion}
f(x,y^a)^2 &= \frac{L_{\pm}^2}{4}\left( e^{\pm 2x+2c_{\pm}} + f^{(-1)}_{\pm}(y^a) e^{\pm x} + f^{(0)}_{\pm}(y^a) + \hdots\right),
\\
\rho(x,y^a)^2 & = L_{\pm}^2\left(1 + \rho^{(1)}_{\pm}(y^a) e^{\mp x} + \rho^{(2)}_{\pm}(y^a)e^{\mp 2x} + \hdots\right),
\\
G(x,y^a) & = G_{\pm}^{(0)}(y^a) + G_{\pm}^{(1)}(y^a)e^{\mp x} + G_{\pm}^{(2)}(y^a)e^{\mp 2 x} + \hdots,
\end{split}
\end{align}
where $c_{\pm}$ are constants and the $\cdots$ denote terms sub-leading in $e^{\pm x}$ compared to those shown. We use $\pm$ subscripts to indicate that the expansion coefficients $f^{(-1)}_{\pm}(y^a)$, $\rho^{(1)}_{\pm}(y^a)$, $G_{\pm}^{(0)}(y^a)$, etc., may approach different values in the $x \to \pm \infty$ limits. The leading terms in the expansions of eq.~\eqref{eq:expansion} are fixed such that the metric approaches the asymptotic form of the $AdS_{d+1} \times \mathcal{M}_y$ metric as $x \to \pm \infty$, where the $AdS_{d+1}$ metric is in the $AdS_d$ slicing of eq.~\eqref{eq:AdSslicing} with radius of curvature $L_{\pm}$, and $\mathcal{M}_y$ is a compact internal space with metric $G_{\pm}^{(0)}(y^a)$. The two asymptotically $AdS_{d+1}$ regions are glued together at the $AdS_d$ boundary in a fashion similar to the $AdS_d$ slicing of $AdS_{d+1}$, though now with a genuine defect in the field theory located at the plane along which the two pieces are glued.

Given a metric of the form in eq.~\eqref{E:generalMetric}, we want to compute holographically the spherical EE. The minimal-area surface we want is a codimension-two surface sitting at a constant time and wrapping the $\mathbb{S}^{d-3}$ inside $AdS_d$, and is thus a hypersurface in the $\{r,u,x,y^a\}$ space. Parameterizing that hypersurface as $u=u(r,x,y^a)$, the area functional becomes (temporarily ignoring the bounds of integration)
\beq
\label{E:areaDefect1}
\mathcal{A} = \text{vol}(\mathbb{S}^{d-3}) \int dy^a \,dx\,dr \, r^{d-3}\rho \left(\frac{f}{u}\right)^{d-2}\sqrt{\text{det}\, G} \sqrt{1+ (\partial_r u)^2 + \frac{f^2}{u^2}\left( \rho^{-2} (\partial_x u)^2 + G^{ab} \partial_a u \partial_b u\right)},
\eeq
where $\int dy^a$ represents integration over all the internal directions $y^a$, $\partial_a u\equiv \partial u/\partial{y^a}$, and we have used the fact that $G_{bc}(x,y^a)$ is positive-definite to define its inverse $G^{bc}(x,y^a)$.

The Euler-Lagrange equation for $u(r,x,y^a)$ that arises from eq.~\eqref{E:areaDefect1} is a complicated second-order partial differential equation. Remarkably, the solution that describes a spherical $\mathcal{M}$ centered on the defect is simple: it is given by~\cite{Jensen:2013lxa}
\beq
\label{eq:minsurface2}
u^2 + r^2 = R^2.
\eeq
In other words, although $u$ could depend on all of $\{r,x,y^a\}$, the $u$ that describes the minimal-area surface depends only on $r$, and in fact is identical in form to the minimal-area solution in the $AdS_d$-slicing of pure $AdS_{d+1}$, eq.~\eqref{eq:minsurface}. A proof that eq.~\eqref{eq:minsurface2} is the global minimum of the area functional, for metrics of the form in eq.~\eqref{E:generalMetric} but without the internal directions $y^a$, appears in appendix A of ref.~\cite{Jensen:2013lxa}. We can easily generalize that proof to include the internal directions $y^a$, as follows. First, in the $(u,r)$ plane we switch to polar coordinates,
\beq
\label{eq:polar}
u = \zeta \sin\varphi, \qquad r = \zeta \cos\varphi,
\eeq
where $\zeta \in [0,\infty)$ and $\varphi\in [0,\pi/2]$. Next, we re-parameterize the hypersurface $u(r,x,y^a)$ as $\zeta = \zeta(\varphi,x,y^a)$, so that the area functional becomes
\beq
\mathcal{A} = \text{vol}(\mathbb{S}^{d-3}) \int dy^a\, dx \, d\varphi \, \rho \,f^{d-2}\frac{\cot^{d-3}\varphi }{\sin\varphi}\sqrt{\text{det}\,G}\sqrt{1+\frac{(\partial_{\varphi}\zeta)^2}{\zeta^2}+\frac{f^2}{\zeta^2}\frac{ \rho^{-2}(\partial_x\zeta)^2 + G^{ab}\partial_a \zeta\partial_b \zeta}{\sin^2\varphi}}.
\eeq
The crucial observation is that $\zeta$ appears only in the terms under the square root, in a sum of squares where each term is proportional to a derivative of $\zeta$. As a result, the area functional attains its global minimum only when $\zeta$ is constant in all variables. Eq.~\eqref{eq:polar} then implies $u^2 + r^2 =\zeta^2$ is constant. To describe a spherical $\mathcal{M}$ of radius $R$ centered on the defect, we choose $\zeta=R$. Eq.~\eqref{eq:minsurface2} is therefore the global minimum of the area functional, among surfaces that asymptotically approach the entangling surface we want.

Plugging the minimal area solution eq.~\eqref{eq:minsurface} into the area functional eq.~\eqref{E:areaDefect1}, changing integration variables from $u(r)$ to $r(u)$, and multiplying by $1/(4G_N)$, we find for the EE
\beq
\label{E:defectSEE}
S = \frac{\mathcal{A}_{min}}{4G_N} = \frac{\text{vol}(\mathbb{S}^{d-3})R}{4G_N}\int dy^a\, dx\, du \, \sqrt{\text{det}\, G}\,\rho f^{d-2} \frac{ (R^2-u^2)^{(d-4)/2}}{u^{d-2}}.
\eeq
The integrand of eq.~\eqref{E:defectSEE} exhibits divergences near the asymptotic boundary, for example the integrand diverges exponentially in $x$ in each asymptotically $AdS_{d+1} \times \mathcal{M}_y$ region at large $|x|$. From the DCFT perspective, these are the expected short-distance divergences of highly-entangled modes near the entangling surface. To obtain a finite EE we must introduce a regulator. As discussed in subsection~\ref{S:systems}, to compute the defect entropy via the background subtraction in eq.~\eqref{E:defDefBdyS}, we must use the same regulator in the DCFT as in the parent CFT. In the previous subsection, for the parent CFT dual to $AdS_{d+1}$ we chose a FG regulator $z = \varepsilon$, which we must therefore also use here.

Any asymptotically $AdS_{d+1}$ metric may be written in FG form, at least locally, in the asymptotically $AdS_{d+1}$ region. Similar to the change of coordinates in $AdS_{d+1}$ from Poincar\'e to $AdS_d$ slicing, eq.~\eqref{AdSFG}, to switch from the form in eq.~\eqref{E:generalMetric} to FG form we must replace the coordinates $\{x,u\}$ with the FG coordinates $\{z,x_{\perp}\}$, where $x_{\perp}$ is the field theory direction normal to the defect. After that change of coordinates, the FG form of the metric in eq.~\eqref{E:generalMetric} will be, in an asymptotically $AdS_{d+1}$ region with radius $L$,
\begin{align}
\begin{split}
\label{E:FGpatch}
g =& \frac{L^2}{z^2}\left( dz^2 + g_1\left( \frac{x_{\perp}}{z},\tilde{y}^a\right) \left(-dt^2+\sum_{i=1}^{d-2}(dx^i)^2\right) + g_2\left( \frac{x_{\perp}}{z},\tilde{y}^a\right)dx_{\perp}^2\right)
\\
& \quad+  \mathcal{G}_a\left( \frac{x_{\perp}}{z},\tilde{y}^a\right)\frac{dx_{\perp} d\tilde{y}^a}{z} +\mathcal{G}_{ab}\left( \frac{x_{\perp}}{z},\tilde{y}^c\right)d\tilde{y^a}d\tilde{y}^b\,,
\end{split}
\end{align}
where in general the internal coordinates $\tilde{y}^a$ will be different from the $y^a$ in eq.~\eqref{E:generalMetric}. In eq.~\eqref{E:FGpatch} the dependence on $x_{\perp}$ and $z$ is fixed by the scale invariance of the DCFT: the warp factors $g_1\left( \frac{x_{\perp}}{z},\tilde{y}^a\right)$, $g_2\left( \frac{x_{\perp}}{z},\tilde{y}^a\right)$, $\mathcal{G}_a\left( \frac{x_{\perp}}{z},\tilde{y}^a\right)$ and $\mathcal{G}_{ab}\left( \frac{x_{\perp}}{z},\tilde{y}^c\right)$ can only depend on the ratio $x_{\perp}/z$, rather than on $x_{\perp}$ and $z$ separately. To guarantee that the metric in the DCFT is conformal to the Minkowski metric, we require that $g_1\left( \frac{x_{\perp}}{z},\tilde{y}^a\right)\to 1$ and $g_2\left( \frac{x_{\perp}}{z},\tilde{y}^a\right)\to 1$ as $z \to 0$. We call any region of spacetime where the map to a FG metric eq.~\eqref{E:FGpatch} exists a ``FG patch.'' In a FG patch, we can perform a FG expansion in powers of $z$ about $z = 0$, and then introduce the FG cutoff $z = \varepsilon$.

\begin{figure}
  \begin{center}
    \includegraphics[width=.6\textwidth]{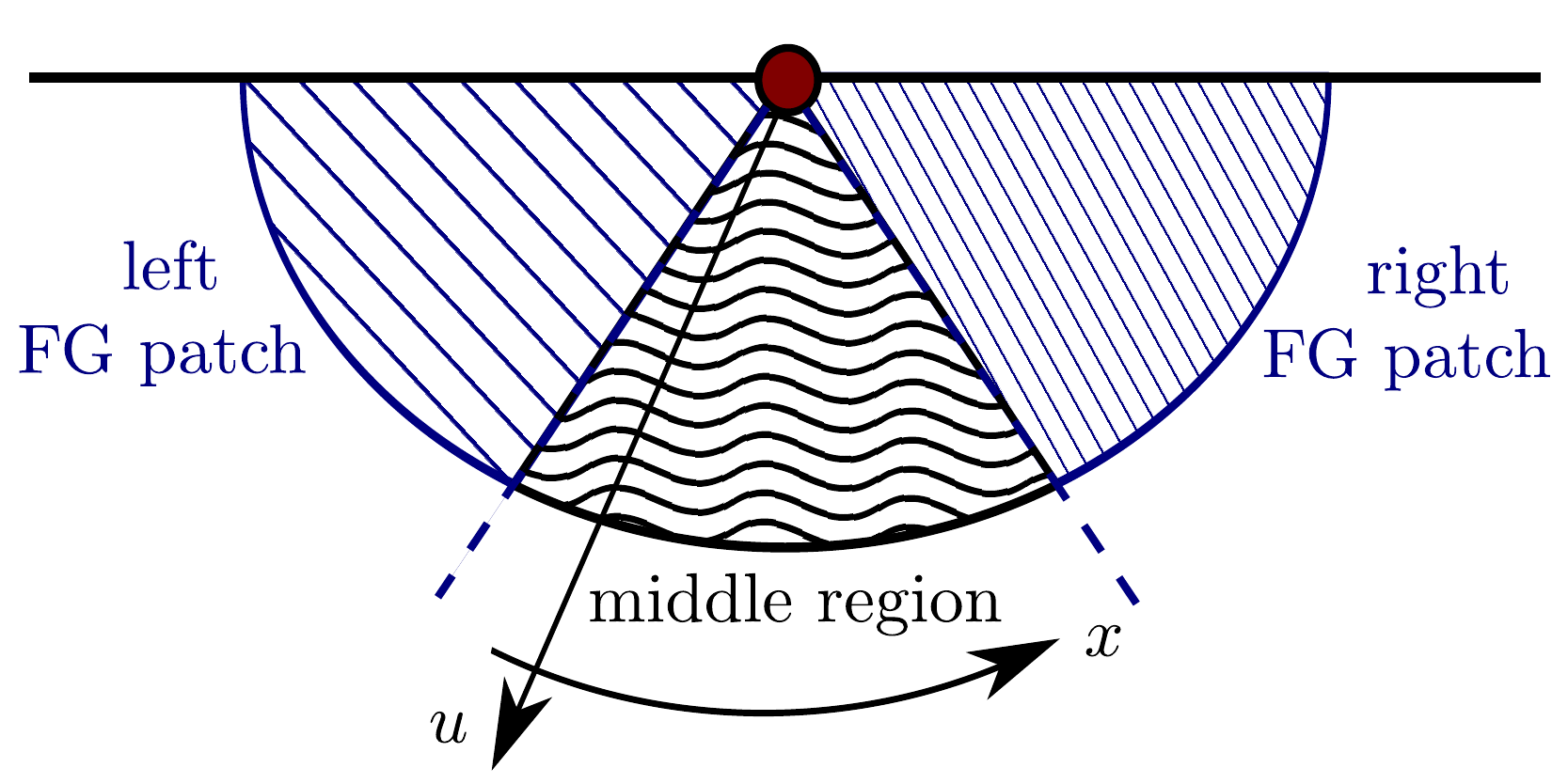}
  \end{center}
  \caption{A schematic depiction of the holographic duals of the $(d+1)$-dimensional DCFTs that we study, which have metrics of the form in eq.~\eqref{E:generalMetric}. We depict the space spanned by the coordinates $u$ and $x$. Fefferman-Graham (FG) patches only exist for some range of $x$ near the asymptotically $AdS_{d+1}$ regions $x \to\pm \infty$. The rest of the geometry is a ``middle region'' between these FG patches.}
\label{fig:FGpatches}
\end{figure}

Crucially, however, the map to the FG patch does not necessarily exist everywhere: the FG expansion may break down~\cite{Papadimitriou:2004rz,Nozaki:2012qd}. For metrics with the FG form in eq.~\eqref{E:FGpatch}, the reason is intuitively obvious: the FG expansion will actually be an expansion in $z/x_{\perp} \ll 1$, so if we ``move too close'' to the defect, $x_{\perp} \to 0$, then $z/x_{\perp}$ will no longer be $\ll 1$, and the expansion may break down. To see how such a breakdown could occur, consider the simple example of a metric of the form in eq.~\eqref{E:generalMetric}, but without any internal $y^a$-directions. For such a metric, we can write the coordinate transformation to FG form eq.~\eqref{E:FGpatch} in closed form:
\begin{equation}
\label{eq:FGtrans1}
z = u \,k_1^{\pm}(x)\,, \qquad x_{\perp} = u\, k_2^{\pm}(x),
\end{equation}
where the $\pm$ correspond to $x \to \pm \infty$, and
\begin{equation}
\label{eq:FGtrans2}
k_1^{\pm}(x) \equiv \exp\left[ \mp \int dx' \frac{\rho}{f}\sqrt{\frac{f^2}{L_{\pm}^2}-1}\right]\,, \qquad k_2^{\pm}(x) \equiv \exp \left[ \mp \int dx'  \frac{\rho}{f}\frac{1}{\sqrt{\frac{f^2}{L_{\pm}^2}-1}}\right].
\end{equation}
In general $f(x)$ decreases as we decrease $x$, moving into the bulk. If $f(x)/L_{\pm}$ becomes $<1$ at some value of $x$, then the square roots in eq.~\eqref{eq:FGtrans2} become imaginary and the coordinate transformation in eq.~\eqref{eq:FGtrans1} ceases to exist. For every geometry we will study in section~\ref{S:examples}, such a breakdown of the FG expansion indeed occurs. As a result, the geometry splits into three regions, two covered by FG patches near $x \to \pm \infty$, which we call the ``right'' ($x \to \infty$) and ``left'' ($x \to -\infty$) FG patches, and a ``middle region'' covering the remaining values of $x$. We illustrate these three regions in fig.~\ref{fig:FGpatches}.

We have not been able to find a closed-form expression for the coordinate transformation that puts the general metric in eq.~\eqref{E:generalMetric} into the FG form of eq.~\eqref{E:FGpatch}, due to the presence of the internal coordinates $y^a$. We have been able to compute the coordinate transformation asymptotically, however, which will suffice for what follows. In other words, we computed the coordinate transformation to FG form order-by-order in large $e^{\pm x}$. The result that will be of most use to us is $x$ in terms of a mix of coordinates from eqs.~\eqref{E:generalMetric} and~\eqref{E:FGpatch}: we will need $x$ in terms of $z/u$ and $y^a$. In terms of the expansion coefficients $\rho^{(1)}_{\pm}(y^a)$, $\rho^{(2)}_{\pm}(y^a)$, and $G^{(0)}_{\pm}(y^a)$ in eq.~\eqref{eq:expansion}, but suppressing their $y^a$-dependence for the sake of brevity, we find
\begin{align}
\label{E:xFromUZ}
x_{\pm}\left( \frac{z}{u},y^a\right) = &\pm \left[ \ln\left( \frac{2u}{z} \right) - c_{\pm} + \frac{e^{c_{\pm}}\rho^{(1)}_{\pm}}{4}\left(\frac{z}{u}\right) \right]
\\
\nonumber
& \pm \left[ \frac{e^{2c_{\pm}}\rho^{(2)}_{\pm}-4}{16}-\frac{e^{2c_{\pm}}}{64}\left( 5 (\rho^{(1)}_{\pm})^2+L_{\pm}^2 (G^{ab}_{\pm})^{(0)}\partial_a \rho^{(1)}_{\pm}\partial_b\rho^{(1)}_{\pm}\right)\right] \left( \frac{z}{u}\right)^2 + \mathcal{O}\left( \frac{z}{u}\right)^3,
\end{align}
where we have fixed some integration constants by demanding that $g_1\left( \frac{x_{\perp}}{z},\tilde{y}^a\right)\to 1$ and $g_2\left( \frac{x_{\perp}}{z},\tilde{y}^a\right)\to 1$ as $z \to 0$.

\begin{figure}
  \begin{center}
    \includegraphics[width=.6\textwidth]{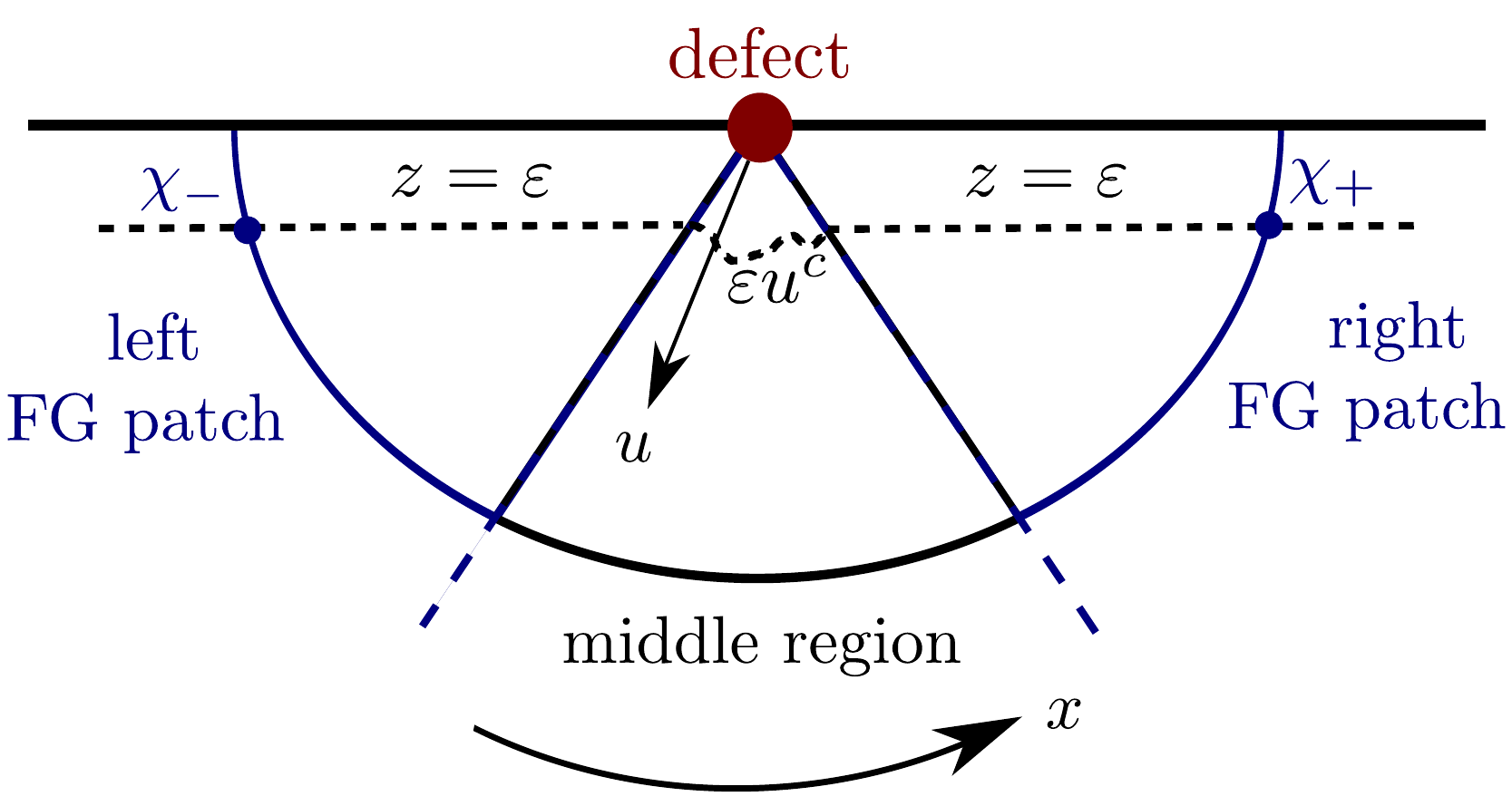}
  \end{center}
  \caption{A schematic depiction of the cutoff surface that we use to regulate the divergences in the spherical EE, eq.~\eqref{E:defectSEE}. In each FG patch, our cutoff surface coincides with the FG cutoff surface $z=\varepsilon$ that we used to regulate the spherical EE when the defect is absent (\textit{i.e.}\ in pure $AdS_{d+1}$). These cutoffs give rise to cutoffs $\chi_{\pm}$ in the $x$ integration in eq.~\eqref{E:defectSEE}. In the middle region between the FG patches, our cutoff surface, parameterized by $u^c(y^a)$, continuously connects the two $z = \varepsilon$ cutoffs, but is otherwise arbitrary. In the appendix we show that our results for the universal terms in the defect or boundary entropy are insensitive to the choice of cutoff surface in the middle region.}
\label{fig:cutoff}
\end{figure}

We can now specify the cutoffs we use to compute the spherical EE in eq.~\eqref{E:defectSEE}. In each FG patch we introduce the FG cutoff surface $z = \varepsilon$. Between the FG patches we will demand that the cutoff surface is continuous, and connects the two $z = \varepsilon$ surfaces of the two FG patches, but is otherwise unconstrained. In practice, in eq.~\eqref{E:defectSEE} we integrate in $x$ only up to cutoffs $\chi_{\pm}$ whose values will depend on $\varepsilon$ as well as $\{u,y^a\}$. Indeed, the scale invariance of the DCFT constrains $\chi_{\pm}$ to be of the form $\chi_{\pm}(\frac{\varepsilon}{u},y^a)$. To implement the FG cutoffs $z = \varepsilon$ in the two FG patches, we take $\chi_{\pm}(\frac{\varepsilon}{u},y^a)$ to be given by eq.~\eqref{E:xFromUZ}, evaluated at $z = \varepsilon$. We integrate in $u$ from a cutoff $u=\varepsilon u^c(y^a)$ to $u=R$. Our only constraint on the $y^a$-dependent cutoff $u^c(y^a)$ is that it continuously connects the $z=\varepsilon$ cutoffs in the FG patches. We summarize these choices as
\beq
\label{E:theBounds}
\chi_{\pm}\left( \frac{\varepsilon}{u},y^a\right) =\left\{ \begin{array}{ll} x_{\pm}\left( \frac{\varepsilon}{u},y^a\right), & \left\{\frac{\varepsilon}{u},y^a\right\}\in \text{ FG patches}, \\ \text{arbitrary but continuous},\quad \phantom{\,} & \left\{\frac{\varepsilon}{u},y^a\right\}\in \text{ middle region},\end{array}\right.
\eeq
where in the second line we mean that the cutoff surface in the middle region must continuously connect the $z = \varepsilon$ surfaces in the two FG patches, but is otherwise arbitrary. We schematically depict our choice of cutoff surface in fig.~\ref{fig:cutoff}.

With our choice of cutoff surface, the integral for the spherical EE in eq.~\eqref{E:defectSEE} becomes
\beq
\label{E:defectSEEfinal}
S= \frac{\text{vol}(\mathbb{S}^{d-3})R}{4G_N} \int dy^a \int_{\varepsilon u^c(y^a)}^R du \int_{\chi_-\left(\frac{\varepsilon}{u},y^a\right)}^{\chi_+\left( \frac{\varepsilon}{u},y^a\right)}dx \,\sqrt{\text{det}\,G}\rho f^{d-2} \frac{(R^2-u^2)^{(d-4)/2}}{u^{d-2}},
\eeq
where the order of the integrations is important: we integrate over $x$ first because $\chi_{\pm}\left(\frac{\varepsilon}{u},y^a\right)$ depend on $u$ and $y^a$, we integrate over $u$ second because $u^c(y^a)$ depends on $y^a$, and we integrate over $y^a$ last. Eq.~\eqref{E:defectSEEfinal}, with the integration bounds eq.~\eqref{E:theBounds}, is the first of the three major results in this section.

Starting from eq.~\eqref{E:defectSEEfinal}, in the appendix we show that the defect entropy, as defined in eq.~\eqref{E:defDefBdyS}, takes the form in eq.~\eqref{E:Sdefect},
\beq
\label{E:Sdefect2}
S_{\textrm{defect}} = \left\{ \begin{array}{lc} D_{log}^{(d=3)} \ln\left( \frac{2R}{\varepsilon}\right) + \tilde{D}_0^{(d=3)}, & \quad d=3, \\ D_1^{(d=4)} \frac{R}{\varepsilon} + D_0^{(d=4)}, & \quad d=4.\end{array}\right.
\eeq
In the appendix we also show that $D_{log}^{(d=3)}$ and $D_0^{(d=4)}$ are universal. In particular, we show that $D_{log}^{(d=3)}$ and $D_0^{(d=4)}$ are independent of our choice of cutoff surface in the middle region between the FG patches, including our choice of $u^c(y^a)$. In the appendix we also show that $D_{log}^{(d=3)}$ and $D_0^{(d=4)}$ are insensitive to terms in $\chi_{\pm}(\frac{\varepsilon}{u},y^a)$ of order $(\varepsilon/u)^3$ or higher, which is why we did not bother to compute any terms of order $(z/u)^3$ or higher in eq.~\eqref{E:xFromUZ}. The take-away message is that $D_{log}^{(d=3)}$ and $D_0^{(d=4)}$ provide physically meaningful information that characterizes the defect. Eq.~\eqref{E:Sdefect2} is the second of the three main results of this subsection.

Let us now turn to BCFTs whose holographic duals have metrics of the form in eq.~\eqref{E:generalMetric}. The analysis here is very similar to the DCFT case above, so we will be brief. For the dual of a BCFT, the bulk geometry will have only a single asymptotic $AdS_{d+1} \times {\mathcal{M}}_y$ region. We will choose the $x$ coordinate so that $x \in (-\infty,\infty)$, with the asymptotic $AdS_{d+1} \times {\mathcal{M}}_y$ region at $x \to \infty$. The geometry will cap off smoothly as $x \to -\infty$. We can cover the asymptotic $AdS_{d+1} \times {\mathcal{M}}_y$ region with a single FG patch, although at some $x$ the FG expansion will break down. The minimal area surface that asymptotically approaches a hemi-spherical entangling surface centered is given by eq.~\eqref{eq:minsurface}, $u^2 + r^2 = R^2$, and the integral for the EE is of the form in eq.~\eqref{E:defectSEE}. That integral diverges, and requires a cutoff. We choose a cutoff surface that agrees with the FG cutoff surface $z = \varepsilon$ in the single FG patch, and that extends continuously outside the FG patch. In practice, we integrate $x$ over $\left(-\infty,\chi_+\left( \frac{\varepsilon}{u},y^a\right)\right]$, where we choose $\chi_+\left( \frac{\varepsilon}{u},y^a\right)=x_+\left( \frac{\varepsilon}{u},y^a\right)$ inside the FG patch. We integrate $u$ over $[\varepsilon u^c(y^a),R]$. The integral for the EE is then identical in form to that in eq.~\eqref{E:defectSEEfinal}, but with $\chi_-\left( \frac{\varepsilon}{u},y^a\right) \to -\infty$, where this $x \to -\infty$ endpoint of the $x$ integration does not produce a divergence. Starting from this integral, in the appendix we compute the boundary entropy $S_{\partial}$, as defined in eq.~\eqref{E:defDefBdyS}, which takes the form in eq.~\eqref{E:Sbdy}:
\beq
\label{E:Sbdy2}
S_{\partial} =  \left\{ \begin{array}{lc} B_{log}^{(d=3)} \ln\left( \frac{2R}{\varepsilon}\right) + \tilde{B}_0^{(d=3)}, & \quad d=3, \\ B_1^{(d=4)} \frac{R}{\varepsilon} + B_0^{(d=4)}, & \quad d=4. \end{array}\right.
\eeq
In the appendix we show that $B_{log}^{(d=3)}$ and $B_0^{(d=4)}$ are universal. In particular, $B_{log}^{(d=3)}$ and $B_0^{(d=4)}$ are independent of our choice of cutoff surface, including our choice of $u^c(y^a)$, and are insensitive to terms in $\chi_+\left( \frac{\varepsilon}{u},y^a\right)$ of order $(\varepsilon/u)^3$ or higher. The take-away message is that  $B_{log}^{(d=3)}$ and $B_0^{(d=4)}$ provide physically meaningful information that characterizes the boundary. Eq.~\eqref{E:Sbdy2} is the third of the three main results of this subsection.

%%%%%%%%%%%%%%%%%%%%%%%%%%%%%%%%%%%%%%%%%%%%%%%%%%%%%%%%%%%%%%%%%%%%%%%%%%%%%%%%%%%%%%%%%%%%%%%%%%%%%%%%%%%%%%%%%%
\section{Examples}
\label{S:examples}
%%%%%%%%%%%%%%%%%%%%%%%%%%%%%%%%%%%%%%%%%%%%%%%%%%%%%%%%%%%%%%%%%%%%%%%%%%%%%%%%%%%%%%%%%%%%%%%%%%%%%%%%%%%%%%%%%%

In this section we compute the defect or boundary entropy, $S_{\textrm{defect}}$ or $S_{\partial}$, for several examples of DCFTs and BCFTs in $d=3$ and $d=4$ holographically dual to type IIB string theory or M-theory. Actually, we will compute only the universal terms in $S_{\textrm{defect}}$ or $S_{\partial}$: in the $S_{\textrm{defect}}$ of eq.~\eqref{E:Sdefect} these are $D_{log}^{(d=3)}$ and $D_0^{(d=4)}$, and in the $S_{\partial}$ of eq.~\eqref{E:Sbdy} this is $B_0^{(d=4)}$. Our examples do not include a BCFT in $d=3$, so we present no examples of $B_{log}^{(d=3)}$ in eq.~\eqref{E:Sbdy}. In each example we also discuss the physics of our results. In particular, in one class of examples, the D3/D5 DCFTs~\cite{Karch:2000gx,DeWolfe:2001pq,Erdmenger:2002ex,Gomis:2006cu,D'Hoker:2007xy,D'Hoker:2007xz} and BCFTs~\cite{Aharony:2011yc}, we will show that $-D_0^{(d=4)}$ or $-B_0^{(d=4)}$ decreases monotonically under a certain class of defect or boundary RG flows, and may either increase or decrease under a certain class of RG flows in the ambient CFT.

In all of our examples, the bulk metric is of the form in eq.~\eqref{E:generalMetric}. Our task is to evaluate the integral for the (hemi-)spherical EE, eq.~\eqref{E:defectSEEfinal}, with the cutoffs described in subsection~\ref{S:generalDefect}, or at least to extract from the integral the universal terms in $S_{\textrm{defect}}$ or $S_{\partial}$.

%%%%%%%%%%%%%%%%%%%%%%%%%%%%%%%%%%%%%%%%%%%%%%%%%%%%%%%%%%%%%%%%%%%%%%%%%%%%%%%%%%%%%%%%%%%%%%%%%%%%%%%%%%%%%%%%%%
\subsection{D3/D5 DCFT and BCFT}
\label{S:D3D5}
%%%%%%%%%%%%%%%%%%%%%%%%%%%%%%%%%%%%%%%%%%%%%%%%%%%%%%%%%%%%%%%%%%%%%%%%%%%%%%%%%%%%%%%%%%%%%%%%%%%%%%%%%%%%%%%%%%

Our first example of a DCFT is $\N=4$ SYM theory with gauge group $SU(N_3)$ coupled to a number $N_5$ of $(2+1)$-dimensional hypermultiplets in the fundamental representation of $SU(N_3)$~\cite{Karch:2000gx}. We take these flavor fields to be restricted to a planar defect, which we take to be at $x^3=0$ without loss of generality. The classical Lagrangian for this theory appears in refs.~\cite{DeWolfe:2001pq,Erdmenger:2002ex}. The hypermultiplets preserve eight real supercharges, $SO(3,2)$ defect conformal symmetry, and an $SO(3) \times SO(3)$ subgroup of the original $SO(6)$ R-symmetry~\cite{DeWolfe:2001pq,Erdmenger:2002ex}. In other words, the hypermultiplets preserve $OSp(4|4,\mathbb{R})$ superconformal symmetry.

A novel feature of this DCFT is a Higgs branch of vacua including a subset of vacua in which the rank of the gauge group is different on the two sides of the defect~\cite{Hanany:1996ie}. More precisely, this subset of Higgs vacua describe $\N=4$ SYM coupled to defect hypermultiplets with gauge group $SU(N_3^+)$ on one side of the defect ($x^3>0$) and gauge group $SU\left(N_3^-\right)$ on the other side ($x^3<0$), with $N_3^+ \neq N_3^-$. Detailed discussions of these vacua appear in refs.~\cite{Gaiotto:2008sa,Gaiotto:2008ak}. At large $N_3^{\pm}$ and large coupling, the holographic duals (discussed below) indicate that this subset of Higgs vacua preserve defect conformal symmetry, which is perhaps counter-intuitive, since normally a scalar expectation value breaks scale invariance. As explained in ref.~\cite{McAvity:1995zd}, however, defect conformal symmetry allows a primary scalar operator of dimension $\Delta$ to have a non-zero one-point function $\propto (x^3)^{-\Delta}$. To our knowledge, whether this subset of Higgs vacua preserves defect conformal symmetry for all values of $N_3^{\pm}$ and 't Hooft coupling is an open question.

These DCFTs appear in string theory as the low-energy field theory living at the $(2+1)$-dimensional intersection of $N_3^{\pm}$ D3-branes and $N_5$ D5-branes, with $\Delta N_3\equiv N_3^+-N_3^-$ D3-branes ending on the D5-branes. When $N_3^{\pm}$ and $N_5$ are small, so that the D3- and D5-branes are probes in $(9+1)$-dimensional Minkowski space, the intersection is that of table~\ref{T:D3D5}, with the D5-branes at $x^3=0$ and with $N_3^+$ or $N_3^-$ D3-branes located in the half-spaces $x^3>0$ or $x^3<0$, respectively.
\begin{table}[ht]
\begin{center}
\begin{tabular}{c|c|c|c|c|c|c|c|c|c|c}
 & $x^0$ & $x^1$ & $x^2$ & $x^3$ & $x^4$ & $x^5$ & $x^6$ & $x^7$ & $x^8$ & $x^9$ \\
 \hline
 $N_3^{\pm}$ D3 & X & X & X & X  & & & & & &\\
 \hline
 $N_5$ D5 & X & X & X & & X & X & X & &  \\
 \hline
\end{tabular}
\caption{\label{T:D3D5} The $(2+1)$-dimensional D3/D5 intersection that we study in this subsection. An `X' denotes a direction in which the corresponding brane is extended. We introduce $N_5$ D5-branes at $x^3=0$, with $N_3^+$ D3-branes in the $x^3>0$ region and $N_3^-$ D3-branes in the $x^3<0$ region, where $\Delta N_3\equiv N_3^+-N_3^-$ D3-branes end on the D5-branes.}
\end{center}
\end{table}
The brane intersection preserves the $ISO(1,2) \times SO(3) \times SO(3)$ subgroup of $SO(9,1)$ along with eight real supercharges. At the IR fixed point the symmetry is enhanced to $OSp(4|4,\mathbb{R})$ superconformal symmetry.

In the decoupling and Maldacena limits, the D3/D5 intersection gives rise to an Einstein-frame metric of the form in eq.~\eqref{E:generalMetric}~\cite{Gomis:2006cu,D'Hoker:2007xy,D'Hoker:2007xz}
\begin{align}
\label{susymet}
g = f_4^2 \, g_{AdS_4} + \rho^2 \, dv d\bar{v} + f_1^2 \, g_{\mathbb{S}^2} + f_2^2 \, g_{\bar{\mathbb{S}}^2}\,,
\end{align}
where $g_{\mathbb{S}^2}$ and $g_{\bar{\mathbb{S}}^2}$ denote unit-radius metrics of two different $\mathbb{S}^2$'s and $v = x + iy$ is a complex coordinate on an infinite strip, $x \in (-\infty,\infty)$ and $y \in [0,\pi/2]$. Thanks to SUSY, the warp factors $f_4^2$, $\rho^2$, $f_1^2$, and $f_2^2$ are completely determined by two real functions $h_1(v,\bar{v})$ and $h_2(v,\bar{v})$ that are harmonic on the two-dimensional space spanned by $v$ and $\bar{v}$~\cite{Gomis:2006cu,D'Hoker:2007xy}, via
\begin{subequations}
\label{susydef}
\begin{align}
f_4^8 &= 16 \, \frac{F_1 F_2}{w^2}\,, &  \rho^8 &= \frac{2^8 \, F_1 F_2 \, w^2}{h_1^4 \, h_2^4}\,,
\\
f_1^8& = 16 \, h_1^8 \, \frac{F_2 \, w^2}{F_1^3}\,, & f_2^8 &= 16 \, h_2^8 \, \frac{F_1 \, w^2}{F_2^3}\,,
\end{align}
\beq
F_i \equiv 2 h_1 h_2 \, |\partial_v h_i |^2 - h_i^2 \, w\,, \quad (i=1,2) \qquad w \equiv \partial_v\partial_{\bar{v}}(h_1h_2)\,.
\eeq
\end{subequations}
As shown in refs.~\cite{Gomis:2006cu,D'Hoker:2007xy}, the type IIB SUGRA solution also includes a non-trivial dilaton $\phi$ and non-trivial Ramond-Ramond (RR) three- and five-forms, which are also completely determined by $h_1(v,\bar{v})$ and $h_2(v,\bar{v})$. To compute EE we will only need the metric in eq.~\eqref{susymet} and, to translate our results to field theory quantities, the dilaton, which is given by
\beq
\label{E:dilaton}
e^{4\phi} = \frac{F_2}{F_1}\,.
\eeq
Invoking standard arguments, we expect type IIB SUGRA in this background to be holographically dual to the D3/D5 DCFT. In particular, the $SO(3,2)$ defect conformal symmetry is dual to the isometry of the $AdS_4$ slice and the $SO(3) \times SO(3)$ global symmetry is dual to the isometry of the two $\mathbb{S}^2$'s.

Actually, the D3/D5 BCFT that we will study later in this subsection and the SUSY DCFTs that we will study in subsections~\ref{ssec:tsun} and~\ref{ssec:Janus} also have $SO(3,2) \times SO(3) \times SO(3)$ symmetry and are dual to type IIB SUGRA, with $g$ and $\phi$ of the forms given in eqs.~\eqref{susymet}, \eqref{susydef}, and~\eqref{E:dilaton}. What distinguishes the various solutions are the harmonic functions $h_1(v,\bar{v})$ and $h_2(v,\bar{v})$, as we will see.

For the dual of the D3/D5 DCFT, the harmonic functions are~\cite{D'Hoker:2007xz}
\begin{align}
\begin{split}
\label{E:D3D5h}
	h_1(v,\bar{v}) =& \alpha' \left[ - i\, \alpha \sinh(v) - \frac{N_5}{4} \ln \left( \tanh \left( \frac{i \pi}{4} - \frac{v-\delta}{2} \right) \right) \right] +  \text{c.c.}\,,
\\
	h_2(v,\bar{v}) =& \alpha' \hat \alpha \cosh(v) + \text{c.c.}\,,
\end{split}
\end{align}
where $\alpha$, $\hat{\alpha}$, and $\delta$ are real parameters whose meaning we discuss below.  Crucially, we must have $\alpha \geq 0$ and $\hat{\alpha}\geq0$. Taking $N_5 = 0$ reproduces $AdS_5 \times \mathbb{S}^5$ supported by $N_3 = \frac{1}{2} \left(N_3^+ + N_3^-\right)$ units of RR five-form flux sourced by the D3-branes, with $N_3^+=N_3^-$. When $N_5 \neq 0$, the geometry has two asymptotically $AdS_5\times \mathbb{S}^5$ regions at $x\to\pm\infty$, that is, as $x \to \pm \infty$ the metric approaches the form in eq.~\eqref{eq:expansion},
\begin{align}
\begin{split}
\label{eq:frhoGasymp}
f(x,y^a)^2 &= \frac{L_{\pm}^2}{4}\left[ e^{\pm 2x+2c_{\pm}} + {\mathcal{O}}\left(e^{\pm x}\right)\right]\,,
\\
\rho(x,y^a)^2 & = L_{\pm}^2 \left[1 + \rho^{(1)}_{\pm}(y^a) \, e^{\mp x} + \rho^{(2)}_{\pm}(y^a) \, e^{\mp 2x} + {\mathcal{O}}\left(e^{\mp 3x}\right)\right]\,,
\\
G(x,y^a) & = G_{\pm}^{(0)}(y^a) + G_{\pm}^{(1)}(y^a) \, e^{\mp x} + {\mathcal{O}}\left(e^{\mp 2x}\right)\,,
\end{split}
\end{align}
with the specific values
\begin{subequations}
\beq
\label{E:D3D5Match1}
L_{\pm}^4 = 8 \hat{\alpha}(e^{\pm \delta} N_5 + 2\alpha)(\alpha')^2\,, \qquad e^{2c_{\pm}} = \frac{2\alpha}{e^{\pm \delta}N_5+2\alpha}\,,
\eeq
\beq
\rho_{\pm}^{(1)}(y^a) = 0\,, \qquad \rho_{\pm}^{(2)}(y^a) = \frac{e^{\pm 2\delta}N_5(N_5\mp 4\alpha\sinh\delta)}{2\alpha(e^{\pm\delta}N_5+2\alpha)}\, \cos(2y)\,,
\eeq
\beq
G_{\pm}^{(0)}(y^a) = L_{\pm}^2 \left[ dy^2 + \sin^2(y) \, g_{\mathbb{S}^2} + \cos^2(y) \, g_{\bar{\mathbb{S}}^2}  \right] = L_{\pm}^2 \, g_{\mathbb{S}^5}\,, \qquad G_{\pm}^{(1)}(y^a) = 0\,.
\eeq
\end{subequations}
As $x \to \pm \infty$, the dilaton approaches
\beq
\label{eq:D3D5dilaton}
e^{2\phi} = \frac{\hat{\alpha}}{\alpha} +  \mathcal{O}\left(e^{\mp x}\right)\,,
\eeq
so in each asymptotically $AdS_5 \times \mathbb{S}^5$ region we identify the string coupling as $g_s = \hat{\alpha}/\alpha$.\footnote{We follow the conventions of ref.~\cite{D'Hoker:2007xy}, where $\phi$ is related to the standard dilaton by a factor of two, so that the string coupling $g_s$ is given by the asymptotic value of $e^{2\phi}$.}

We can determine the bulk parameters $\{\alpha,\hat{\alpha},\delta\}$ in terms of the field theory parameters $\{g_{YM}^2,N_3^{\pm},N_5\}$ as follows. Using $g_s = \hat{\alpha}/\alpha$ and $g_{YM}^2 =4\pi g_s$, we find $\hat{\alpha} = g_{YM}^2 \alpha/(4\pi)$. From $L_{\pm}^4$ in eq.~\eqref{E:D3D5Match1}, and using $L^4_{\pm} = 4\pi N_3^{\pm} (\alpha')^2$, $N_3 = \frac{1}{2} \left(N_3^++N_3^-\right)$, and $\hat{\alpha} = g_{YM}^2 \alpha/(4\pi)$, we find
\beq
\label{E:alpha}
4\pi N_3 = \frac{2g_{YM}^2\alpha}{\pi}(N_5\cosh(\delta) +2\alpha)\,,
\eeq
which we can solve for $\alpha$ as a function of $\{g_{YM}^2,N_3^{\pm},N_5\}$ and $\delta$,
\beq
\label{E:alphasol}
\alpha = - \frac{N_5}{4}\cosh(\delta) + \sqrt{\frac{\pi^2 N_3}{g_{YM}^2} +\frac{N_5^2}{16}\cosh^2(\delta)}\,,
\eeq
and we chose the positive branch of the square root to guarantee $\alpha \geq 0$. Returning to the $L_{\pm}^4$ in eq.~\eqref{E:D3D5Match1} and using $\Delta N_3 = N_3^+ - N_3^-$, we find
\beq
\label{E:D3D5deltaMatch}
\pi^2 \Delta N_3 = g_{YM}^2\alpha N_5\sinh(\delta)\,,
\eeq
which leads to four branches of solutions for $e^{\delta}$. The physical branch is
\beq
\label{eq:deltadef}
e^{\delta} = \sqrt{\frac{2g_{YM}^2N_3N_5^2+4\pi^2 \Delta N_3^2+\sqrt{(2g_{YM}^2N_3N_5^2+4\pi^2 \Delta N_3^2)^2-  g_{YM}^4N_5^4 (4N_3^2-\Delta N_3^2)}}{g_{YM}^2N_5^2(2N_3-\Delta N_3)}}\,,
\eeq
where we have chosen the positive branches of both square roots to guarantee $e^{\delta} > 0$, and so that $e^{\delta}\geq 1$ for $\Delta N_3\geq 0$ while $e^{\delta}\in (0,1)$ for $\Delta N_3 < 0$, as dictated by eq.~\eqref{E:D3D5deltaMatch}. The bulk parameters $\{\alpha,\hat{\alpha},\delta\}$ are thus uniquely determined by the field theory parameters: given $\{g_{YM}^2,N_3^{\pm},N_5\}$, eq.~\eqref{eq:deltadef} gives us $\delta$, which we then insert into eq.~\eqref{E:alphasol} to determine $\alpha$, and from that $\hat{\alpha} = g_{YM}^2 \alpha/(4\pi)$. The explicit expressions for $\{\alpha,\hat{\alpha},\delta\}$ in terms of $\{g_{YM}^2,N_3^{\pm},N_5\}$ are cumbersome and unilluminating, so we will omit writing them in full generality. We will only present their explicit forms at leading order in the $\Delta N_3 \ll 1$ or equivalently $\delta \ll 1$ limit,
\begin{align}
\begin{split}
\label{E:D3D5simpleBulk}
\delta &= \frac{4\pi^2 \Delta N_3}{N_5}\frac{1}{  \xi-g_{YM}^2N_5 } + \mathcal{O}\left(\frac{\Delta N^3}{N_5^3\xi^3}\right)\,,
\\
\alpha &=\frac{\xi-g_{YM}^2N_5}{4g_{YM}^2} - \frac{2\pi^3 \Delta N_3^2}{N_5}\frac{1}{\xi(\xi-g_{YM}^2N_5)} + \mathcal{O}\left( \frac{\Delta N_3^4}{N_5^3\xi^4}\right)\,,
\\
\hat{\alpha} &= \frac{g_{YM}^2}{4\pi} \alpha\,,
\end{split}
\end{align}
where for notational convenience we have defined
\beq
\label{E:xiDef}
\xi^2 \equiv 16\pi^2 g_{YM}^2 N_3 + (g_{YM}^2N_5)^2\,.
\eeq

Our one and only example of a BCFT is the D3/D5 BCFT, obtained in string theory as the low-energy theory on $N_3$ coincident D3-branes that end on $N_5$ D5-branes. This BCFT is $\N=4$ SYM with gauge group $SU(N_3)$ on a half space $x^3 \geq 0$ coupled to $N_5$ hypermultiplets localized at the boundary $x^3=0$. The D3/D5 BCFT preserves eight real supercharges and $SO(3,2) \times SO(3) \times SO(3)$ bosonic symmetry, and at large $N_3$ and large 't Hooft coupling is dual to type IIB SUGRA in a background of the form in eqs.~\eqref{susymet},~\eqref{susydef}, and~\eqref{E:dilaton}, with harmonic functions~\cite{Aharony:2011yc}
\begin{align}
\begin{split}
\label{E:D3D5boundaryHarmonic}
h_1(v,\bar{v}) & = \alpha'\left[- \frac{i\underline{\alpha}}{2}e^v - \frac{N_5}{4}\ln\left( \tanh\left( \frac{i\pi}{4}-\frac{v}{2}\right)\right)\right] + \text{c.c.}\,,
\\
h_2(v,\bar{v}) & = \alpha' \frac{\hat{\underline{\alpha}}}{2}e^v + \text{c.c.}\,,
\end{split}
\end{align}
with real parameters $\{\underline{\alpha},\hat{\underline{\alpha}}\}$.

In the bottom-up holographic models of BCFTs of refs.~\cite{Takayanagi:2011zk,Fujita:2011fp,Nozaki:2012qd}, the field theory's spatial boundary gives rise in the holographic dual to a ``brane'' on which the bulk spacetime ends. In contrast, in the dual of the D3/D5 BCFT the bulk spacetime does not end on a ``brane,'' but caps off smoothly~\cite{Assel:2011xz}: if in eq.~\eqref{E:D3D5boundaryHarmonic} we change coordinates as
\beq
r^2 = \frac{2 N_5 e^{2(x-\delta)}}{N_5 + e^{\delta}\underline{\alpha}},
\eeq
then as $x \to -\infty$ or equivalently $r \to 0$, the metric approaches
\beq
ds^2 = L_+^2 \left [ g_{AdS_4} \, + \, dr^2 \, + \, r^2 \left( dy^2 + \sin^2(y) \, g_{\mathbb{S}^2} + \cos^2(y) \, g_{\mathbb{S}^2}\right) \right],
\eeq
with $L_+^4 =   8 \hat{\underline{\alpha}} e^{\delta} N_5 (\alpha')^2$. Clearly the spacetime caps off smoothly as $r \to 0$.

We can obtain the D3/D5 BCFT from the D3/D5 DCFT by sending the number of D3-branes on one side of the D5-branes to zero. To be concrete, we will take $N_3^-\to0$ while keeping $N_3^+$ fixed. In that limit the harmonic functions corresponding to the D3/D5 DCFT, eq.~\eqref{E:D3D5h}, reduce to those of the D3/D5 BCFT, eq.~\eqref{E:D3D5boundaryHarmonic}, as we will now show. The radius $L_-$ of the asymptotically $AdS_5 \times \mathbb{S}^5$ region at $x \to -\infty$ is related to the number of D3-branes there as $L_-^4 = 4\pi N_3^-(\alpha')^2$. The $N_3^- \to 0$ limit thus implies $L_-\to 0$, which by eqs.~\eqref{E:D3D5Match1} and~\eqref{E:alpha} means we must take $\alpha \to 0$ and $\hat{\alpha}\to 0$ while keeping fixed
\begin{equation*}
\alpha e^{\delta}=\frac{2\pi^2 N_3^+}{g_{YM}^2 N_5}\,, \qquad \textrm{and} \qquad \frac{\hat{\alpha}}{\alpha}=\frac{g_{YM}^2}{4\pi}\,.
\end{equation*}
In this limit, $\delta \to \infty$, and upon defining $\tilde{v}\equiv v - \delta = (x - \delta) + i y$, the harmonic functions corresponding to the D3/D5 DCFT, eq.~\eqref{E:D3D5h}, become
\begin{align}
\begin{split}
\label{E:harmoniclimit}
h_1(\tilde{v},\bar{\tilde{v}}) & = \alpha'\left[- \frac{i(\alpha e^{\delta})}{2}e^{\tilde{v}} -\frac{N_5}{4}\ln\left( \tanh\left( \frac{i\pi}{4}-\frac{\tilde{v}}{2}\right)\right) \right]+ \frac{i (\alpha e^{\delta})\alpha'}{2}e^{-\tilde{v}-2\delta}+\text{c.c.}\,,
\\
h_2(\tilde{v},\bar{\tilde{v}}) & = \alpha'\frac{g_{YM}^2(\alpha e^{\delta})}{8\pi}e^{\tilde{v}} + \alpha' \frac{g_{YM}^2 (\alpha e^{\delta})}{8\pi}e^{-\tilde{v}-2\delta} + \text{c.c.}\,.
\end{split}
\end{align}
Dropping the $e^{-\tilde{v}-2\delta}$ terms, which are exponentially suppressed as $\delta \to \infty$, and identifying
\beq
\alpha e^{\delta} = \underline{\alpha}, \qquad \frac{g_{YM}^2 \alpha e^{\delta} }{4\pi}= \hat{\underline{\alpha}}, \qquad \tilde{v} = v\,,
\eeq
we see that the harmonic functions in eq.~\eqref{E:harmoniclimit} are precisely those corresponding to the D3/D5 BCFT, eq.~\eqref{E:D3D5boundaryHarmonic}, as advertised. In what follows we will thus obtain results for the D3/D5 BCFT by working with the D3/D5 DCFT and then taking the limit above.

The SUGRA duals of the D3/D5 DCFT and BCFT exhibit characteristic D5-brane singularities: both $\exp(2\phi)$ and the Einstein-frame metric go to zero at the D5-branes. As a result, near the D5-branes stringy corrections remain small but curvature corrections must become important. Currently the form of these curvature corrections is unknown, so for now we will simply work within the SUGRA approximation. Because the Einstein-frame metric vanishes at the D5-branes, the area density of the minimal surface (\textit{i.e.}\ the integrand in eq.~\eqref{E:defectSEE}) is integrable at the D5-branes, so in practice the curvature singularity presents no obstruction to our holographic calculation of the EE. We hasten to emphasize, however, that we do not understand what role the curvature singularity plays, if any, when accounting for higher-derivative corrections in the holographic calculation of the EE.

%%%%%%%%%%%%%%%%%%%%%%%%%%%%%%%%%%%%%%%%%%%%%%%%%%%%%%%%%%%%%%%%%%%%%%%%%%%%%%%%%%%%%%%%%%%%%%%%%%%%%%%%%%%%%%%%%%
\subsubsection{The Defect and Boundary Entropies}
%%%%%%%%%%%%%%%%%%%%%%%%%%%%%%%%%%%%%%%%%%%%%%%%%%%%%%%%%%%%%%%%%%%%%%%%%%%%%%%%%%%%%%%%%%%%%%%%%%%%%%%%%%%%%%%%%%

For geometries of the form in eq.~\eqref{susymet} the integral for the spherical EE, eq.~\eqref{E:defectSEEfinal}, is
\beq
\label{E:susyEE}
S = \frac{\text{vol}(\mathbb{S}^1)\text{vol}(\mathbb{S}^2)^2 R}{4G_N}\int_0^{\frac{\pi}{2}}dy \int_{\varepsilon u^c(y)}^R \frac{du}{u^2} \, \int_{\chi_-\left(\frac{\varepsilon}{u},y\right)}^{\chi_+\left(\frac{\varepsilon}{u},y\right)} dx \, (f_4f_1f_2\rho)^2\,,
\eeq
where the ten-dimensional Newton's constant $G_N$ is given by $4 G_N = 2^5 \pi^6 (\alpha')^4$. The integrand of eq.~\eqref{E:susyEE} takes a simple form when written in terms of the harmonic functions $h_1(v,\bar{v})$ and $h_2(v,\bar{v})$,
\beq
\label{E:susyAreaDensity}
(f_4f_1f_2\rho)^2 = - 2^5 \, h_1 h_2 \, w = - 2^5 \, h_1 h_2 \, \partial_v\partial_{\bar{v}}(h_1h_2)\,.
\eeq
For the dual of the D3/D5 DCFT the harmonic functions are those in eq.~\eqref{E:D3D5h}, which give
\beq
\label{E:D3D5integrand}
(f_4f_1f_2\rho)^2 = \mathcal{F}_0 + N_5 \mathcal{F}_1 + N_5^2 \mathcal{F}_2\,,
\eeq
\begin{align}
\begin{split}
\mathcal{F}_0 &\equiv 2^8 \alpha^2\hat{\alpha}^2(\alpha')^4 \cosh^2(x) \cos^2(y) \sin^2(y)\,,
\\
\mathcal{F}_1 & \equiv\! 2^5 \alpha\hat{\alpha}^2(\alpha')^4\!\cosh (x)\! \cos^2(y)\!\sin(y)\!\!\left[\! \frac{4\cosh(2x)\!\cosh(2x\!-\!\delta)\!\sin(y)}{ \cos(2y) + \cosh(2(x-\delta))} \!- \!\ln\! \left| \tanh \!\left( \!\frac{i\pi}{4}\!-\!\frac{v-\delta}{2}\right)\!\right|^2\right],
\\
\mathcal{F}_2 &\equiv - \frac{2^4 \hat{\alpha}^2(\alpha')^4\cosh(x)\cosh(2x-\delta)\cos^2(y)\sin(y)}{\cos(2y)+\cosh(2(x-\delta))}\ln\left|\tanh\left( \frac{i\pi}{4}-\frac{v-\delta}{2}\right)\right|^2\,. \nonumber
\end{split}
\end{align}

As explained in subsection~\ref{S:generalDefect}, we obtain the $x$-cutoffs $\chi_{\pm}\left(\frac{\varepsilon}{u},y^a\right)$ by inserting $c_{\pm}$, $\rho_{\pm}^{(1)}(y^a)$, $\rho_{\pm}^{(2)}(y^a)$, and $G_{\pm}^{(0)}(y^a)$ from eq.~\eqref{E:D3D5Match1} into eq.~\eqref{E:xFromUZ} and taking $z = \varepsilon$:
\bea
\label{E:D3D5Match2}
\chi_{\pm}\left(\frac{\varepsilon}{u},y^a\right) &=& \pm \left[ \ln\left( \frac{2u}{\varepsilon}\right) - c_{\pm} + \frac{e^{2c_{\pm}}\rho_{\pm}^{(2)}(y^a)-4}{16}\left(\frac{\varepsilon}{u}\right)^2+{\mathcal{O}}\left( \frac{\varepsilon^4}{u^4}\right)\right] \nonumber \\
& \equiv & \pm \left[  \ln\left( \frac{2u}{\varepsilon}\right) - \frac{1}{2} \ln \left(\frac{2 \alpha}{e^{\pm \delta} N_5 + 2 \alpha}\right) +\mathcal{C}_{\pm}^{(2)}(y)\left(\frac{\varepsilon}{u}\right)^2+{\mathcal{O}}\left( \frac{\varepsilon^4}{u^4}\right)\right],
\eea
where for later convenience we have defined
\beq
\label{E:D3D5Match3}
\mathcal{C}_{\pm}^{(2)}(y) \equiv \frac{e^{\pm 2\delta}N_5 (N_5\mp 4\alpha\sinh \delta)}{16(e^{\pm\delta}N_5+2\alpha)^2} \cos(2y) - \frac{1}{4}\,.
\eeq

We now proceed to evaluate the integral for $S$ in eq.~\eqref{E:susyEE}. The integral exhibits divergences in the $\varepsilon\to 0$ limit, which from the bulk perspective are infinite volume divergences from the large-$|x|$, asymptotically $AdS_5 \times \mathbb{S}^5$ regions, and from the SYM perspective are the divergences in the EE from highly-entangled modes near the entangling surface. To isolate the divergences, in the large-$|x|$ regions we split the integrand in eq.~\eqref{E:D3D5integrand} as
\beq
\label{eq:split1}
(f_4f_1f_2\rho)^2 = A_{\pm}^{(-2)}(y) e^{\pm 2x + 2c_{\pm}} + A_{\pm}^{(0)} (y) + A_{\pm}(x,y)\,,
\eeq
where the only information we will need about $A_{\pm}(x,y)$ is its leading asymptotic behavior at large $|x|$, which is $\exp(\mp 2x)$, and
\begin{align}
\begin{split}
\label{eq:split2}
A_{\pm}^{(-2)}(y) &= \frac{L_{\pm}^8}{4}\cos^2(y)\,\sin^2(y)\,,
\\
 A_{\pm}^{(0)}(y) &= \frac{1}{2} \, \cos^2(y)\,\sin^2(y) \, \left(L_{\pm}^8 - 2^7 e^{\pm 3\delta}N_5 \, \alpha \, \hat{\alpha}^2(\alpha')^4 \cos(2y)\right)\,,
\end{split}
\end{align}
which will ultimately give rise to $R^2/\varepsilon^2$ and $\ln \left(\frac{\varepsilon}{2R}\right)$ divergences in $S$, respectively. Next we split the integration over $x$ into three domains: $[\chi_-,x^c_-]$, $[x^c_-,x^c_+]$, and $[x^c_+,\chi_+]$, where $x^c_{\pm}$ are arbitrary, and may be set to any convenient values. Obviously the final result for $S$ cannot depend on the choices of $x^c_{\pm}$. The integral for $S$ correspondingly splits into three terms,
\beq
S = S_- + S_0 + S_+\,.
\eeq

For $S_{\pm}$ we find, using eqs.~\eqref{eq:split1} and~\eqref{eq:split2},
\begin{align}
\label{eq:spm}
\nonumber
S_{\pm} & \equiv\pm  \frac{\text{vol}(\mathbb{S}^1)\text{vol}(\mathbb{S}^2)^2 R}{4G_N}\int_0^{\frac{\pi}{2}}dy \int_{\varepsilon u^c(y)}^R \frac{du}{u^2} \, \int_{x^c_{\pm}}^{\chi_{\pm}\left(\frac{\varepsilon}{u}\right)} dx \, (f_4f_1f_2\rho)^2 \\
\nonumber
& =\! \frac{\text{vol}(\mathbb{S}^1)\text{vol}(\mathbb{S}^2)^2}{4G_N}\! \! \int_0^{\frac{\pi}{2}} \!\! dy \left [ A_{\pm}^{(-2)}(y) \frac{2R^2}{\varepsilon^2}  - A_{\pm}^{(0)}(y) \ln\left( \frac{2R}{\varepsilon}\right) \! -\! \left( A_{\pm}^{(0)}(y) + 4 A_{\pm}^{(-2)}(y)\mathcal{C}_{\pm}^{(2)}(y)\right) \right] \\
&\qquad + \mathcal{O}\left( \frac{R}{\varepsilon}\right) + \mathcal{S}_{\pm} + \mathcal{O}\left( \frac{\varepsilon}{R}\right),
\end{align}
where we will not bother to compute the non-universal $\mathcal{O}\left( \frac{R}{\varepsilon}\right)$ term, and where
\beq
\label{eq:mathcalspm}
\mathcal{S}_{\pm} \equiv \frac{\text{vol}(\mathbb{S}^1)\text{vol}(\mathbb{S}^2)^2}{4G_N}\int_0^{\frac{\pi}{2}} dy \left[ \frac{1}{2} A_{\pm}^{(-2)}(y) \, e^{\pm 2x^c_{\pm}+2c_{\pm}} \pm A_{\pm}^{(0)}(y)\left(x^c_{\pm}\pm c_{\pm}\right) \pm \mathcal{A}_{\pm}(x^c_{\pm},y)\right],
\eeq
where $\mathcal{A}_{\pm}(x,y)$ is the indefinite integral of $A_{\pm}(x,y)$, subject to the condition that $\mathcal{A}_{\pm}(x,y)$ has leading asymptotic behavior $\exp(\mp 2x)$ at large $|x|$. The integration over $y$ in eq.~\eqref{eq:mathcalspm} is straightforward, but the result is too cumbersome to write explicitly.

For $S_0$ we find, using eqs.~\eqref{eq:split1} and~\eqref{eq:split2},
\begin{align}
\begin{split}
\label{E:S0def}
S_0 &\equiv \frac{\text{vol}(\mathbb{S}^1)\text{vol}(\mathbb{S}^2)^2R}{4G_N} \int_0^{\frac{\pi}{2}} dy \int_{\varepsilon u^c(y)}^R \frac{du}{u^2} \int_{x^c_-}^{x^c_+} dx \, (f_4f_1f_2\rho)^2
\\
&= \mathcal{O}\left( \frac{R}{\varepsilon}\right) + \mathcal{S}_0+ \mathcal{O}\left(\frac{\varepsilon}{R}\right)\,,
\end{split}
\end{align}
where once again we will not bother to compute the non-universal $\mathcal{O}\left( \frac{R}{\varepsilon}\right)$ term, and where $\mathcal{S}_0$ comes entirely from the $u=R$ endpoint of the integration over $u$,
\beq
\label{eq:mathcals0}
\mathcal{S}_0\equiv -\frac{\text{vol}(\mathbb{S}^1)\text{vol}(\mathbb{S}^2)^2}{4G_N}\int_0^{\frac{\pi}{2}}dy\int_{x^c_-}^{x^c_+}dx \, (f_4f_1f_2\rho)^2.
\eeq
The simplest way we have to found to perform the integrations in eq.~\eqref{eq:mathcals0} is the following. For any finite $x_{\pm}^c$, the integration over $x$ in eq.~\eqref{eq:mathcals0} yields a finite result, allowing us to exchange the order of the $x$ and $y$ integrations. We then expand $(f_4f_1f_2\rho)^2$ as a convergent power series in $\exp(\delta - x)$ for $x>\delta$, and in $\exp(x - \delta)$ for $x<\delta$. We next exchange the sum of the expansion with the $y$ integral, and then integrate in $y$ term-by-term. Finally, we re-sum the expansion, obtaining, for $x > \delta$,
\begin{align}
\label{eq:mathcals0int}
\int_0^{\frac{\pi}{2}}dy \frac{(f_4f_1f_2\rho)^2}{2\pi \hat{\alpha}^2(\alpha')^4} = & 8 \alpha^2 \cosh^2(x) + N_5 \alpha\left[ e^{2x+\delta}+4e^{\delta}+\left( 3+\frac{e^{2\delta}}{3}\right)e^{-2x+\delta}+e^{-6x+3\delta}\right]
\\
\nonumber
& + 4N_5^2 \cosh (x)\cosh(2x-\delta)\left[ e^{-x+\delta}-2\text{arctanh}\left( e^{-2(x-\delta)}\right)\sinh (x-\delta)\right],
\end{align}
where we included the factor $1/(2\pi \hat{\alpha}^2(\alpha')^4)$ on the left-hand-side for convenience. The integration over $x$ is then straightforward.\footnote{In practice, for the integration over $x$ we found the choice $x_{\pm}^c \to \pm \infty$ the most convenient. We hasten to repeat, however, that the result for $S$ is independent of the choice of $x_{\pm}^c$, as mentioned below eq.~\eqref{eq:split2}.} For $x<\delta$, we find the same result as eq.~\eqref{eq:mathcals0int}, but with $\{x,\delta\}\to\{-x,-\delta\}$.

Upon summing our results for $S_{\pm}$ and $S_0$, we find (ignoring terms that vanish as $\varepsilon \to 0$)
\begin{align}
\nonumber
S =&  \frac{\text{vol}(\mathbb{S}^1)\text{vol}(\mathbb{S}^2)^2}{4G_N} \int_0^{\frac{\pi}{2}} dy \left [ \left( A_{+}^{(-2)}(y)+A_-^{(-2)}(y)\right)\frac{2R^2}{\varepsilon^2} - \left( A_{+}^{(0)}(y)+A_-^{(0)}(y)\right) \ln\left( \frac{2R}{\varepsilon}\right) \right.
\\
\label{E:SEEforD3D5}
&\qquad\qquad  \left. - \left(A_{+}^{(0)}(y)+A_-^{(0)}(y)  + 4 A_{+}^{(-2)}(y)\mathcal{C}_{+}^{(2)}(y)+4A_-^{(-2)}(y)\mathcal{C}_-^{(2)}(y)\right)\right ]
\\
\nonumber
& \qquad\qquad \qquad \qquad + D_1 \frac{R}{\varepsilon} + \mathcal{S}_-+\mathcal{S}_0+\mathcal{S}_+\,,
\end{align}
where the term $D_1 \frac{R}{\varepsilon}$ is the sum of the ${\mathcal{O}}\left(\frac{R}{\varepsilon}\right)$ terms in eqs.~\eqref{eq:spm} and~\eqref{E:S0def}. We did not bother to compute $D_1$, which is non-universal. Upon performing the integration over $y$ in the first and second lines of eq.~\eqref{E:SEEforD3D5}, we find
\beq
\label{E:SEED3D5part1}
S = \frac{\left(N_3^+\right)^2+\left(N_3^-\right)^2}{2}\left[\frac{R^2}{\varepsilon^2} -\ln\left( \frac{2R}{\varepsilon}\right) -\frac{1}{2}\right] + D_1 \frac{R}{\varepsilon} + \mathcal{S}_-+\mathcal{S}_0+\mathcal{S}_+\,.
\eeq
The term in brackets in eq.~\eqref{E:SEED3D5part1} is precisely half the spherical EE for $\mathcal{N}=4$ SYM theory with gauge group $SU(N_3^+)$ plus half of the spherical EE for $\mathcal{N}=4$ SYM theory with gauge group $SU(N_3^-)$. Following eqs.~\eqref{E:defDefBdyS} and~\eqref{E:Sdefect}, we thus identify the universal contribution to the defect entropy,
\begin{align}
\nonumber
D_ 0 = & \, \mathcal{S}_-+\mathcal{S}_0+\mathcal{S}_+ \\ = & \frac{\pi^4}{4G_N}\left\{  \left(L_{+}^8c_++L_{-}^8c_-\right) + \frac{128 }{3} N_5\alpha\hat{\alpha}^2(\alpha')^4 \left[ \cosh(3\delta)-6\cosh(\delta)+12\delta\sinh(\delta)\right]\right. \nonumber \\
\label{E:SEED3D5part2}
&\left. \phantom{\frac{1}{G}}+32N_5^2 \hat{\alpha}^2(\alpha')^4\left[(4\delta \sinh(2\delta) -3\cosh(2\delta)+8\ln 2 \sinh^2(\delta))\right]\right\}.
\end{align}
Using eq.~\eqref{E:D3D5Match1}, $L_{\pm}^4 = 4\pi N_3^{\pm}(\alpha')^2$, and $\hat{\alpha} = g_{YM}^2 \alpha/(4\pi)$, we can write our result for $D_0$ in terms of $g_{YM}^2$, $N_3^{\pm}$, $N_5$, $\alpha$, and $\delta$,
\begin{align}
\begin{split}
\label{E:D3D5fullS}
D_0 = &  \frac{1}{4}\left[ \left(N_3^+\right)^2 \ln \left(\frac{g_{YM}^2\alpha^2}{\pi^2 N_3^+}\right) + \left(N_3^-\right)^2 \ln\left( \frac{g_{YM}^2\alpha^2}{\pi^2 N_3^-}\right)\right]
\\
& \qquad + \frac{1}{12\pi^4}g_{YM}^4N_5\alpha^3 \left[  \cosh(3\delta) - 6 \cosh(\delta) + 12 \delta \sinh(\delta) \right]
\\
& \qquad \qquad+ \frac{1}{16\pi^4}g_{YM}^4N_5^2 \alpha^2 \left[ 4\delta \sinh(2\delta) - 3\cosh(2\delta) + 8\ln2 \sinh^2(\delta)\right]\,,
\end{split}
\end{align}
which is the main result of this subsection. In eq.~\eqref{E:D3D5fullS} we can translate $\alpha$ and $\delta$ to field theory quantities easily, using eqs.~\eqref{E:alphasol} and~\eqref{eq:deltadef}, but the result is cumbersome and unilluminating, so we will not present it in full generality. Instead, we will present the result in a few simplifying limits. When $N_5=0$, which via eq.~\eqref{E:D3D5deltaMatch} implies $\Delta N_3=0$, we find $D_0=0$, as expected. When $N_5 \neq 0$ and $\Delta N_3=0$, using eq.~\eqref{E:D3D5simpleBulk} we find
\beq
\label{E:D3D5simpleS}
D_0 = \frac{N_3^2}{2}\ln\left(  \frac{(\xi-g_{YM}^2N_5)^2}{16\pi^2 g_{YM}^2N_3} \right) - \frac{N_5(\xi-g_{YM}^2N_5)^2(5\xi+4g_{YM}^2N_5)}{768\pi^4g_{YM}^2}\,,
\eeq
where we recall $\xi^2 \equiv 16\pi^2 g_{YM}^2 N_3 + (g_{YM}^2N_5)^2$ from eq.~\eqref{E:xiDef}. If we additionally take the probe limit $N_5 \ll N_3$, then we find
\beq
\label{eq:D3D5EEprobe}
D_0 = - \frac{2}{3\pi} \, \sqrt{\lambda} \, N_5 N_3 + \mathcal{O}\left( \lambda N_5^2\right)\,,
\eeq
where $\lambda\equiv g_{YM}^2N_3$ is the 't Hooft coupling. The order-$\sqrt{\lambda}$ term in eq.~\eqref{eq:D3D5EEprobe} agrees perfectly with that computed in refs.~\cite{Jensen:2013lxa,Chang:2013mca} using probe D5-branes in $AdS_5 \times \mathbb{S}^5$.

As explained above, if we take $N_3^- \to 0$ with $N_3^+$ fixed, then the D3/D5 DCFT becomes the D3/D5 BCFT. In that limit the universal part of the defect entropy, $D_0$ in eq.~\eqref{E:D3D5fullS}, becomes the universal part of the boundary entropy, $B_0$,
\beq
\label{E:D3D5defectToBdy}
\lim_{N_3^-\to 0} D_0 = B_0 = \frac{N_3^2}{8}\left( 2\ln\left(\frac{16\pi^2 N_3}{g_{YM}^2N_5^2}\right) -3\right) + \frac{\pi^2 N_3^3}{3g_{YM}^2N_5^2}\,.
\eeq
We also obtained the $B_0$ in eq.~\eqref{E:D3D5defectToBdy} directly, by plugging the harmonic functions corresponding to the D3/D5 BCFT, eq.~\eqref{E:D3D5boundaryHarmonic}, into eq.~\eqref{E:susyEE} and performing the integrations.

%%%%%%%%%%%%%%%%%%%%%%%%%%%%%%%%%%%%%%%%%%%%%%%%%%%%%%%%%%%%%%%%%%%%%%%%%%%%%%%%%%%%%%%%%%%%%%%%%%%%%%%%%%%%%%%%%%
\subsubsection{Monotonicity of the Defect and Boundary Entropies}
\label{S:mono}
%%%%%%%%%%%%%%%%%%%%%%%%%%%%%%%%%%%%%%%%%%%%%%%%%%%%%%%%%%%%%%%%%%%%%%%%%%%%%%%%%%%%%%%%%%%%%%%%%%%%%%%%%%%%%%%%%%

With access only to the gravity dual of the D3/D5 DCFT or BCFT, rather than the dual of an RG flow between DCFTs or BCFTs, \textit{a priori} we seem unable to say anything about any putative higher-dimensional $g$-theorem. In fact, however, we can provide indirect evidence that the defect or boundary entropy, $D_0$ in eq.~\eqref{E:D3D5fullS} or $B_0$ in eq.~\eqref{E:D3D5defectToBdy}, changes monotonically under a certain class of defect or boundary RG flows, and may either increase or decrease under a certain class of bulk RG flows.

In the D3/D5 field theory, we will consider a defect or boundary RG flow triggered by the maximally-SUSY mass term for the hypermultiplets, and we will consider a bulk RG flow that arises from moving onto the Higgs branch. Each of these deformations preserves eight real supercharges, which will be essential for identifying the IR DCFT or BCFT.

In the D3/D5 system, we introduce the maximally-SUSY hypermultiplet mass deformation for some number $\Delta N_5 \leq N_5$ of the hypermultiplets by separating $\Delta N_5$ of the D5-branes from the D3-branes in a mutually transverse direction. Such a mass preserves eight real supercharges and an $SU(2) \times U(1)$ subgroup of the $SO(3) \times SO(3)$ R-symmetry. At the IR fixed point, the SUSY will be enhanced to the sixteen real supercharges of the superconformal symmetry, and the R-symmetry will be enhanced back to $SO(3) \times SO(3)$. Assuming the ambient CFT remains unchanged during the RG flow, so that $g_{YM}$ and $N_3^{\pm}$ remain unchanged, the only DCFT or BCFT with the given symmetries is the D3/D5 theory, now with $N_5 - \Delta N_5$ flavors~\cite{DeWolfe:2001pq,Erdmenger:2002ex}. Our prediction is thus that, to be consistent with a putative higher-dimensional $g$-theorem, $D_0$ or $B_0$ should be monotonic as a function of $N_5$, with $g_{YM}$ and $N_3^{\pm}$ fixed.

To test our prediction, we can simply take the partial derivative $\partial/\partial N_5$ of our result $D_0$ in eq.~\eqref{E:D3D5fullS}, with $g_{YM}$ and $N_3^{\pm}$ fixed. Since $D_0$ in eq.~\eqref{E:D3D5fullS} is most simply written as a function of $\alpha$ and $\delta$, rather than $g_{YM}$ and $N_3^{\pm}$, we will combine eqs.~\eqref{E:alpha} and~\eqref{E:D3D5deltaMatch} to write
\begin{equation*}
2\pi^2 N_3^{\pm} =  g_{YM}^2\alpha(e^{\pm\delta}N_5 + 2\alpha)\,,
\end{equation*}
and then use the chain rule,
\begin{align}
\nonumber
\left.\frac{\partial}{\partial N_5}\right|_{g_{YM},N_3^{\pm}} =& \left.\frac{\partial}{\partial N_5}\right|_{g_{YM},N_3^{\pm},\alpha,\delta} -\frac{\alpha}{N_5+4\alpha\cosh\delta} \left.\frac{\partial}{\partial\alpha}\right|_{g_{YM},N_3^{\pm},N_5,\delta}
\\
\label{E:D3D5chainRule}
& \qquad - \frac{4\alpha\sinh\delta}{N_5(N_5+4\alpha\cosh\delta)}\left.\frac{\partial}{\partial \delta}\right|_{g_{YM},N_3^{\pm},N_5,\alpha}\,.
\end{align}
For the $D_0$ in eq.~\eqref{E:D3D5fullS}, we then find
\beq
\label{eq:D3D5D0N5deriv}
\left.\frac{\partial D_0}{\partial N_5}\right|_{g_{YM},N_3^{\pm}} \!\!= - \frac{3\pi^4\left((N_3^+)^2\!+(N_3^-)^2\right) + 2g_{YM}^4\alpha^3(4\alpha\cosh(2\delta)+2N_5 \cosh(3\delta)+\alpha\cosh(4\delta))}{6\pi^4(N_5+4\alpha\cosh\delta)},
\eeq
so that, after recalling that $\alpha \geq 0$, we find $\partial{D_0}/\partial N_5 \leq 0$. The $N_3^- \to 0$ limit then immediately implies $\partial{B_0}/\partial N_5 \leq 0$. (Bear in mind, however, that in the D3/D5 BCFT we cannot reduce $N_5$ to zero, since then the D3-branes would have no D5-branes on which to end.) We have thus shown that $-D_0$ or $-B_0$ always monotonically decreases as we decrease $N_5$, when $g_{YM}$ and $N_3^{\pm}$ are fixed, consistent with our expectation for a higher-dimensional g-theorem.

In the D3/D5 intersection, to move onto the Higgs branch we allow some D3-branes to move away from the rest of the D3-brane stack in a direction along the D5-branes (the $(x^4,x^5,x^6)$ directions in table~\ref{T:D3D5}). Like the maximally-SUSY flavor mass, these Higgs branch states preserve eight real supercharges and an $SU(2)$ subgroup of the R-symmetry. Unlike the maximally-SUSY flavor mass, however, moving onto the Higgs branch is not a relevant deformation. Nevertheless, these states will exhibit an RG flow from one ambient CFT to another. At the IR fixed point, the SUSY will be enhanced to the sixteen real supercharges of the superconformal symmetry, and the R-symmetry will be enhanced back to $SO(3) \times SO(3)$. Once again the IR DCFT or BCFT must therefore be the D3/D5 theory, now with a gauge group of smaller rank. Indeed, the D3/D5 intersection makes clear that if we separate $\Delta N_3^{\pm}$ D3-branes on one side or the other of the D5-branes, then the IR DCFT will involve $\N=4$ SYM with gauge groups $SU\left(N_3^{\pm}-\Delta N_3^{\pm}\right)$ on one side or the other of the defect, with $g_{YM}$ and $N_5$ unchanged. Under such a bulk RG flow, presumably a higher-dimensional $g$-theorem places no constraint on the monotonicity of $D_0$ or $B_0$. Our prediction is thus that $D_0$ or $B_0$ may either increase or decrease as a function of either of $N_3^{\pm}$, with $g_{YM}$ and $N_5$ fixed.

The simplest way we have found to test this prediction is for the D3/D5 BCFT: taking $\partial/\partial N_3$ of eq.~\eqref{E:D3D5defectToBdy}, we find
\beq
\label{E:D3D5counter}
\left.\frac{\partial B_0}{\partial N_3}\right|_{g_{YM},N_5} = \frac{N_3}{2}\left( \ln\left( \frac{16\pi^2 N_3}{g_{YM}^2N_5}\right)-1\right) + \frac{\pi^2 N_3^2}{g_{YM}^2N_5^2}\,,
\eeq
which is positive when $(g_{YM}^2N_5)/\sqrt{\lambda} \ll 1$ and negative when $(g_{YM}^2N_5)/\sqrt{\lambda} \gg 1$. We have thus shown that $B_0$ can either increase or decrease as we decrease $N_3$ with $g_{YM}$ and $N_5$ fixed, consistent with our expectation for a higher-dimensional $g$-theorem.

To summarize, eq.~\eqref{eq:D3D5D0N5deriv} shows that the universal part of the defect or boundary entropy changes monotonically under RG flows triggered by a maximally-SUSY hypermultiplet mass. In particular, as $N_5$ decreases under the RG flow, we found that $-D_0$ or $-B_0$ monotonically decreases, and hence could potentially act as a measure of defect or boundary degrees of freedom. Eq.~\eqref{E:D3D5counter} shows that the universal part of the boundary entropy, $B_0$ in eq.~\eqref{E:D3D5defectToBdy}, can either increase or decrease under RG flows on a subspace of the Higgs branch of the ambient CFT. These results are consistent with our expectations for a higher-dimensional $g$-theorem, namely that the universal part of the defect or boundary entropy should change monotonically under a defect or boundary RG flow, but may either increase or decrease under a bulk RG flow.

%%%%%%%%%%%%%%%%%%%%%%%%%%%%%%%%%%%%%%%%%%%%%%%%%%%%%%%%%%%%%%%%%%%%%%%%%%%%%%%%%%%%%%%%%%%%%%%%%%%%%%%%%%%%%%%%%%
\subsection{{$T[SU(N)]$} Defect}
\label{ssec:tsun}
%%%%%%%%%%%%%%%%%%%%%%%%%%%%%%%%%%%%%%%%%%%%%%%%%%%%%%%%%%%%%%%%%%%%%%%%%%%%%%%%%%%%%%%%%%%%%%%%%%%%%%%%%%%%%%%%%%

Our next example of a DCFT is $(3+1)$-dimensional $\N=4$ SYM with gauge group $SU(N_3)$ coupled to a $(2+1)$-dimensional CFT, the so-called $T[SU(N)]$ CFT of ref.~\cite{Gaiotto:2008ak} (the simplest of the CFTs introduced in ref.~\cite{Gaiotto:2008ak}). We will first compute the spherical EE in the $T[SU(N)]$ CFT itself, and then in the DCFT obtained by coupling the $T[SU(N)]$ CFT to (3+1)-dimensional $\N=4$ SYM as a defect.

\begin{figure}[ht!]
\begin{center}
\includegraphics[width=0.75\textwidth]{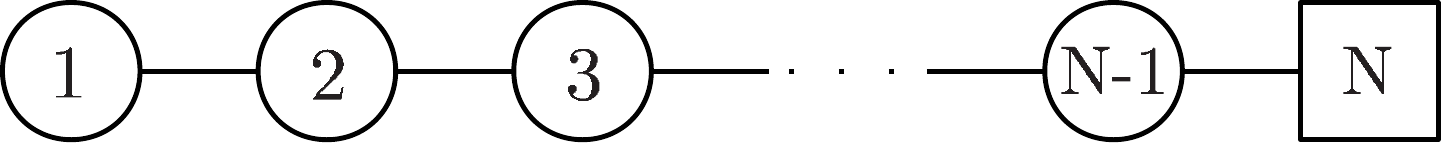}
\end{center}
\caption{A $T[SU(N)]$ CFT arises as the low-energy limit of the $(2+1)$-dimensional ${\cal N}=4$ SYM theory with the quiver above. The $i$-th node represents an ${\cal N}=4$ SYM theory with gauge group $SU(N_i)$. The line connecting the $i$-th node to the $i+1$-th node represents an ${\cal N}=4$ hypermultiplet in the bi-fundamental representation of $SU(N_i)\times SU(N_{i+1})$. The box represents a collection of $N$ hypermultiplets in the fundamental representation of $SU(N-1)$.}
\label{fig:TSUNquiver}
\end{figure}

\begin{figure}[ht!]
\begin{center}
\includegraphics[width=0.55\textwidth]{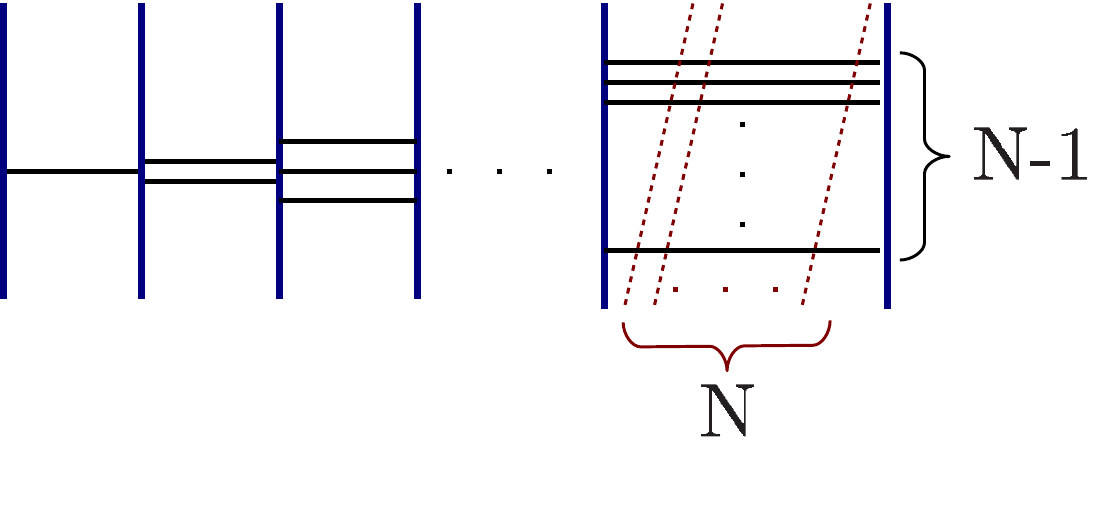}
\end{center}
\vskip-0.5in
\caption{The $(2+1)$-dimensional $T[SU(N)]$ CFT arises in type IIB string theory as the low-energy theory of the above D3/D5/NS5-brane intersection.  The solid blue vertical lines represent NS5-branes, the solid black horizontal lines represent D3-branes, and the dashed red slanted lines represent D5-branes.}
\label{fig:TSUNbraneconst}
\end{figure}

The $T[SU(N)]$ theory is specified by a choice of integer $N \geq 0$, and arises as the low-energy theory of the $(2+1)$-dimensional $\N=4$ SYM theory with field content given by the quiver diagram in fig.~\ref{fig:TSUNquiver}. In type IIB string theory the $T[SU(N)]$ CFT arises as the low-energy theory living on the intersection of D3-, D5-, and NS5-branes shown in fig.~\ref{fig:TSUNbraneconst}.

%%%%%%%%%%%%%%%%%%%%%%%%%%%%%%%%%%%%%%%%%%%%%%%%%%%%%%%%%%%%%%%%%%%%%%%%%%%%%%%%%%%%%%%%%%%%%%%%%%%%%%%%%%
%%%%%%%%%%%%%%%%%%%%%%%%%%%%%%%%%%%%%%%%%%%%%%%%%%%%%%%%%%%%%%%%%%%%%%%%%%%%%%%%%%%%%%%%%%%%%%%%%%%%%%%%%%

%%%%%%%%%%%%%%%%%%%%%%%%%%%%%%%%%%%%%%%%%%%%%%%%%%%%%%%%%%%%%%%%%%%%%%%%%%%%%%%%%%%%%%%%%%%%%%%%%%%%%%%%%%
%%%%%%%%%%%%%%%%%%%%%%%%%%%%%%%%%%%%%%%%%%%%%%%%%%%%%%%%%%%%%%%%%%%%%%%%%%%%%%%%%%%%%%%%%%%%%%%%%%%%%%%%%%

A $T[SU(N)]$ theory has $OSp(4|4,\mathbb{R})$ superconformal symmetry, just like the D3/D5 DCFT and BCFT. As a result, the holographic dual is type IIB SUGRA in the background with metric and dilaton given by eqs.~\eqref{susymet}, \eqref{susydef}, and~\eqref{E:dilaton}, with a particular set of harmonic functions $h_1(v,\bar{v})$ and $h_2(v,\bar{v})$~\cite{Assel:2011xz}. To obtain the harmonic functions for the dual of $T[SU(N)]$, we begin with a more general solution of ref.~\cite{Assel:2011xz} given by the harmonic functions
\begin{align}
\label{eq:tsunharmonic}
h_1(v,\bar{v}) =& \alpha' \left[ - \frac{N_{\text{D5}}}{4} \ln \left( \tanh \left( \frac{i \pi}{4} - \frac{v-\delta}{2} \right) \right) \right] + \text{c.c.}, \cr
h_2(v,\bar{v}) =& \alpha' \left[ - \frac{N_{\text{NS5}}}{4}  \ln \left( \tanh \left( \frac{v-\hat \delta}{2} \right) \right) \right] +\text{c.c},
\end{align}
where $v = x + i y$ with $x \in (-\infty,\infty)$ and $y \in [0,\pi/2]$, and where $\delta$ and $\hat{\delta}$ are real-valued. These harmonic functions produce an $AdS_4 \ltimes \mathcal{M}_6$ spacetime, where $\mathcal{M}_6$ is a compact six-dimensional manifold describing specific D5- and NS5-brane sources~\cite{Assel:2011xz}, namely $N_{\text{D5}}$ D5-branes at $v = \delta + i \pi/2$ and $N_{\text{NS5}}$ NS5-branes at $v = \hat \delta$, with $N_{D3}$ D3-branes ending on the D5-brane stack and $\hat{N}_{D3} = - N_{D3}$ D3-branes ending on the NS5-brane stack, where
\beq
%\label{linking}
N_{D3} = - N_{\text{D5}} N_{\text{NS5}} \frac{2}{\pi} \arctan(e^{\hat \delta - \delta}).
\eeq
To obtain the dual of $T[SU(N)]$, in eq.~\eqref{eq:tsunharmonic} we set $N_{\text{D5}}=N$, $N_{\text{NS5}}=N$, and $N_{D3}=-\hat{N}_{D3}=N$~\cite{Assel:2011xz}.

The integral for the spherical EE is eq.~\eqref{E:susyEE}, where in this case we do not need the cutoffs $\chi_{\pm}(\frac{\varepsilon}{u},y)$ because $\mathcal{M}_6$ is compact and hence has finite volume $\vol(\mathcal{M}_6) = (4\pi)^2 \int dx \, dy (f_4 f_1 f_2 \rho)^2$. Using $4 G_N= 2^5 \pi^6 (\alpha')^4$ we find for the spherical EE
\begin{subequations}
\beq
\label{eq:tsunEE}
S = \frac{\vol(\mathcal{M}_6)}{2^4 \pi^5 (\alpha')^4} \left[ \frac{R}{\varepsilon} - 1 \right] = C_1 \frac{\ell}{\varepsilon} + C_0^{(d=3)},
\eeq
\beq
\label{eq:tsununiv}
C_0^{(d=3)} = - \frac{1}{2} N^2 \ln N + {\mathcal{O}}(N^2),
\eeq
\end{subequations}
where we did not bother to compute the non-universal constant $C_1^{(d=3)}$, and where we used the result of ref.~\cite{Assel:2012cp} for $\vol(\mathcal{M}_6)$ to extract the leading large-$N$ behavior.

For the $T[SU(N)]$ CFT, the free energy on Euclidean $\mathbb{S}^3$, $F_{\mathbb{S}^3}$, was computed using SUSY localization in ref.~\cite{Nishioka:2011dq}. In the large-$N$ limit, $F_{\mathbb{S}^3} = \frac{1}{2} N^2 \ln N + {\mathcal{O}}(N^2)$, where the leading term agrees with the holographic calculation of $F_{\mathbb{S}^3}$ using the SUGRA solution above~\cite{Assel:2012cp}. The leading large-$N$ contribution to the universal constant in the spherical EE, $C_0^{(d=3)}$ in eq.~\eqref{eq:tsununiv}, is precisely the leading large-$N$ contribution to $-F_{\mathbb{S}^3}$, as expected~\cite{Myers:2010xs,Myers:2010tj,Casini:2011kv}.

Now let us consider the DCFT obtained by introducing the $T[SU(N)]$ CFT as a defect in $(3+1)$-dimensional $\N=4$ SYM. That DCFT has $OSp(4|4,\mathbb{R})$ superconformal symmetry and is dual to type IIB SUGRA in a background with metric and dilaton given by eqs.~\eqref{susymet}, \eqref{susydef}, and~\eqref{E:dilaton}. To obtain the harmonic functions for the dual of this DCFT, we once again begin with a more general solution of ref.~\cite{Assel:2011xz}, given by the harmonic functions
\begin{align}
\label{eq:hforTSUN}
h_1(v,\bar{v}) =& \alpha' \left[ - i\, \alpha \sinh(v) - \frac{N_{\text{D5}}}{4} \ln \left( \tanh \left( \frac{i \pi}{4} - \frac{v-\delta}{2} \right) \right) \right] + \text{c.c.}, \cr
h_2(v,\bar{v}) =& \alpha' \left[ \hat{\alpha} \cosh(v) - \frac{N_{\text{NS5}}}{4}  \ln \left( \tanh \left( \frac{v-\hat{\delta}}{2} \right) \right) \right] + \text{c.c.},
\end{align}
where $v = x + i y$ with $x \in (-\infty,\infty)$ and $y\in[0,\pi/2]$, and where $\alpha$, $\hat{\alpha}$, $\delta$, and $\hat{\delta}$ are real-valued. The only difference between the harmonic functions in eqs.~\eqref{eq:tsunharmonic} and~\eqref{eq:hforTSUN} are the $\sinh(v)$ and $\cosh(v)$ terms in the latter, which lead to two asymptotically $AdS_5 \times \mathbb{S}^5$ regions as $x \to \pm \infty$. Following eq.~\eqref{eq:frhoGasymp}, we extract the asymptotic $AdS_5 \times \mathbb{S}^5$ radii of curvature $L_{\pm}$ from the behavior of $\rho(x,y^a)^2$ as $x \to \pm \infty$, and we extract the string coupling $g_s$ from the behavior of $e^{2 \phi}$ as $x \to \pm \infty$,
\begin{align}
\label{tsurad}
&\frac{L_{\pm}^4}{(\alpha')^2} = 16 \, \alpha \hat{\alpha} + 8 \, \hat \alpha e^{\pm \delta}  N_{\text{D5}} + 8 \, \alpha e^{\pm \hat{\delta}}  N_{\text{NS5}}, & &g_s= \left| \frac{\hat{\alpha}}{\alpha} \right|,&
\end{align}
where again we identify $g_{YM}^2=4\pi g_s$. The number of D3-branes ending on the D5-brane stack, $N_{D3}$, and the number of D3-branes ending on the NS5-brane stack, $\hat{N}_{D3}$, are now
\begin{align}
\begin{split}
\label{linking}
N_{D3} =& N_{\text{D5}} \left( \frac{4 \hat{\alpha}}{\pi} \sinh(\delta) -  N_{\text{NS5}} \frac{2}{\pi} \arctan(e^{\hat{\delta} - \delta}) \right), \cr
\hat{N}_{D3} =& N_{\text{NS5}}\left( \frac{4 \alpha}{\pi} \sinh(\hat{\delta}) + N_{\text{D5}} \frac{2}{\pi} \arctan(e^{\hat{\delta} - \delta}) \right).
\end{split}
\end{align}
To obtain the $(3+1)$-dimensional $\N=4$ SYM with gauge group $SU(N_3)$ coupled to a $T[SU(N)]$ defect, we take $L_+^4 = L_-^4 = L^4 = 4 \pi  N_3(\alpha')^2$, which via eq.~\eqref{tsurad} leads to the constraint
\begin{align}
\label{Lconst}
\hat \alpha \sinh(\delta) N_{\text{D5}} + \alpha \sinh(\hat \delta) N_{\text{NS5}} = 0.
\end{align}
As a consequence of eq.~\eqref{Lconst}, we can identify $N = \hat N_{D3} = -N_{D3}$. To obtain a $T[SU(N)]$ defect, we take $N_{\text{D5}} = N$, $\hat{N}_{\text{NS5}} = N$, and $\hat{N}_{D3} = -N_{D3} = N$. The constraint in eq.~\eqref{Lconst} is then trivially satisfied. We use these values of $N_{\text{D5}}$, $N_{\text{NS5}}$, $N_{D3}$, and $\hat{N}_{D3}$ throughout the rest of this subsection.

We can determine the four bulk parameters $\{\alpha,\hat{\alpha},\delta,\hat{\delta}\}$, subject to the constraint in eq.~\eqref{Lconst}, in terms of the three field theory parameters $\{g_{YM}^2,N_3,N\}$ as follows. First we solve eq.~\eqref{linking} for $\alpha$ and $\hat{\alpha}$ in terms of $N$, $\delta$ and $\hat{\delta}$. We then insert those values of $\alpha$ and $\hat{\alpha}$ into the expression for $g_s$ in eq.~\eqref{tsurad} to find $\hat{\delta} = -\text{arcsinh}\left(g_s \sinh(\delta)\right)$. We then insert $\alpha$, $\hat{\alpha}$, and $\hat{\delta}$, all in terms of $g_s$, $N$, and $\delta$, into the expression for $L_{\pm}$ in eq.~\eqref{tsurad}, which gives us an equation for $\delta$. Solving that equation in full generality is difficult, so we will restrict to the limit $\delta \gg 1$, which implies $\hat{\delta} \ll -1$. In that case, we can expand eq.~\eqref{tsurad} as
\begin{align}
\label{E:TSUNparams}
4 \pi N_3 = - 4 \pi N + \frac{16\pi}{g_{YM}^2}\left [2 N^2 + \pi^2 - N \pi \left(\frac{g_{YM}^2}{4\pi}+\frac{4\pi}{g_{YM}^2}\right) \right] e^{-2 \delta} + {\cal O} \left(e^{-4 \delta}\right),
\end{align}
where we used $g_{YM}^2=4\pi g_s$. We will further take $N \gg 1$ such that we can neglect all of the terms in the square brackets in eq.~\eqref{E:TSUNparams} except $2 N^2$. Of course we also take the usual Maldacena limits, $g_{YM}^2 \to 0$ and $N_3 \gg 1$ with $g_{YM}^2 N_3 \gg 1$. In that case $g_{YM}^{-2} \gg 1$, so to guarantee that $N^2$ dominates all other terms in the square brackets in eq.~\eqref{E:TSUNparams}, we must take $N^2 \gg N/g_{YM}^2$ or equivalently $g_{YM}^2 N \gg 1$. Taking these limits, and dropping the ${\cal O} \left(e^{-4 \delta}\right)$ terms, we solve eq.~\eqref{E:TSUNparams} for $e^{-2\delta}$, which then also gives us $\hat{\delta}$, $\alpha$, and $\hat{\alpha}$ as explained above:
\begin{align}
\label{eq:tsunparamsmap}
&e^{2 \delta}
=  \frac{8}{ g_{YM}^2} \frac{N^2}{N+N_3},&
&e^{-2 \hat{\delta}}
=  \frac{g_{YM}^2}{2\pi^2} \frac{N^2}{N+N_3},& \cr
&\alpha = \frac{\pi^2}{\sqrt{2}} \frac{N_3 \sqrt{N_3 +N}}{g_{YM} N^2},&
&\hat{\alpha} = \frac{\pi}{2^{5/2}}\frac{g_{YM} N_3\sqrt{N_3 +N}}{N^2}.&
\end{align}
The expression for $e^{2 \delta}$ in eq.~\eqref{eq:tsunparamsmap} shows that the $\delta \gg 1$ limit is only consistent if $N^2 \gg N + N_3$, which because $N \gg1$ implies $N^2 \gg N_3$. We still have freedom to specify how $N \gg 1$ compares to $N_3 \gg 1$, however.

In this case the integral for the spherical EE is again eq.~\eqref{E:susyEE}, where now we need the $x$ cutoffs $\chi_{\pm}\left(\frac{\varepsilon}{u},y\right)$. Using eqs.~\eqref{eq:expansion} and~\eqref{E:xFromUZ} with $z=\varepsilon$, we find
\begin{align}
&\chi_{\pm}\left( \frac{\varepsilon}{u},y\right) = \pm \ln \left( \frac{2u}{\varepsilon}\right) \pm \frac{1}{2} \ln \left( \frac{2 \alpha \hat \alpha + \hat \alpha e^{\pm \delta} N + \alpha e^{\pm \hat \delta} N}{2 \alpha \hat \alpha} \right) \\ &\pm \left( - \frac{1}{4} + N \frac{\hat{\alpha}^2 e^{\pm 2 \delta}[N\mp4 \alpha \sinh(\delta)]
-\alpha^2 e^{\pm 2 \hat \delta}[N\mp4 \hat \alpha \sinh(\hat{\delta})]
}{16(\alpha e^{\pm \hat{\delta}} N + \hat{\alpha} e^{\pm \delta} N + 2 \alpha \hat{\alpha})^2} \cos(2y)  \right) \left(\frac{\varepsilon}{u}\right)^2+\mathcal{O}\left( \frac{\varepsilon^4}{u^4}\right). \nonumber
\end{align}
As mentioned in the previous subsection, the integrand of eq.~\eqref{E:susyEE} takes a simple form when written in terms of the harmonic functions $h_1(v,\bar{v})$ and $h_2(v,\bar{v})$, namely the form in eq.~\eqref{E:susyAreaDensity}, which we repeat here for convenience:
\begin{align}
\label{TSUNintg}
(f_4 f_1 f_2 \rho)^2 = -32 \, h_1 h_2 \, \p_v \bar \p_v (h_1 h_2).
\end{align}
The $x$ integration in eq.~\eqref{E:susyEE} is difficult to do exactly. In the limit $\delta \gg 1$ and $\hat{\delta} \ll -1$, the $\tanh$ functions appearing in the harmonic functions in eq.~\eqref{eq:hforTSUN} can be well-approximated by a step function, in which case
\begin{align}
\label{e:tsunh1h2}
\frac{h_1(v,\bar{v})}{\alpha'}
=
\begin{cases}
   \alpha \sin(y) e^{+x} + (\alpha + N e^{\delta}) \sin(y) e^{-x} - \frac{N}{3} \sin(3y) e^{-3(x-\delta)} + {\cal{O}}\left(e^{-5(x-\delta)}\right)
   &  x > \delta, \\
   \alpha \sin(y) e^{-x} + (\alpha + N e^{-\delta}) \sin(y) e^{+x} - \frac{N}{3} \sin(3y) e^{-3(\delta-x)} + {\cal{O}}\left(e^{-5(\delta -x)}\right) &x < \delta, \\
  \end{cases} \cr
\frac{h_2(v,\bar{v})}{\alpha'}
=
\begin{cases}
   \hat{\alpha} \cos(y) e^{+x} + (\hat{\alpha} + N e^{\hat{\delta}}) \cos(y) e^{-x} + \frac{N}{3} \cos(3y) e^{3(\hat{\delta}-x)} + {\cal{O}}\left(e^{5(\hat{\delta}-x)}\right)
   & x > \hat{\delta},  \\
   \hat{\alpha} \cos(y) e^{-x} + (\hat{\alpha} + N e^{-\hat{\delta}}) \cos(y) e^{+x} + \frac{N}{3} \cos(3y) e^{3(x-\hat{\delta})} + {\cal{O}}\left(e^{-5(x-\hat{\delta})}\right) &  x < \hat{\delta}.\\
  \end{cases}
\end{align}
We can argue that the terms of ${\cal{O}}(e^{\pm5(x-\delta)})$ and ${\cal{O}}(e^{\pm5(x-\hat{\delta})})$ and higher (henceforth the ``neglected terms'') do not contribute to the divergent or constant terms in the spherical EE, as follows. In the appendix we show explicitly that the $R^2/\varepsilon^2$ and $\ln(2R/\varepsilon)$ terms in the spherical EE receive contributions only from terms in eq.~\eqref{TSUNintg} that are non-vanishing in the $|x| \to \infty$ limit. The neglected terms vanish in that limit and hence do not contribute to the $R^2/\varepsilon^2$ and $\ln(2R/\varepsilon)$ terms in the spherical EE. The constant term in the spherical EE receives contributions of order $N^2$ and $N_3^2$ from the neglected terms, however these are not the leading contributions to the constant term: the biggest contribution comes from a term proportional to $N^2 \ln N$ or $N_3^2 \ln \left(N^2/N_3\right)$, as we will see below. These logarithmic contributions come from terms in eq.~\eqref{TSUNintg} that are independent of $x$. The neglected terms cannot contribute to a term independent of $x$, simply because eq.~\eqref{TSUNintg} involves a product of four harmonic functions, and so a term of order ${\cal{O}}(e^{\pm5(x-\delta)})$ or ${\cal{O}}(e^{\pm5(x-\hat{\delta})})$ would multiply a term of at most ${\cal{O}}(e^{\pm3x})$, coming from a product of the ${\cal{O}}(e^{\pm x})$ terms of three harmonic functions. In short, to obtain the leading divergent and constant contributions to the spherical EE, we only need the leading terms shown explicitly in eq.~\eqref{e:tsunh1h2}.

Using eq.~\eqref{e:tsunh1h2} in eq.~\eqref{TSUNintg} and then performing the integrations in eq.~\eqref{E:susyEE}, we find that the universal part of the defect entropy, $D_0$, in the $N \gg 1$ limit depends on how we scale $N_3$ as we take $N \gg 1$:
\begin{subequations}
\label{eq:tsunresult}
\begin{align}
S &= N_3^2 \left[ \frac{R^2}{\varepsilon^2}-\ln\left( \frac{2R}{\varepsilon}\right)-\frac{1}{2}\right] + D_1 \frac{R}{\varepsilon} + D_0,
\\
D_0 &=
\begin{cases}
- \frac{1}{2} N^2 \ln N + {\cal O}(N^2) & N \gg N_3 \gg 1,
\\
- \frac{1}{2} N^2 \left(1 + 2 \frac{N_3}{N} + 2 \frac{N_2^3}{N^2} \right) \ln N + {\cal O}(N^2) & N \propto N_3 \gg 1\,,
\\
-  N_3^2 \ln \left(\frac{N^2}{N_3} \right) + {\cal O}(N_3^2) & N^2 \gg N_3 \gg N \gg 1\,,
\end{cases}
\end{align}
\end{subequations}
where once again we did not bother to compute the non-universal constant $D_1$.

Our result for $D_0$ in eq.~\eqref{eq:tsunresult} offers a big hint for a higher-dimensional $g$-theorem: in the limit $N \gg N_3 \gg 1$ the leading contribution to $D_0$ is clearly minus the leading large-$N$ contribution to the free energy of the $T[SU(N)]$ CFT on $\mathbb{S}^3$, $-F_{\mathbb{S}^3} = -\frac{1}{2} N^2 \ln N + {\cal{O}}(N^2)$~\cite{Nishioka:2011dq}, precisely the quantity that obeys the F-theorem. Can the proof of the F-theorem in ref.~\cite{Casini:2012ei}, based primarily on the strong sub-additivity of EE, be adapted to prove a higher-dimensional $g$-theorem? We will leave this important question for future research.

%%%%%%%%%%%%%%%%%%%%%%%%%%%%%%%%%%%%%%%%%%%%%%%%%%%%%%%%%%%%%%%%%%%%%%%%%%%%%%%%%%%%%%%%%%%%%%%%%%%%%%%%%%%%%%%%%%
\subsection{Non-SUSY Janus}
\label{ssec:NSJanus}
%%%%%%%%%%%%%%%%%%%%%%%%%%%%%%%%%%%%%%%%%%%%%%%%%%%%%%%%%%%%%%%%%%%%%%%%%%%%%%%%%%%%%%%%%%%%%%%%%%%%%%%%%%%%%%%%%%

The non-SUSY Janus solution of type IIB SUGRA is a one-parameter deformation of the $AdS_5 \times \mathbb{S}^5$ solution in which only the metric, dilaton, and RR five-form are non-trivial, and all SUSY is broken~\cite{Bak:2003jk,D'Hoker:2006uu}. The solution is most easily written in terms of elliptic functions. In particular, we will need the Weierstrass elliptic function $\We(\xi)$, defined by the equation
\beq
\label{Wieredef}
\left( \partial_{\xi} \We \right)^2 = 4 \We^3 - g_2 \We - g_3\,,
\eeq
where $g_2$ and $g_3$ are determined by the ratio of periods.  We will also need the Weierstrass $\zeta$-function and $\sigma$-function, which are related to $\We(\xi)$ via
\beq
\We(\xi) = - \zeta'(\xi)\,, \qquad \zeta(\xi) = \frac{\sigma'(\xi)}{\sigma(\xi)}\,.
\eeq
The Einstein-frame metric of the non-SUSY Janus solution is
\beq
\label{eq:nonsusyjanusmetric}
g = L^2 \left( \gamma^{-1} h(\xi)^2 d\xi^2 + h(\xi) \, g_{AdS_4} \right)+ L^2 g_{\mathbb{S}^5}\,,
\eeq
where $L^4 = 4 \pi N \alpha'^2$, with $N$ the number of D3-branes, $\gamma$ is a real parameter obeying $3/4 \leq \gamma \leq 1$, and the warp factor $h(\xi)$ is
\beq
\label{eq:nsjwarp}
h(\xi) = \g \left( 1 + \frac{4 \g - 3}{\We(\xi) + 1 - 2\g} \right)\,, \qquad g_2 = 16 \g (1-\g)\,, \qquad g_3 = 4(\g-1)\,.
\eeq
The dilaton of the non-SUSY Janus solution, $\phi(\xi)$, is
\beq
\label{eq:nonsusyjanusdilaton}
\phi(\xi) =  \phi_0 + \sqrt{6(1-\gamma)} \left(\xi + \frac{4 \gamma - 3}{\We'(\xi_1)} \left( \ln \frac{\sigma(\xi+\xi_1)}{\sigma(\xi-\xi_1)} - 2 \zeta(\xi_1) \xi\right)\right)\,,
\eeq
where $\phi_0$ is a real constant and $\xi_1$ is defined by $\We(\xi_1) = 2(1-\g)$. When $\g=1$, the solution reduces to $AdS_5 \times \mathbb{S}^5$ with constant dilaton $\phi(\xi)=\phi_0$, while $\g=3/4$ leads to a linear dilaton solution. Let $\xi_0$ denote the positive solution of $\We(\xi)=2\g-1$. Clearly $h(\xi)$ in eq.~\eqref{eq:nsjwarp} has poles at $\xi=\pm \xi_0$. As $\xi \to \pm \xi_0$, the non-SUSY Janus solution asymptotes to $AdS_5 \times \mathbb{S}^5$ with constant dilaton $\phi_{\pm} = \phi(\pm \xi_0)$, where $\phi_+ \neq \phi_-$ unless $\gamma =1$. In other words, for generic $\g$ the non-SUSY Janus solution has two asymptotically $AdS_5 \times \mathbb{S}^5$ regions in which the dilaton takes two different values. Notice that to put the non-SUSY Janus metric into the form of eq.~\eqref{E:generalMetric} in each asymptotically $AdS_5 \times \mathbb{S}^5$ region, we must take $\xi = \xi_0 \tanh(x)$.

We can obtain new solutions from the non-SUSY Janus solution using the $SL(2,\mathbb{R})$ duality of type IIB supergravity. Combining the dilaton and axion (RR zero-form) $C_{(0)}$ into the single complex field $\tau \equiv C_{(0)} + i e^{-2\phi}$, an $SL(2,\mathbb{R})$ transformation acts as
\beq
\label{eq:sl2r}
\tau \to \frac{a \tau + b}{c \tau + d},
\eeq
where $a,b,c,d \in \mathbb{R}$ and $ad-bc=1$, while the metric and RR five-form are unchanged. In general, one of $(a,b,c,d)$ can be absorbed into the choice of $\phi_0$, so an $SL(2,\mathbb{R})$ transformation introduces only two additional parameters. These determine the two asymptotic values of the axion, $C_{(0)}^{\pm} \equiv C_{(0)}(\pm \xi_0)$. A solution obtained via an $SL(2,\mathbb{R})$ transformation of non-SUSY Janus is thus completely determined by five real parameters: $N$, $\phi_{\pm}$, and $C_{(0)}^{\pm}$.

The field theory dual to non-SUSY Janus is a deformation of $\N=4$ SYM in which $g_{YM}$ takes two different values on the two sides of a $(2+1)$-dimensional interface, \textit{i.e.}\ ``jumps'' across an interface. An $SL(2,\mathbb{R})$ transformation can then generate a jumping $\theta$-angle. A jumping $g_{YM}$ is analogous to a dielectric interface in ordinary electromagnetism, while $\N=4$ SYM with a constant (non-jumping) $g_{YM}$ but a jumping $\theta$-angle describes a fractional topological insulator~\cite{Estes:2012nx}. The $g_{AdS_4}$ and $g_{\mathbb{S}^5}$ factors in eq.~\eqref{eq:nonsusyjanusmetric} indicate that in the field theory a jumping $g_{YM}$ and/or $\theta$ preserves $(2+1)$-dimensional conformal symmetry and the $SO(6)$ global symmetry. The non-SUSY Janus solution breaks all SUSY~\cite{Bak:2003jk}, so the $SO(6)$ global symmetry is no longer an R-symmetry. For more details about the field theories dual to non-SUSY Janus and its $SL(2,\mathbb{R})$ cousins, see refs.~\cite{Clark:2004sb,D'Hoker:2006uv,Gaiotto:2008sd}. We will choose normalizations in the $\N=4$ SYM action such that the $SL(2,\mathbb{R})$-covariant coupling is
\beq
\label{eq:nsjtau}
\tau \equiv \frac{\theta}{2\pi} +  \frac{4 \pi i}{g_{YM}^2}\,.
\eeq
By matching to the dual $SL(2,\mathbb{R})$-covariant bulk field $\tau$, we identify
\beq
\label{eq:janusdictionary}
C_{(0)}^\pm = \frac{\theta^{\pm}}{2\pi}\,, \qquad e^{\phi_\pm} = \frac{g_{YM}^{\pm}}{\sqrt{4\pi}}\,.
\eeq
For non-SUSY Janus, $\g$ completely determines, via eqs.~\eqref{eq:nonsusyjanusdilaton} and~\eqref{eq:janusdictionary}, $\delta \phi \equiv \phi_+ - \phi_- = \ln (g_{YM}^+/g_{YM}^-)$. In what follows we will also consider an especially simple $SL(2,\mathbb{R})$ transform of non-SUSY Janus where $\theta$ jumps but $g_{YM}$ does not, with $\delta \theta \equiv \theta^+ - \theta^- = \frac{16 \pi^2}{g_{YM}^2} \sinh (\delta \phi)$ determined completely by the original $\delta \phi$, and hence by $\g$.

The integral for the EE is simplest when written as in eq.~\eqref{E:defectSEEfinal}, but with $\xi$ instead of $x$,
\beq
\label{eq:nsjEE1}
S = \frac{\text{vol}(\mathbb{S}^1)\text{vol}(\mathbb{S}^5)R}{4G_N} \, L^8  \int_{\varepsilon}^R \frac{du}{u^2}  \int_{\xi_-(\frac{\varepsilon}{u})}^{\xi_+(\frac{\varepsilon}{u})}d\xi \, \frac{h(\xi)^2}{\sqrt{\gamma}}.
\eeq
To compute the cutoffs $\xi_{\pm}(\varepsilon/u)$, in each asymptotically $AdS_5 \times \mathbb{S}^5$ region we take $\xi = \xi_0 \tanh(x)$ and then use eqs.~\eqref{eq:expansion} and~\eqref{E:xFromUZ} with $z=\varepsilon$ to find
\beq
\chi_{\pm}\left(\frac{\varepsilon}{u}\right) = \pm \ln \left( \frac{2u}{\varepsilon} \right) \mp \frac{1}{4} \ln \left( \frac{\gamma}{\xi_0^2} \right) \mp \frac{\xi_0 + \sqrt{\gamma}}{8 \xi_0} \, \left(\frac{\varepsilon}{u}\right)^2 + \mathcal{O}\left(\frac{\varepsilon^3}{u^3}\right),
\eeq
so that again using $\xi = \xi_0 \tanh(x)$ we find
\beq
\xi_{\pm}\left(\frac{\varepsilon}{u}\right) = \pm \xi_0 \mp \frac{\sqrt{\gamma}}{2} \left(\frac{\varepsilon}{u}\right)^2 \mp \frac{\sqrt{\gamma}}{8} \left(\frac{\varepsilon}{u}\right)^4 + \mathcal{O}\left(\frac{\varepsilon^6}{u^6}\right).
\eeq
To perform the integration over $\xi$ in eq.~\eqref{eq:nsjEE1}, we use
\beq
\label{eq:nsjint}
\int  \frac{d\xi}{\wp(\xi) - \wp(\xi_0)} = \frac{1}{\wp'(\xi_0)} \left [ \ln \left( \frac{\sigma(\xi_0-\xi)}{\sigma(\xi_0+\xi)} \right) + 2 \zeta(\xi_0) \, \xi \right],
\eeq
as well as $\frac{\partial}{\partial \xi_0}$ of eq.~\eqref{eq:nsjint}, with the result
\begin{multline}
\!\!\!\!\!\!\int_{\xi_-(\frac{\varepsilon}{u})}^{\xi_+(\frac{\varepsilon}{u})} d\xi \frac{h(\xi)^2}{\sqrt{\gamma}} \!=\!\! \left [ - \frac{1}{2} \left( \zeta(\xi_0) - \sqrt{\gamma} \right) \xi - \frac{1}{4} \ln  \left(\frac{\sigma(\xi_0 - \xi)}{\sigma(\xi_0+\xi)}\right)
+ \frac{\sqrt{\gamma}}{4} \left( \zeta(\xi_0 - \xi) - \zeta(\xi_0 + \xi) \right) \right ]_{\xi_-(\frac{\varepsilon}{u})}^{\xi_+(\frac{\varepsilon}{u})} \\ = \frac{u^2}{\varepsilon^2} + \ln \left(\frac{2 u}{\varepsilon}\right) - \frac{1}{4} - \left(\zeta(\xi_0) - \sqrt{\gamma}\right) \xi_0 + \frac{1}{2} \ln \left(\frac{\sigma(2 \xi_0)}{2 \sqrt{\gamma}}\right) - \frac{\sqrt{\gamma}}{2} \zeta(2\xi_0) + {\cal O} \left(\frac{\varepsilon^2}{u^2}\right). \nn
\end{multline}
The integration over $u$ in eq.~\eqref{eq:nsjEE1} is then straightforward, with the result
\begin{subequations}
\label{eq:nsjEE2}
\beq
S = N^2 \left [ \frac{R^2}{\varepsilon^2} - \ln \left( \frac{2R}{\varepsilon} \right) - \frac{1}{2} \right ]+ D_1 \, \frac{R}{\varepsilon}   + D_0,
\eeq
\beq
\label{eq:nsjuniversal}
D_0 = N^2 \left [  -\frac{1}{4} + \left(\zeta(\xi_0) - \sqrt{\gamma} \right) \xi_0 - \frac{1}{2} \ln \left(\frac{\sigma(2 \xi_0)}{2 \sqrt{\gamma}} \right) + \frac{\sqrt{\gamma}}{2} \zeta(2 \xi_0) \right],
\eeq
\end{subequations}
where once again we did not bother to compute the non-universal constant $D_1$.
\begin{figure}
\begin{center}
\begin{tabular}{c @{\hspace{0.3in}} c}
\includegraphics[height=1.3in]{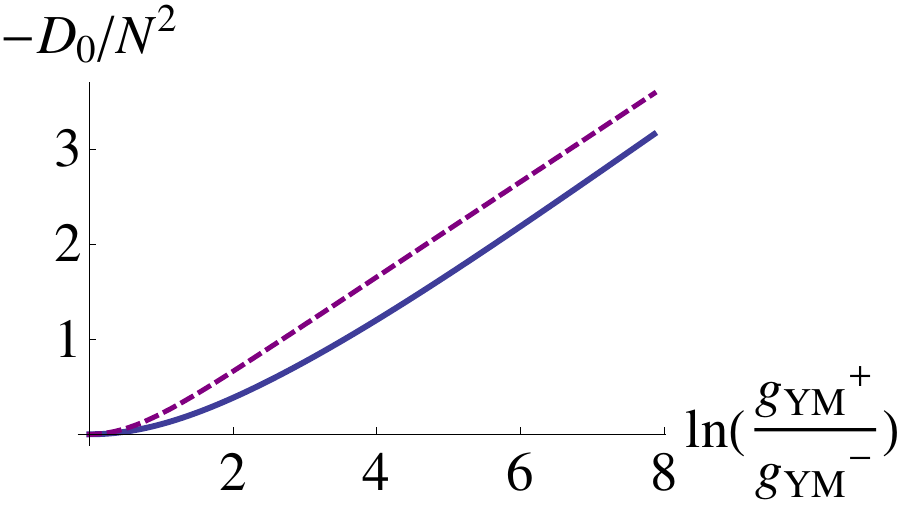}
&
\includegraphics[height=1.3in]{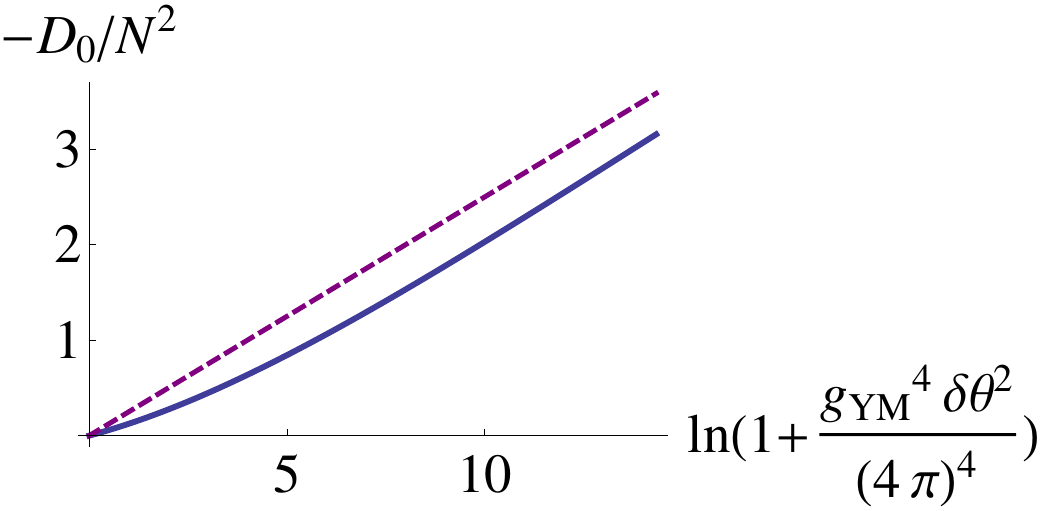}
\end{tabular}
\caption{\label{fig:janus}\textbf{Left:} We plot $-D_0/N^2$, minus the universal constant contribution to the defect EE divided by $N^2$, as a function of $\ln(g_{YM}^+/g_{YM}^-)$ with $\theta^+=\theta^-$ for non-SUSY Janus, eq.~\eqref{eq:nsjuniversal} (solid blue curve), and for SUSY Janus, eq.~\eqref{eq:sjuniversal} (dashed purple curve). \textbf{Right:} We plot $-D_0/N^2$ as a function of $\ln (1+ g_{YM}^2 \delta \theta/(4\pi)^4)$ with $\delta \theta \equiv \theta^+ - \theta^-$ and $g_{YM}^+=g_{YM}^-=g_{YM}$ for non-SUSY Janus (solid blue curve) and SUSY Janus (dashed red curve).}
\end{center}
\end{figure}

Presumably a higher-dimensional $g$-theorem would require that $D_0$ change monotonically under a defect RG flow, and may either increase or decrease under a bulk RG flow. At the moment, we can say little about the behavior of the $D_0$ in eq.~\eqref{eq:nsjuniversal} under defect RG flows. The non-SUSY Janus metric, eq.~\eqref{eq:nonsusyjanusmetric}, depends only on $N$ and $\gamma$, hence the $D_0$ in eq.~\eqref{eq:nsjuniversal} depends only on $N$ and $\gamma$, or equivalently on $N$ and the size of the jump in the complex coupling $\tau$. A defect RG flow cannot change $N$ or the size of the jump in $\tau$: correlators at points arbitrarily far from the defect depend on the values of $N$ and $\tau$, so only a bulk RG flow can change $N$ or the size of the jump. In the next subsection we will provide some speculation about how $D_0$ might change under a certain class of possible defect RG flows in this DCFT.

Under a bulk RG flow in which the only change is the size of the jump in $\tau$ (if such a bulk RG flow exists), the $D_0$ in eq.~\eqref{eq:nsjuniversal} would in fact change monotonically. For example, in figure~\ref{fig:janus} we plot $-D_0/N^2$ first with jumping $g_{YM}$ and non-jumping $\theta$, as a function of $\ln \left(g_{YM}^+/g_{YM}^-\right)$, and then with non-jumping $g_{YM}$ and jumping $\theta$, as a function of $\ln (1+g_{YM}^2 \delta \theta^2/(4\pi)^4)$. In each case we find that $-D_0/N^2$ increases monotonically as the size of the jump in $g_{YM}$ or $\theta$ increases. Presumably, such behavior would be consistent with, but not required by, a higher-dimensional $g$-theorem.

As argued in ref.~\cite{Estes:2012nx}, $\N=4$ SYM with $g_{YM}^+ = g_{YM}^-$ and $\theta^+ = \theta^- + n \pi$ with $n$ an odd integer may be interpreted as the low-energy effective description of a certain $(3+1)$-dimensional time-reversal-invariant fractional topological insulator, which in the Maldacena limits will additionally be strongly-coupled. As proposed in ref.~\cite{Grover:2011fa}, the universal constant contribution to EE may provide one way to detect topological order in $(3+1)$ dimensions. For the case where $g_{YM}^+ = g_{YM}^-$ and $\theta^+ = \theta^- + n \pi$, our result for $D_0$ in eq.~\eqref{eq:nsjuniversal} may be, or at least may contain a contribution from, such topological EE. To what extent our result in eq.~\eqref{eq:nsjuniversal} ``knows'' about topological order is a question we leave for future research.

%%%%%%%%%%%%%%%%%%%%%%%%%%%%%%%%%%%%%%%%%%%%%%%%%%%%%%%%%%%%%%%%%%%%%%%%%%%%%%%%%%%%%%%%%%%%%%%%%%%%%%%%%%%%%%%%%%
\subsection{SUSY Janus}
\label{ssec:Janus}
%%%%%%%%%%%%%%%%%%%%%%%%%%%%%%%%%%%%%%%%%%%%%%%%%%%%%%%%%%%%%%%%%%%%%%%%%%%%%%%%%%%%%%%%%%%%%%%%%%%%%%%%%%%%%%%%%%

In the field theory dual to non-SUSY Janus, the jumping $g_{YM}$ breaks all the SUSY of $\N=4$ SYM. Various amounts of SUSY can be restored by adding to the Lagrangian appropriate defect-localized operators~\cite{D'Hoker:2006uv}. Here we will only consider the maximally SUSY case, preserving eight real supercharges, where the R-symmetry is broken from $SO(6)$ to $SO(4)$. In this case, explicit forms for the defect-localized operators appear in refs.~\cite{D'Hoker:2006uv,Gaiotto:2008sd}.

The holographic dual is the maximally-SUSY Janus solution of type IIB SUGRA~\cite{D'Hoker:2007xy}, which like non-SUSY Janus has two asymptotically $AdS_5 \times \mathbb{S}^5$ regions, each of radius $L$, in which the dilaton can take two distinct values, $\phi_{\pm}$. The metric of maximally-SUSY Janus is of the form in eq.~\eqref{susymet} with the warp factors in eq.~\eqref{susydef} and with the particular harmonic functions \cite{D'Hoker:2007xy}
\begin{align}
h_1(v,\bar{v}) = - i \alpha_1 \sinh \left(v - \frac{\delta \phi}{2}\right) + \text{c.c.}\,, \qquad h_2(v,\bar{v}) = \alpha_2 \cosh\left(v+\frac{\delta \phi}{2}\right) + \text{c.c.}\,,
\end{align}
where $v = x + i y$ with $x \in (-\infty,\infty)$ and $y \in [0,\pi/2]$, and where the real constants $\alpha_1$, $\alpha_2$, and $\delta \phi$ are related to the radius of curvature $L$ and the Yang-Mills coupling $g_{YM}^{\pm}$ as
\beq
L^4 = 16 \, |\alpha_1 \alpha_2| \cosh (\delta \phi)\,, \qquad \frac{(g_{YM}^{\pm})^2}{4\pi} = e^{2 \phi_{\pm}} = \left| \frac{\alpha_2}{\alpha_1}\right| e^{\pm \delta \phi}\,.
\eeq
The $SO(4)$ R-symmetry is dual to the $SO(3) \times SO(3) \simeq SO(4)$ isometry of the two $\mathbb{S}^2$'s in the metric of eq.~\eqref{susymet}. We can obtain new solutions from the maximally-SUSY Janus solution using the $SL(2,\mathbb{R})$ duality of type IIB supergravity. As with non-SUSY Janus, a generic $SL(2,\mathbb{R})$ transformation will generate a non-trivial axion $C_{(0)}$, and so add two additional parameters to the solution, the asymptotic values $C_{(0)}^{\pm}$. Generically, the dual field theory will have jumping $g_{YM}$ and $\theta$, where we again identify $g_{YM}^{\pm}$ and $\theta^{\pm}$ as in eq.~\eqref{eq:janusdictionary}.

In this case the integral for the EE, eq.~\eqref{E:defectSEEfinal} or equivalently eq.~\eqref{E:susyEE}, is
\beq
\label{eq:susyJanarea}
S = \frac{ \vol(\mathbb{S}^1) \vol(\mathbb{S}^2)^2 R L^8}{4G_N} \int_0^{\frac{\pi}{2}} dy \sin^2(y) \cos^2(y) \int_{\varepsilon u^c(y)}^R \frac{du}{2 u^2} \int_{\chi_-\left(\frac{\varepsilon}{u},y\right)}^{\chi_+\left(\frac{\varepsilon}{u},y\right)} dx \left(1 + \frac{\cosh(2x)}{\cosh(\delta \phi)}\right). \nonumber
\eeq
Using eqs.~\eqref{eq:expansion} and~\eqref{E:xFromUZ} with $z=\varepsilon$, we find for the $x$-cutoffs
\begin{align}\label{eq:xcexpansion}
\chi_{\pm}\left(\frac{\varepsilon}{u},y\right) = \pm \ln \left(\frac{2u}{\varepsilon}\right) \pm \frac{1}{2} \ln \cosh (\delta \phi) \mp \frac{4 - \cos(2y) \tanh(\delta \phi)}{16} \, \left(\frac{\varepsilon}{u}\right)^2 + \mathcal{O}\left(\frac{\varepsilon^3}{u^3}\right).
\end{align}
The $x$, $u$, and $y$ integrations in eq.~\eqref{eq:susyJanarea} are then straightforward to perform, with the result
\begin{subequations}
\beq
S = N^2 \left [ \frac{R^2}{\varepsilon^2} - \ln \left( \frac{2R}{\varepsilon} \right) - \frac{1}{2} \right ]+ D_1 \frac{R}{\varepsilon} + D_0,
\eeq
\beq
\label{eq:sjuniversal}
D_0 = - \frac{1}{2} \, N^2 \ln \left(\cosh(\delta \phi)\right),
\eeq
\end{subequations}
where we used $4 G_N= \pi \vol(\mathbb{S}^1) \vol(\mathbb{S}^2) \vol(\mathbb{S}^2) L^8/(16 N^2)$, and once again we did not bother to compute the non-universal constant $D_1$. Using $\delta \phi = \ln (g_{YM}^+/g_{YM}^-)$, and also considering an $SL(2,\mathbb{R})$ transformation to the case with jumping $\theta$ and non-jumping $g_{YM}$, with $g_{YM}^2 \delta \theta = 16 \pi^2 \sinh(\delta \phi)$ as explained below eq.~\eqref{eq:janusdictionary}, we find
\beq
D_0 = \left\{ \begin{array}{lc} - \frac{N^2}{2} \ln \left(1 + \frac{\left(g_{YM}^+ - g_{YM}^-\right)^2}{2 \, g_{YM}^+\, g_{YM}^-}  \right), & \theta^+=\theta^-, \\
 - \frac{N^2}{4} \ln \left(1 + \frac{\delta \theta^2 g_{YM}^4}{256 \pi^4} \right), & g_{YM}^+=g_{YM}^-. \end{array}\right.
\eeq
Clearly $-D_0/N^2$ increases monotonically with the size of the jump in $g_{YM}$ or $\theta$, as we also show in fig.~\ref{fig:janus}.

When we compare the DCFTs dual to non-SUSY and SUSY Janus, the only differences are certain defect-localized operators~\cite{D'Hoker:2006uv,Gaiotto:2008sd}. For the sake of argument, imagine that some defect RG flows between these DCFTs exist, triggered by these defect operators. Furthermore, imagine that a higher-dimensional $g$-theorem exists. Our results for $D_0$ from non-SUSY and SUSY Janus, eqs.~\eqref{eq:nsjuniversal} and~\eqref{eq:sjuniversal}, respectively, would place constraints on the allowed defect RG flows. For example, consider the case where $g_{YM}$ jumps while $\theta^+ = \theta^-$. The left plot in fig.~\ref{fig:janus} shows that the only defect RG flow allowed under these circumstances is from the SUSY DCFT to the non-SUSY DCFT. The right plot in fig.~\ref{fig:janus} shows the same for the case where $\theta$ jumps while $g_{YM}^+ = g_{YM}^-$. Whether these speculations are in fact realized is a question we leave for future research.

As argued in ref.~\cite{Estes:2012nx}, $\N=4$ SYM with $g_{YM}^+ = g_{YM}^-$ and $\theta^+ = \theta^- + n \pi$ with $n$ an odd integer may be interpreted as the low-energy effective description of a certain $(3+1)$-dimensional time-reversal-invariant fractional topological insulator, which with appropriate defect-localized terms will additionally be SUSY. Our statements about topological EE at the end of the previous subsection therefore apply here as well, and in particular, for the case where $g_{YM}^+ = g_{YM}^-$ and $\theta^+ = \theta^- + n \pi$ with $n$ an odd integer, our result for $D_0$ in eq.~\eqref{eq:sjuniversal} may ``know about'' topological order.

%%%%%%%%%%%%%%%%%%%%%%%%%%%%%%%%%%%%%%%%%%%%%%%%%%%%%%%%%%%%%%%%%%%%%%%%%%%%%%%%%%%%%%%%%%%%%%%%%%%%%%%%%%%%%%%%%%
\subsection{M-Theory Janus}
%%%%%%%%%%%%%%%%%%%%%%%%%%%%%%%%%%%%%%%%%%%%%%%%%%%%%%%%%%%%%%%%%%%%%%%%%%%%%%%%%%%%%%%%%%%%%%%%%%%%%%%%%%%%%%%%%%

The original M-theory Janus solution~\cite{D'Hoker:2009gg} is a one-parameter deformation of the $AdS_4 \times \mathbb{S}^7$ solution of eleven-dimensional SUGRA that preserves half the SUSY: the $AdS_4 \times \mathbb{S}^7$ vacuum of M-theory preserves $OSp(8|4,\mathbb{R})$ SUSY, of which M-theory Janus preserves an $OSp(4|2,\mathbb{R}) \times OSp(4|2,\mathbb{R})$ subgroup. In particular, the bosonic subgroup (the isometry) breaks from $SO(3,2) \times SO(8)$ down to $SO(2,2) \times SO(4) \times SO(4)$. M-theory Janus has two asymptotically $AdS_4 \times \mathbb{S}^7$ regions separated by a localized source for the four-form. The dual field theory is ABJM theory with Chern-Simons level $k=1$ deformed by an interface-localized conformal primary operator of dimension two in the $\mathbf{15}$ of $SU(4) \subset SO(8)$~\cite{D'Hoker:2009gg}. Notice that in contrast to the Janus solutions of subsections~\ref{ssec:NSJanus} and~\ref{ssec:Janus}, in this case the coupling does not jump across the interface.

Here we will consider a two-parameter, SUSY-preserving deformation of the original M-theory Janus solution. First, we introduce the deformation of refs.~\cite{Estes:2012vm,Bobev:2013yra}, involving one new parameter\footnote{The $\gamma$ here should not be confused with the $\gamma$ of subsection~\ref{ssec:NSJanus}.}, $\gamma \in [0,\infty)$, which generically deforms $OSp(4|2,\mathbb{R}) \times OSp(4|2,\mathbb{R})$ to $D(2,1;\gamma,0) \times D(2,1;\gamma,0)$. Only when $\gamma=1$, where $D(2,1;\gamma,0) = OSp(4|2,\mathbb{R})$, is the super-isometry a subgroup of $OSp(8|4,\mathbb{R})$. When $\gamma \neq 1$ the dual field theory is thus not merely ABJM theory deformed by a defect-localized operator, although exactly what deformation $\gamma$ represents is currently unknown. We will additionally orbifold, producing solutions with two asymptotically $AdS_4 \times \mathbb{S}^7/\mathbb{Z}_k$ regions, where $k \in \mathbb{Z}$ is our second deformation parameter. When $\gamma=1$, the dual field theory is then ABJM theory with Chern-Simons level $k\geq1$ deformed by a defect-localized operator.

The metric for this two-parameter deformation of M-theory Janus is~\cite{D'Hoker:2009gg,Estes:2012vm,Bobev:2013yra}
\beq
g = f_1^2 \, g_{AdS_3} + \rho^2 \, dw d\bar w + f_2^2 \, g_{\mathbb{S}^3} + f_3^2 \, g_{\bar{\mathbb{S}}^3}\,,
\eeq
where $g_{\mathbb{S}^3}$ and $g_{\bar{\mathbb{S}}^3}$ are metrics on two unit-radius $\mathbb{S}^3$'s, $w = x/2 + i y$ is a complex coordinate defined on a strip, $x \in (-\infty,\infty)$ and $y \in [0,\pi/2]$, and the warp factors $f_1^2$, $\rho^2$, $f_2^2$, and $f_3^2$ depend only on $w$ and $\bar{w}$. For the solution with super-isometry $D(2,1;\gamma,0) \times D(2,1;\gamma,0)$, the warp factors are specified by a harmonic function $h(w,\bar{w})$ and a complex function $H(w,\bar{w})$:
\begin{align}
&f_1 ^6 =  \frac{h^2 W_+ W_-}{ C_1 ^6 \, ( H \bar{H} -1)^2}\,, & &\rho^6 =  \frac{ |\p_w h |^6}{C_2^3 C_3^3\,  h^4 }  ( H \bar{H} -1) W_+ W_-\,,& \cr
&f_2 ^6 =  \frac{ h^2 ( H \bar{H} -1) W_-}{C_2 ^3 C_3^3\,  W_+^2}\,, & &f_3 ^6 =  \frac{ h^2 ( H \bar{H} -1) W_+ }{ C_2 ^3 C_3^3\, W_-^2}\,,& \nonumber
\end{align}
\beq
W_{\pm} \equiv |H \pm i|^2 + \gamma^{\pm1} (H \bar{H})-1\,, \nonumber
\eeq
\beq
h = -2 i \alpha [\sinh(2w) - \sinh(2 \bar w)]\,, \qquad H =i \frac{\cosh(w+\bar w) + \lambda \sinh(w-\bar w)}{\cosh(2 \bar w)}\,, \nonumber
\eeq
where $2 \p_w H = (H+\bar{H}) \p_w \ln h$. The real constants $\{C_1,C_2,C_3\}$ obey $C_1 + C_2 + C_3 = 0$, with $\gamma = C_2/C_3$. Specifying $\gamma$ thus uniquely determines $\{C_1,C_2,C_3\}$ up to an overall scale, which we can absorb into the normalization of $h$. The solution is also invariant under $\gamma \to 1/\gamma$. The parameter $\lambda \in (-\infty,\infty)$ is that of the original M-theory Janus solution~\cite{D'Hoker:2009gg}. The real constant $\alpha$, along with $\gamma$ and $\lambda$, determines the radius of the asymptotically $AdS_4 \times \mathbb{S}^7$ regions as $L^6 = \alpha^2 (1+ \lambda^2)/|C_3|^6 \gamma^3$. To recover exactly $AdS_4 \times \mathbb{S}^7$, we simply take $\lambda =0$ and $\gamma=1$.

To orbifold, we first embed the $\mathbb{S}^7$ into $\mathbb{R}^8$ with complex coordinates $Z_i$ with $i=1,2,3,4$, with $\mathbb{S}^7$ the set of points obeying $|Z_1|^2 + |Z_2|^2 + |Z_3|^2 + |Z_4|^2 = 1$. Writing $Z_i = e^{i \psi} z_i$ gives us the Hopf fibration of $\mathbb{S}^7$, with $\psi$ the coordinate of the $U(1)$ fiber over $\mathbb{CP}^3$. The ABJM orbifold~\cite{Aharony:2008ug} consists of shifting $\psi$'s periodicity from $2\pi$ to $2\pi/k$. For M-theory Janus we describe $\mathbb{S}^7$ as $\mathbb{S}^3 \times \mathbb{S}^3$ fibered over a line segment $y \in [0,\pi/2]$. To implement the ABJM orbifold we use the Hopf fibration of each $\mathbb{S}^3$, with $U(1)$ Hopf fiber coordinates $\theta_1$ and $\theta_2$, define $\theta_2 \equiv \psi$ and $\theta_1 \equiv \psi + \phi$, and then shift $\psi$'s periodicity from $2\pi$ to $2\pi/k$ while keeping the periodicity of $\phi = \theta_1 - \theta_2$ fixed at $2\pi$. The result is a smooth geometry with two asymptotically $AdS_4 \times \mathbb{S}^7/\mathbb{Z}_k$ regions.

For these solutions, the integral for the EE, eq.~\eqref{E:defectSEEfinal}, is
\begin{align}
%\label{eq:abjmjanusEE}
S =  \frac{R\, \vol(\mathbb{S}^3)^2}{4 G_Nk} \frac{2^8 L^9 \sqrt{\gamma}}{(1+\gamma)\sqrt{1+\lambda^2}}\int_0^{\frac{\pi}{2}} dy \sin^3(y) \cos^3(y) \int_{\varepsilon u^c(y)}^R \frac{du}{u \sqrt{R^2 - u^2}}\int_{\chi_-\left(\frac{\varepsilon}{u},y\right)}^{\chi_+\left(\frac{\varepsilon}{u},y\right)} dx \cosh(x),
\end{align}
where we used $\vol(\mathbb{S}^0) = 2$, as mentioned below eq.~\eqref{eq:sphereee1}. Using eqs.~\eqref{eq:expansion} and~\eqref{E:xFromUZ} with $z=\varepsilon$, we find for the $x$-cutoffs
\beq
\label{eq:xcABJM}
\chi_{\pm}\left(\frac{\varepsilon}{u},y\right) = \pm \ln \left( \frac{2 u}{\varepsilon} \right) \pm \ln \left( \frac{(1+\gamma)\sqrt{1 + \lambda^2}}{2 \sqrt{\gamma}} \right) + \frac{2 \sqrt{\gamma} \lambda \cos(2y)}{3(1+\gamma)\sqrt{1+\lambda^2}} \frac{\varepsilon}{u} + \mathcal{O}\left(\frac{\varepsilon^2}{u^2}\right).
\eeq
The $x$, $u$, and $y$ integrations are then straightforward to perform. As we show in the appendix, generically for a holographic DCFT in $d=3$, $S$ will take the form in eq.~\eqref{E:Sdefect}, including a contribution $D_{log} \ln\left(\frac{2R}{\varepsilon}\right)$ with universal coefficient $D_{log}$. In the current case we find $D_{log}=0$. To see why, we simply perform the $x$ integration: if a term constant in $u$ appears, then multiplying by $\frac{1}{u\sqrt{R^2 - u^2}}$ and integrating in $u$ will produce a logarithm. The $x$-integration does not produce a term constant in $u$, however:
\beq
\label{eq:abjmjanusxint}
\int_{x_-\left(\frac{\varepsilon}{u},y\right)}^{x_+\left(\frac{\varepsilon}{u},y\right)} dx \cosh(x) = \frac{(1+\gamma)\sqrt{1+\lambda^2}}{\sqrt{\gamma}} \, \frac{u}{\varepsilon} + \mathcal{O}\left(\frac{\varepsilon}{u}\right),
\eeq
so no logarithm will appear in $S$. Indeed, using $4 G_N = 2^6 \pi^7 l_p^9$ and $L^6 = \frac{\pi}{2} k N l_P^6$ we find
\beq
\label{eq:ABJMJanus}
S = \frac{\pi \sqrt{2}}{3} k^{1/2} N^{3/2}  \frac{\ell}{\varepsilon}+ \tilde{D}_0,
\eeq
where we did not bother to compute the non-universal constant $\tilde{D}_0$.\footnote{The absence of the $\ln\left(\frac{2R}{\varepsilon}\right)$ contribution to $S$ suggests that perhaps $\tilde{D}_0$ is universal. In the appendix we show that this is not the case. $\tilde{D}_0$ is sensitive to the contributions to  $\chi_{\pm}\left(\frac{\varepsilon}{u},y\right)$ of higher order in $\varepsilon/u$ and moreover depends on the choice of $u$-cutoff $u^c(y^a)$.} We thus find $D_{log}=0$, as advertised. The bottom-up models of holographic $d=3$ DCFTs in ref.~\cite{Korovin:2013gha} also had $D_{log}=0$. For an example where $D_{log} \neq 0$, see section 3.2.3 of ref.~\cite{Jensen:2013lxa}.

We may be tempted to think of $D_{log}$ as a ``central charge'' counting degrees of freedom localized to the defect. As discussed in refs.~\cite{Jensen:2013lxa,Korovin:2013gha}, however, we must be careful what we mean by ``central charge.'' For a CFT on a curved manifold in $d=2$, we can extract the central charge from the coefficient of the Ricci scalar in the Weyl anomaly. For a CFT on a curved manifold in $d=3$ with a codimension-one defect along some curve, the Weyl anomaly will be a delta function at the curve times a linear combination of various terms involving not only the Ricci scalar but also the second fundamental form of the embedding. The coefficients of these terms provide a set of central charges that characterize the DCFT. Wess-Zumino consistency fixes some of these central charges in terms of others. Presumably $D_{log}$ is some linear combination of these central charges. Our result $D_{log}=0$ indicates that, for the DCFTs above, obtained as deformations of large-$N$, strongly-coupled ABJM theory, that particular linear combination vanishes. Whether $D_{log}$ counts defect degrees of freedom and changes monotonically along a defect RG flow remain open questions.

\acknowledgments

We would like to thank K.~Balasubramanian, C.~Herzog, A.~Karch, Y.~Korovin, and T. Takayanagi for reading and commenting on the manuscript. We also thank V.~Ker\"anen and R.~Myers for illuminating discussions. K.J. was supported in part by NSERC, Canada as well as by the National Science Foundation under grant PHY-0969739. A.O'B. was supported by a University Research Fellowship from the Royal Society. T.W. was supported by a Research Fellowship (Grant number WR 166/1-1) from the German Research Foundation (DFG).

%%%%%%%%%%%%%%%%%%%%%%%%%%%%%%%%%%%%%%%%%%%%%%%%%%%%%%%%%%%%%%%%%%%%%%%%%%%%%%%%%%%%%%%%%%%%%%%%%%%%%%%%%%%%%%%%%%%
\appendix
\section*{Appendix: Cutoff Prescription}
\addcontentsline{toc}{section}{Appendix: Cutoff Prescription}
%%%%%%%%%%%%%%%%%%%%%%%%%%%%%%%%%%%%%%%%%%%%%%%%%%%%%%%%%%%%%%%%%%%%%%%%%%%%%%%%%%%%%%%%%%%%%%%%%%%%%%%%%%%%%%%%%%%

\setcounter{equation}{0}
\renewcommand{\theequation}{A.\arabic{equation}}

In this appendix, we study in detail the integral for the EE of a sphere centered on the defect in a holographic DCFT, eq.~\eqref{E:defectSEEfinal},
\beq
\label{E:defectSEEfinalapp}
S= \frac{\text{vol}(\mathbb{S}^{d-3})R}{4G_N} \int dy^a \int_{\varepsilon u^c(y^a)}^R du \int_{\chi_-\left(\frac{\varepsilon}{u},y^a\right)}^{\chi_+\left( \frac{\varepsilon}{u},y^a\right)}dx \,\sqrt{\text{det}\,G}\rho f^{d-2} \frac{(R^2-u^2)^{(d-4)/2}}{u^{d-2}}.
\eeq
We leave the $u$-cutoff, $u^c(y^a)$, an arbitrary function of the internal coordinates $y^a$. We choose the $x$-cutoffs $\chi_{\pm}\left(\frac{\varepsilon}{u},y^a\right)$ to reproduce the FG cutoff $z = \varepsilon$ in the two FG patches by taking $\chi_{\pm}\left(\frac{\varepsilon}{u},y^a\right) = x_{\pm}\left(\frac{\varepsilon}{u},y^a\right)$ using the $x_{\pm}\left(\frac{z}{u},y^a\right)$ in eq.~\eqref{E:xFromUZ}. Although we will only explicitly discuss DCFTs in this appendix, our results are straightforward to generalize to BCFTs by taking $\chi_-\left(\frac{\varepsilon}{u},y^a\right) \to -\infty$, as mentioned at the end of subsection~\ref{S:generalDefect}. With these choices, we will prove four things about the integral in eq.~\eqref{E:defectSEEfinalapp}. First, we show that the integral takes the form
\begin{align}
\begin{split}
\label{app:sphericalEE}
	S = \left\{
		\begin{array}{lc}
			C_1\frac{R}{\varepsilon}+C^{(d=3)}_{log}\ln\left( \frac{2R}{\varepsilon}\right)+C^{(d=3)}_0, & d=3,
			\\
			C_2 \frac{R^2}{\varepsilon^2}+C_1\frac{R}{\varepsilon}+C^{(d=4)}_{log}\ln\left( \frac{2R}{\varepsilon}\right) + C_0^{(d=4)}, & d=4,
\end{array}\right.
\end{split}
\end{align}
where the various $C$'s are $R$- and $\varepsilon$-independent constants, and we have neglected terms that vanish as $\varepsilon \rightarrow 0$.  Second, we show that $C^{(d=3)}_0$ and $C_1$ both depend on the choice of $u^c(y^a)$, whereas $C^{(d=3)}_{log}$, $C^{(d=4)}_{log}$ and $C_0^{(d=4)}$ are all independent of the choice of $u^c(y^a)$. Third, we show that $C^{(d=3)}_{log}$ only depends on terms up to and including order $\frac{\varepsilon}{u}$ in $\chi_{\pm}\left(\frac{\varepsilon}{u},y^a\right)$, while $C^{(d=4)}_{log}$ and $C_0^{(d=4)}$ only depend on terms up to and including order $(\frac{\varepsilon}{u})^2$. In other words, we may deform the cutoffs $\chi_{\pm}\left(\frac{\varepsilon}{u},y^a\right)$ in any way that we like at higher orders in $\frac{\varepsilon}{u}$ without changing $C^{(d=3)}_{log}$, $C^{(d=4)}_{log}$, or $C_0^{(d=4)}$. Fourth, we show that $C^{(d=3)}_{log}$, $C^{(d=4)}_{log}$ and $C_0^{(d=4)}$ all depend on the entire bulk geometry, not just the asymptotic regions or the part near the defect.

As discussed in subsection~\ref{S:generalDefect}, the FG patches do not cover the entire space (recall fig.~\ref{fig:FGpatches}), so we begin by splitting the $x$-integration into three regions,
\bea
\label{E:intsplit}
S &=& \frac{\vol(\mathbb{S}^{d-3}) R}{4G_N} \int dy^a \int_{\varepsilon u^c(y^a)}^R du \frac{(R^2-u^2)^\frac{d-4}{2}}{u^{d-2}} \\
&\times & \ls \int_{\chi_-\left(\frac{\varepsilon}{u},y^a\right)}^{x_-^c} dx \sqrt{\text{det}\,G} \, \rho f^{d-2} + \int_{x_-^c}^{x_+^c} dx \sqrt{\text{det}\,G} \, \rho f^{d-2} + \int_{x_+^c}^{\chi_+\left(\frac{\varepsilon}{u},y^a\right)} dx \sqrt{\text{det}\,G} \, \rho f^{d-2} \rs. \nonumber
\eea
where $x_{\pm}^c$ are arbitrary parameters. Since the integrand $\sqrt{\text{det}\,G} \rho f^{d-2}$ is smooth, the final value of $S$ is independent of the choice of $x_{\pm}^c$. Next we expand the integrand in $e^{\pm x}$ in the asymptotic regions $x \to \pm \infty$, using the expansions of the metric functions in eq.~\eqref{eq:expansion},
\beq
\label{eq:detexpansion}
\sqrt{\text{det}\,G}\, \rho f^{d-2} = \sum_{n=2-d}^\infty A^{(n)}_{\pm} e^{\mp n x}\,,
\eeq
where the coefficients $A^{(n)}_{\pm}$ depend on $u$ and the $y^a$, and are straightforward to determine from eq.~\eqref{eq:expansion}. A similar expansion is possible in the middle region between the two FG patches, but is not necessary to evaluate the integral. Integrating in $x$,
% and canceling some contributions from the antiderivative of $\sqrt{\text{det}\,G}\,\rho f^{d-2}$ at $x_{\pm}^c$, 
we find
\begin{align}
\label{E:EEexpand}
S = \frac{\vol(\mathbb{S}^{d-3}) R}{4G_N} \int dy^a \int_{\varepsilon u^c(y^a)}^R du \frac{(R^2-u^2)^\frac{d-4}{2}}{u^{d-2}} & \left[ A^{(0)}_{+} \chi_+ - A^{(0)}_{-} \chi_- +\mathcal{C} \right . \\
& \left . - \sum_{\substack{n\neq0 \\ n=2-d}}^\infty \frac{A^{(n)}_{+}e^{-n \chi_+}+A^{(n)}_{-}e^{n \chi_-}}{n}  \right],\nn
\end{align}
where $\mathcal{C}$ is the leftover contribution from the middle region after canceling the $x_{\pm}^c$-dependent terms from the left and right regions. Next we need the $x$-cutoffs,
\beq
\label{eq:xpmexpansion}
\chi_{\pm}\left( \frac{\varepsilon}{u},y^a\right) = \pm \left[ \ln\left( \frac{2u}{\varepsilon} \right) - c_{\pm} + \sum_{m=1}^\infty X_{\pm}^{(m)} \left(\frac{\varepsilon}{u}\right)^m \right],
\eeq
where the coefficients $X_{\pm}^{(m)}$ are functions of the $y^a$ only. The first few $X_{\pm}^{(m)}$ appear explicitly in eq.~\eqref{E:xFromUZ}. Plugging these $\chi_{\pm}\left( \frac{\varepsilon}{u},y^a\right)$ into eq.~\eqref{E:EEexpand}, we find
\beq
\label{E:EEexpand2}
S = \frac{\vol(\mathbb{S}^{d-3}) R}{4G_N} \int dy^a \int_{\varepsilon u^c(y^a)}^R du \frac{(R^2-u^2)^\frac{d-4}{2}}{u^{d-2}}  \left[  \left ( A^{(0)}_{+}+A^{(0)}_{-} \right )  \ln\left( \frac{2u}{\varepsilon} \right) +\sum_{l=2-d}^\infty \coe_l \lp\frac{\varepsilon}{2 u}\rp^{l}  \right],
\eeq
where the coefficients $\coe_l$ depend only on the $y^a$,
\ba
\label{eq:Cns}
\sum_{l=2-d}^\infty \coe_l \lp\frac{\varepsilon}{2 u}\rp^{l} &=& - \sum_{\substack{n\neq0 \\ n=2-d}}^\infty \frac{A^{(n)}_+e^{n c_+ - n ( \sum_{m=1}^\infty X_+^{(m)} \left(\frac{\varepsilon}{u}\right)^m)} + A^{(n)}_-e^{n c_- - n ( \sum_{m=1}^\infty X_-^{(m)} \left(\frac{\varepsilon}{u}\right)^m )}}{n}\lp \frac{\varepsilon}{2 u}\rp^{n} \cr
&& +\,\mathcal{C} -A^{(0)}_+ c_+- A^{(0)}_- c_-+ \sum_{m=1}^\infty \lp A^{(0)}_+  X_+^{(m)} +A^{(0)}_-  X_-^{(m)} \rp\left(\frac{\varepsilon}{u}\right)^m.
\ea
Explicitly, for $d=4$ and $d=3$ the first few $\coe_l$ are
\bea
\label{eq:Cd4}
d=4: \qquad& \coe_{-2} &= \frac{A^{(-2)}_+ e^{-2 c_+}+A^{(-2)}_- e^{-2 c_-}}{2}, \nonumber \\
&\coe_{-1} &= A^{(-1)}_+ e^{- c_+}+A^{(-1)}_- e^{-c_-} +2 \lp A^{(-2)}_+ e^{-2 c_+} X_+^{(1)} +A^{(-2)}_- e^{-2 c_-} X_-^{(1)} \rp, \nonumber \\
&\coe_0 &= \mathcal{C} - A_0^{(+)} c_+-A_0^{(-)} c_- +4 \lp A^{(-2)}_+ e^{-2 c_+} X_+^{(2)} +A^{(-2)}_- e^{-2 c_-} X_-^{(2)} \rp \cr
&&+ 2\left[ A^{(-1)}_+ e^{- c_+} X_+^{(1)} +A^{(-1)}_- e^{-c_-} X_-^{(1)} \right . \\
&& \left . +2 \lp A^{(-2)}_+ e^{-2 c_+} \lp X_+^{(1)}\rp^2 +A^{(-2)}_- e^{-2 c_-} \lp X_-^{(1)}\rp^2 \rp \right],  \cr\cr
d=3: \qquad & \coe_{-1} &= A^{(-1)}_{+} e^{-c_+}+A^{(-1)}_{-} e^{-c_-} \,,\cr
&\coe_{0} &= \mathcal{C} - A_{0}^{(+)} c_+ - A_{0}^{(-)} c_- +2 \lp A^{(-1)}_{+} e^{-c_+} X_+^{(1)} +A^{(-1)}_{-} e^{-c_-}  X_-^{(1)}\rp . \label{eq:Cd3}
\eea

Now we perform the $u$ integration. For $d=4$ we find
\bea
S &=& \frac{\vol(\mathbb{S}^{1}) R}{4G_N} \int dy^a \int_{\varepsilon u^c(y^a)}^R du   \left[  \left ( A^{(0)}_++A^{(0)}_- \right)  u^{-2} \ln\left( \frac{2u}{\varepsilon} \right) +\sum_{l=-2}^\infty \coe_l \left(\frac{\varepsilon}{2}\right)^l u^{-l-2}  \right] \nonumber \\
&=& \frac{\vol(\mathbb{S}^{1})}{4G_N} \int dy^a \left[ 4 \coe_{-2} \left( \frac{R}{\varepsilon} \right)^2 + 2 \coe_{-1} \left( \frac{R}{\varepsilon} \right) \log\left( \frac{R}{\varepsilon u^c(y^a)}\right) \right . \nonumber \\
&& + \lp \sum_{\substack{l\neq-1 \\ l=-2}}^\infty \frac{\coe_l }{(l+1)(u^c(y^a))^{l+1}2^l} + \frac{ A^{(0)}_++A^{(0)}_-}{u^c(y^a)} \lp 1+\ln\left(2 u^c(y^a) \right) \rp\rp \lp\frac{R}{\varepsilon}\rp \nonumber \\
&&\left . - \lp A^{(0)}_++A^{(0)}_- \rp \ln\left( \frac{2R}{\varepsilon} \right) - \lp A^{(0)}_++A^{(0)}_- + \coe_0 \rp\right]+\mathcal{O}(\varepsilon). \nonumber
\eea
We will not perform the integration over $y^a$, since that will not change the $\varepsilon$-dependence of the terms. We thus find the expected leading divergent term, $\propto Y_{-2}\left(\frac{R}{\varepsilon}\right)^2$. A straightforward exercise using eq.~\eqref{eq:Cns} shows that because the metric asymptotically approaches $AdS_5$, the value of $Y_{-2}$ is exactly the same as in $AdS_5$. The first sub-leading term, $\propto Y_{-1} \frac{R}{\varepsilon}\log\left( \frac{R}{\varepsilon u^c(y^a)}\right)$, is unexpected. Fortunately, we can show that in general $Y_{-1}=0$: if we integrate over the $y^a$, then we find an asymptotically $AdS_5$ spacetime coupled to scalars that are massless due to the defect conformal symmetry, and Einstein's equations then force $Y_{-1}=0$. The next sub-leading term, $\propto \frac{R}{\varepsilon}$, depends on the $Y_l$ for all $l$, and explicitly depends on the choice of $u^c(y^a)$. In other words, $C_1$ in eq.~\eqref{app:sphericalEE} depends on $u^c(y^a)$, as advertised. The $\ln\left( \frac{2R}{\varepsilon} \right)$ term and constant term depend on $c_{\pm}$, $A_{\pm}^{(0)}$, $A_{\pm}^{(1)}$, $A_{\pm}^{(2)}$, $X_{\pm}^{(1)}$ and $X_{\pm}^{(2)}$, which ultimately depend only on the expansion coefficients in eq.~\eqref{eq:expansion}, and on $\mathcal{C}$, whose value depends on the entire geometry, not just the asymptotic regions or the part near the defect. We have thus shown that the coefficients $C^{(d=4)}_{log}$ and $C_0^{(d=4)}$ in eq.~\eqref{app:sphericalEE} are independent of the choice of $u^c(y^a)$, and depend on the entire geometry, but depend on the $x$-cutoffs $\chi_{\pm}\left(\frac{\varepsilon}{u},y^a\right)$ only up to order $\left(\frac{\varepsilon}{u}\right)^2$, as advertised.

Next we consider $d=3$. Plugging $d=3$ into eq.~\eqref{E:EEexpand2}, and using our convention that $\vol(\mathbb{S}^{0})=2$, we find
\begin{align}
\label{E:EEexpand3}
S = \frac{R}{2G_N} \int dy^a \int_{\varepsilon u^c(y^a)}^R \frac{du}{u \sqrt{R^2-u^2}}  \left[  \lp A^{(0)}_++A^{(0)}_- \rp  \ln\left( \frac{2u}{\varepsilon} \right) +\sum_{l=-1}^\infty \coe_l \lp\frac{\varepsilon}{2 u}\rp^{l}  \right].
\end{align}
The integral over $u$ of the $\ln\left( \frac{2u}{\varepsilon} \right)$ term leads to a term $\propto  \lp A^{(0)}_++A^{(0)}_- \rp \lp\ln\left( 2 R/\varepsilon \right)\rp^2$, which is unexpected. Fortunately, we can show that in general  $A^{(0)}_++A^{(0)}_-=0$: if we integrate over the $y^a$, then we find an asymptotically $AdS_4$ spacetime coupled to scalars that are massless due to the defect conformal symmetry, and Einstein's equations then force $A^{(0)}_++A^{(0)}_-=0$. To advance further, we must compute the integral
\begin{align}
{\cal I}_{l} = R \int_{\varepsilon u^c(y^a)}^R \frac{du}{u \sqrt{R^2-u^2}}  \lp\frac{\varepsilon}{2 u}\rp^l,
\end{align}
for integer $l$, at least in the small-$\varepsilon$ limit. This integral is straightforward for all $l\geq -1$. For $l=-1,0$, we simply change variables as $u=R \sin(\theta)$. For $l>0$, we use a recursion relation for $\int d\theta \sin^n(\theta)$. We thus find, in the small-$\varepsilon$ limit,
\beq
{\cal I}_{-1} = \frac{\pi R}{\varepsilon} - 2 u^c(y^a) +\mathcal{O}(\varepsilon), \qquad {\cal I}_{0} = - \ln \left(\frac{\varepsilon u^c(y^a)}{2 R} \right) +\mathcal{O}(\varepsilon), \nonumber
\eeq
\beq
{\cal I}_{l} = \frac{1}{l (2 u^c(y^a))^l} +\mathcal{O}(\varepsilon) \qquad (l > 0). \nonumber
\eeq
Using these integrals in eq.~\eqref{E:EEexpand3}, we find
\beq
\label{E:EEd3app}
S = \frac{1}{2G_N} \int dy^a \left[\pi \coe_{-1} \frac{R}{\varepsilon} + \coe_0 \ln \lp \frac{2R}{\varepsilon} \rp - \coe_0 \ln \lp \frac{u^c(y^a)}{2}\rp + \sum_{\substack{l\neq0 \\ l=-1}}^\infty \frac{\coe_l}{l(2 u^c(y^a))^l}\right] + {\mathcal{O}}\left(\varepsilon\right).
\eeq
Here again we will not perform the integration over the $y^a$, since that will not change the $\varepsilon$-dependence of the terms. We thus find the expected leading divergent term, $\propto \coe_{-1} \frac{R}{\varepsilon}$. A straightforward exercise using eq.~\eqref{eq:Cns} shows that because the metric asymptotically approaches $AdS_4$, the value of $Y_{-1}$ is exactly the same as in $AdS_4$. The first sub-leading term, $\propto \coe_0 \ln \lp \frac{2R}{\varepsilon} \rp$, depends on $c_{\pm}$ , $A^{(1)}_{\pm}$, and $X_{\pm}^{(1)}$ via eq.~\eqref{eq:Cd3}, which ultimately depend only on the expansion coefficients in eq.~\eqref{eq:expansion}, and on $\mathcal{C}$, whose value depends on the entire geometry, not just the asymptotic regions or the part near the defect. We have thus shown that the coefficient $C^{(d=3)}_{log}$ in eq.~\eqref{app:sphericalEE} is independent of the choice of $u^c(y^a)$, and depends on the entire geometry, but depends on the $x$-cutoffs $\chi_{\pm}\left(\frac{\varepsilon}{u},y^a\right)$ only up to order $\left(\frac{\varepsilon}{u}\right)$, as advertised. On the other hand, the constant term in eq.~\eqref{E:EEd3app} depends on the $Y_l$ for all $l$, and explicitly depends on the choice of $u^c(y^a)$. In other words, $C^{(d=3)}_0$ in eq.~\eqref{app:sphericalEE} depends on $u^c(y^a)$, as advertised.

\bibliographystyle{JHEP}
\bibliography{defectee}

\providecommand{\href}[2]{#2}\begingroup\raggedright\begin{thebibliography}{10}

\bibitem{Kitaev:2005dm}
A.~Kitaev and J.~Preskill, {\it {Topological Entanglement Entropy}},  {\em
  Phys.Rev.Lett.} {\bf 96} (2006) 110404,
  [\href{http://xxx.lanl.gov/abs/hep-th/0510092}{{\tt hep-th/0510092}}].

\bibitem{Levin:2006zz}
M.~Levin and X.-G. Wen, {\it {Detecting Topological Order in a Ground State
  Wave Function}},  {\em Phys.Rev.Lett.} {\bf 96} (2006) 110405,
  [\href{http://xxx.lanl.gov/abs/cond-mat/0510613}{{\tt cond-mat/0510613}}].

\bibitem{Bombelli:1986rw}
L.~Bombelli, R.~Koul, J.~Lee, and R.~Sorkin, {\it {A Quantum Source of Entropy
  for Black Holes}},  {\em Phys.Rev.} {\bf D34} (1986) 373--383.

\bibitem{Srednicki:1993im}
M.~Srednicki, {\it {Entropy and Area}},  {\em Phys.Rev.Lett.} {\bf 71} (1993)
  666--669, [\href{http://xxx.lanl.gov/abs/hep-th/9303048}{{\tt
  hep-th/9303048}}].

\bibitem{Holzhey:1994we}
C.~Holzhey, F.~Larsen, and F.~Wilczek, {\it {Geometric and Renormalized Rntropy
  in Conformal Field Theory}},  {\em Nucl.Phys.} {\bf B424} (1994) 443--467,
  [\href{http://xxx.lanl.gov/abs/hep-th/9403108}{{\tt hep-th/9403108}}].

\bibitem{Calabrese:2004eu}
P.~Calabrese and J.~Cardy, {\it {Entanglement Entropy and Quantum Field
  Theory}},  {\em J.Stat.Mech.} {\bf 0406} (2004) P06002,
  [\href{http://xxx.lanl.gov/abs/hep-th/0405152}{{\tt hep-th/0405152}}].

\bibitem{Myers:2010xs}
R.~Myers and A.~Sinha, {\it {Seeing a c-theorem with Holography}},  {\em
  Phys.Rev.} {\bf D82} (2010) 046006,
  [\href{http://xxx.lanl.gov/abs/1006.1263}{{\tt arXiv:1006.1263}}].

\bibitem{Myers:2010tj}
R.~Myers and A.~Sinha, {\it {Holographic c-theorems in Arbitrary Dimensions}},
  {\em JHEP} {\bf 1101} (2011) 125,
  [\href{http://xxx.lanl.gov/abs/1011.5819}{{\tt arXiv:1011.5819}}].

\bibitem{Casini:2011kv}
H.~Casini, M.~Huerta, and R.~Myers, {\it {Towards a Derivation of Holographic
  Entanglement Entropy}},  {\em JHEP} {\bf 1105} (2011) 036,
  [\href{http://xxx.lanl.gov/abs/1102.0440}{{\tt arXiv:1102.0440}}].

\bibitem{Zamolodchikov:1986gt}
A.~Zamolodchikov, {\it {Irreversibility of the Flux of the Renormalization
  Group in a 2D Field Theory}},  {\em JETP Lett.} {\bf 43} (1986) 730--732.

\bibitem{Jafferis:2011zi}
D.~Jafferis, I.~Klebanov, S.~Pufu, and B.~Safdi, {\it {Towards the F-Theorem:
  N=2 Field Theories on the Three-Sphere}},  {\em JHEP} {\bf 1106} (2011) 102,
  [\href{http://xxx.lanl.gov/abs/1103.1181}{{\tt arXiv:1103.1181}}].

\bibitem{Casini:2012ei}
H.~Casini and M.~Huerta, {\it {On the RG Running of the Entanglement Entropy of
  a Circle}},  {\em Phys.Rev.} {\bf D85} (2012) 125016,
  [\href{http://xxx.lanl.gov/abs/1202.5650}{{\tt arXiv:1202.5650}}].

\bibitem{Cardy:1988cwa}
J.~Cardy, {\it {Is There a c Theorem in Four-Dimensions?}},  {\em Phys.Lett.}
  {\bf B215} (1988) 749--752.

\bibitem{Komargodski:2011vj}
Z.~Komargodski and A.~Schwimmer, {\it {On Renormalization Group Flows in Four
  Dimensions}},  {\em JHEP} {\bf 1112} (2011) 099,
  [\href{http://xxx.lanl.gov/abs/1107.3987}{{\tt arXiv:1107.3987}}].

\bibitem{Azeyanagi:2007qj}
T.~Azeyanagi, A.~Karch, T.~Takayanagi, and E.~Thompson, {\it {Holographic
  Calculation of Boundary Entropy}},  {\em JHEP} {\bf 0803} (2008) 054--054,
  [\href{http://xxx.lanl.gov/abs/0712.1850}{{\tt arXiv:0712.1850}}].

\bibitem{Cardy:1989ir}
J.~Cardy, {\it {Boundary Conditions, Fusion Rules and the Verlinde Formula}},
  {\em Nucl.Phys.} {\bf B324} (1989) 581.

\bibitem{Affleck:1991tk}
I.~Affleck and A.~Ludwig, {\it {Universal Non-integer 'Ground State Degeneracy'
  in Critical Quantum Systems}},  {\em Phys.Rev.Lett.} {\bf 67} (1991)
  161--164.

\bibitem{Friedan:2003yc}
D.~Friedan and A.~Konechny, {\it {On the Boundary Entropy of One-dimensional
  Quantum Systems at Low Temperature}},  {\em Phys.Rev.Lett.} {\bf 93} (2004)
  030402, [\href{http://xxx.lanl.gov/abs/hep-th/0312197}{{\tt
  hep-th/0312197}}].

\bibitem{Green:2007wr}
D.~R. Green, M.~Mulligan, and D.~Starr, {\it {Boundary Entropy Can Increase
  Under Bulk RG Flow}},  {\em Nucl.Phys.} {\bf B798} (2008) 491--504,
  [\href{http://xxx.lanl.gov/abs/0710.4348}{{\tt arXiv:0710.4348}}].

\bibitem{Maldacena:1997re}
J.~M. Maldacena, {\it {The Large N Limit of Superconformal Field Theories and
  Supergravity}},  {\em Adv. Theor. Math. Phys.} {\bf 2} (1998) 231--252,
  [\href{http://xxx.lanl.gov/abs/hep-th/9711200}{{\tt hep-th/9711200}}].

\bibitem{Ryu:2006bv}
S.~Ryu and T.~Takayanagi, {\it {Holographic Derivation of Entanglement Entropy
  from AdS/CFT}},  {\em Phys.Rev.Lett.} {\bf 96} (2006) 181602,
  [\href{http://xxx.lanl.gov/abs/hep-th/0603001}{{\tt hep-th/0603001}}].

\bibitem{Ryu:2006ef}
S.~Ryu and T.~Takayanagi, {\it {Aspects of Holographic Entanglement Entropy}},
  {\em JHEP} {\bf 0608} (2006) 045,
  [\href{http://xxx.lanl.gov/abs/hep-th/0605073}{{\tt hep-th/0605073}}].

\bibitem{Lewkowycz:2013nqa}
A.~Lewkowycz and J.~Maldacena, {\it {Generalized Gravitational Entropy}},  {\em
  JHEP} {\bf 1308} (2013) 090, [\href{http://xxx.lanl.gov/abs/1304.4926}{{\tt
  arXiv:1304.4926}}].

\bibitem{Yamaguchi:2002pa}
S.~Yamaguchi, {\it {Holographic RG Flow on the Defect and g-theorem}},  {\em
  JHEP} {\bf 0210} (2002) 002,
  [\href{http://xxx.lanl.gov/abs/hep-th/0207171}{{\tt hep-th/0207171}}].

\bibitem{Gutperle:2012hy}
M.~Gutperle and J.~Samani, {\it {Holographic RG-flows and Boundary CFTs}},
  {\em Phys.Rev.} {\bf D86} (2012) 106007,
  [\href{http://xxx.lanl.gov/abs/1207.7325}{{\tt arXiv:1207.7325}}].

\bibitem{Dias:2013bwa}
O.~Dias, G.~Horowitz, N.~Iqbal, and J.~Santos, {\it {Vortices in Holographic
  Superfluids and Superconductors as Conformal Defects}},
  \href{http://xxx.lanl.gov/abs/1311.3673}{{\tt arXiv:1311.3673}}.

\bibitem{Korovin:2013gha}
Y.~Korovin, {\it {First Order Formalism for the Holographic Duals of Defect
  CFTs}},  \href{http://xxx.lanl.gov/abs/1312.0089}{{\tt arXiv:1312.0089}}.

\bibitem{Takayanagi:2011zk}
T.~Takayanagi, {\it {Holographic Dual of BCFT}},  {\em Phys.Rev.Lett.} {\bf
  107} (2011) 101602, [\href{http://xxx.lanl.gov/abs/1105.5165}{{\tt
  arXiv:1105.5165}}].

\bibitem{Fujita:2011fp}
M.~Fujita, T.~Takayanagi, and E.~Tonni, {\it {Aspects of AdS/BCFT}},  {\em
  JHEP} {\bf 1111} (2011) 043, [\href{http://xxx.lanl.gov/abs/1108.5152}{{\tt
  arXiv:1108.5152}}].

\bibitem{Nozaki:2012qd}
M.~Nozaki, T.~Takayanagi, and T.~Ugajin, {\it {Central Charges for BCFTs and
  Holography}},  {\em JHEP} {\bf 1206} (2012) 066,
  [\href{http://xxx.lanl.gov/abs/1205.1573}{{\tt arXiv:1205.1573}}].

\bibitem{Aharony:2011yc}
O.~Aharony, L.~Berdichevsky, M.~Berkooz, and I.~Shamir, {\it {Near-horizon
  Solutions for D3-branes Ending on 5-branes}},  {\em Phys.Rev.} {\bf D84}
  (2011) 126003, [\href{http://xxx.lanl.gov/abs/1106.1870}{{\tt
  arXiv:1106.1870}}].

\bibitem{Bachas:2013vza}
C.~Bachas, E.~D'Hoker, J.~Estes, and D.~Krym, {\it {M-theory Solutions
  Invariant under $D(2,1;\gamma) \oplus D(2,1;\gamma)$}},
  \href{http://xxx.lanl.gov/abs/1312.5477}{{\tt arXiv:1312.5477}}.

\bibitem{Chiodaroli:2011fn}
M.~Chiodaroli, E.~D'Hoker, and M.~Gutperle, {\it {Simple Holographic Duals to
  Boundary CFTs}},  {\em JHEP} {\bf 1202} (2012) 005,
  [\href{http://xxx.lanl.gov/abs/1111.6912}{{\tt arXiv:1111.6912}}].

\bibitem{Chiodaroli:2012vc}
M.~Chiodaroli, E.~D'Hoker, and M.~Gutperle, {\it {Holographic Duals of Boundary
  CFTs}},  {\em JHEP} {\bf 1207} (2012) 177,
  [\href{http://xxx.lanl.gov/abs/1205.5303}{{\tt arXiv:1205.5303}}].

\bibitem{Aharony:2008ug}
O.~Aharony, O.~Bergman, D.~Jafferis, and J.~Maldacena, {\it {N=6 Superconformal
  Chern-Simons-matter Theories, M2-branes and Their Gravity Duals}},  {\em
  JHEP} {\bf 10} (2008) 091, [\href{http://xxx.lanl.gov/abs/0806.1218}{{\tt
  arXiv:0806.1218}}].

\bibitem{Bak:2003jk}
D.~Bak, M.~Gutperle, and S.~Hirano, {\it {A Dilatonic Deformation of AdS(5) and
  its Field Theory Dual}},  {\em JHEP} {\bf 0305} (2003) 072,
  [\href{http://xxx.lanl.gov/abs/hep-th/0304129}{{\tt hep-th/0304129}}].

\bibitem{Clark:2005te}
A.~Clark and A.~Karch, {\it {Super Janus}},  {\em JHEP} {\bf 0510} (2005) 094,
  [\href{http://xxx.lanl.gov/abs/hep-th/0506265}{{\tt hep-th/0506265}}].

\bibitem{D'Hoker:2006uu}
E.~D'Hoker, J.~Estes, and M.~Gutperle, {\it {Ten-dimensional Supersymmetric
  Janus Solutions}},  {\em Nucl.Phys.} {\bf B757} (2006) 79--116,
  [\href{http://xxx.lanl.gov/abs/hep-th/0603012}{{\tt hep-th/0603012}}].

\bibitem{D'Hoker:2007xy}
E.~D'Hoker, J.~Estes, and M.~Gutperle, {\it {Exact Half-BPS Type IIB Interface
  Solutions. I. Local Solution and Supersymmetric Janus}},  {\em JHEP} {\bf
  0706} (2007) 021, [\href{http://xxx.lanl.gov/abs/0705.0022}{{\tt
  arXiv:0705.0022}}].

\bibitem{D'Hoker:2007xz}
E.~D'Hoker, J.~Estes, and M.~Gutperle, {\it {Exact Half-BPS Type IIB Interface
  Solutions II: Flux Solutions and Multi-Janus}},  {\em JHEP} {\bf 06} (2007)
  022, [\href{http://xxx.lanl.gov/abs/0705.0024}{{\tt arXiv:0705.0024}}].

\bibitem{D'Hoker:2009gg}
E.~D'Hoker, J.~Estes, M.~Gutperle, and D.~Krym, {\it {Janus Solutions in
  M-theory}},  {\em JHEP} {\bf 0906} (2009) 018,
  [\href{http://xxx.lanl.gov/abs/0904.3313}{{\tt arXiv:0904.3313}}].

\bibitem{Suh:2011xc}
M.-W. Suh, {\it {Supersymmetric Janus Solutions in Five and Ten Dimensions}},
  {\em JHEP} {\bf 1109} (2011) 064,
  [\href{http://xxx.lanl.gov/abs/1107.2796}{{\tt arXiv:1107.2796}}].

\bibitem{Berdichevsky:2013ija}
L.~Berdichevsky and B.-e. Dahan, {\it {Local Gravitational Solutions Dual to
  M2-branes Intersecting and/or Ending on M5-branes}},  {\em JHEP} {\bf 1308}
  (2013) 061, [\href{http://xxx.lanl.gov/abs/1304.4389}{{\tt
  arXiv:1304.4389}}].

\bibitem{Bobev:2013yra}
N.~Bobev, K.~Pilch, and N.~Warner, {\it {Supersymmetric Janus Solutions in Four
  Dimensions}},  \href{http://xxx.lanl.gov/abs/1311.4883}{{\tt
  arXiv:1311.4883}}.

\bibitem{DeWolfe:2001pq}
O.~DeWolfe, D.~Freedman, and H.~Ooguri, {\it {Holography and Defect Conformal
  Field Theories}},  {\em Phys. Rev.} {\bf D66} (2002) 025009,
  [\href{http://xxx.lanl.gov/abs/hep-th/0111135}{{\tt hep-th/0111135}}].

\bibitem{Erdmenger:2002ex}
J.~Erdmenger, Z.~Guralnik, and I.~Kirsch, {\it {Four-Dimensional Superconformal
  Theories with Interacting Boundaries or Defects}},  {\em Phys. Rev.} {\bf
  D66} (2002) 025020, [\href{http://xxx.lanl.gov/abs/hep-th/0203020}{{\tt
  hep-th/0203020}}].

\bibitem{Clark:2004sb}
A.~Clark, D.~Freedman, A.~Karch, and M.~Schnabl, {\it {Dual of the Janus
  Solution: An Interface Conformal Field Theory}},  {\em Phys.Rev.} {\bf D71}
  (2005) 066003, [\href{http://xxx.lanl.gov/abs/hep-th/0407073}{{\tt
  hep-th/0407073}}].

\bibitem{D'Hoker:2006uv}
E.~D'Hoker, J.~Estes, and M.~Gutperle, {\it {Interface Yang-Mills,
  Supersymmetry, and Janus}},  {\em Nucl.Phys.} {\bf B753} (2006) 16--41,
  [\href{http://xxx.lanl.gov/abs/hep-th/0603013}{{\tt hep-th/0603013}}].

\bibitem{Kim:2008dj}
C.~Kim, E.~Koh, and K.-M. Lee, {\it {Janus and Multifaced Supersymmetric
  Theories}},  {\em JHEP} {\bf 0806} (2008) 040,
  [\href{http://xxx.lanl.gov/abs/0802.2143}{{\tt arXiv:0802.2143}}].

\bibitem{Gaiotto:2008sa}
D.~Gaiotto and E.~Witten, {\it {Supersymmetric Boundary Conditions in N=4 Super
  Yang-Mills Theory}},  {\em J.Statist.Phys.} {\bf 135} (2009) 789--855,
  [\href{http://xxx.lanl.gov/abs/0804.2902}{{\tt arXiv:0804.2902}}].

\bibitem{Honma:2008un}
Y.~Honma, S.~Iso, Y.~Sumitomo, and S.~Zhang, {\it {Janus Field Theories from
  Multiple M2 Branes}},  {\em Phys.Rev.} {\bf D78} (2008) 025027,
  [\href{http://xxx.lanl.gov/abs/0805.1895}{{\tt arXiv:0805.1895}}].

\bibitem{Gaiotto:2008ak}
D.~Gaiotto and E.~Witten, {\it {S-Duality of Boundary Conditions In N=4 Super
  Yang-Mills Theory}},  {\em Adv.Theor.Math.Phys.} {\bf 13} (2009) 721,
  [\href{http://xxx.lanl.gov/abs/0807.3720}{{\tt arXiv:0807.3720}}].

\bibitem{Jensen:2013lxa}
K.~Jensen and A.~O'Bannon, {\it {Holography, Entanglement Entropy, and
  Conformal Field Theories with Boundaries or Defects}},  {\em Phys.Rev.} {\bf
  D88} (2013) 106006, [\href{http://xxx.lanl.gov/abs/1309.4523}{{\tt
  arXiv:1309.4523}}].

\bibitem{Fursaev:2013mxa}
D.~Fursaev, {\it {Quantum Entanglement on Boundaries}},  {\em JHEP} {\bf 1307}
  (2013) 119, [\href{http://xxx.lanl.gov/abs/1305.2334}{{\tt
  arXiv:1305.2334}}].

\bibitem{Karch:2000gx}
A.~Karch and L.~Randall, {\it {Open and Closed String Interpretation of SUSY
  CFT's on Branes with Boundaries}},  {\em JHEP} {\bf 06} (2001) 063,
  [\href{http://xxx.lanl.gov/abs/hep-th/0105132}{{\tt hep-th/0105132}}].

\bibitem{Gomis:2006cu}
J.~Gomis and C.~Romelsberger, {\it {Bubbling Defect CFT's}},  {\em JHEP} {\bf
  08} (2006) 050, [\href{http://xxx.lanl.gov/abs/hep-th/0604155}{{\tt
  hep-th/0604155}}].

\bibitem{Assel:2011xz}
B.~Assel, C.~Bachas, J.~Estes, and J.~Gomis, {\it {Holographic Duals of D=3 N=4
  Superconformal Field Theories}},  {\em JHEP} {\bf 1108} (2011) 087,
  [\href{http://xxx.lanl.gov/abs/1106.4253}{{\tt arXiv:1106.4253}}].

\bibitem{Gaiotto:2008sd}
D.~Gaiotto and E.~Witten, {\it {Janus Configurations, Chern-Simons Couplings,
  and The theta-Angle in N=4 Super Yang-Mills Theory}},  {\em JHEP} {\bf 1006}
  (2010) 097, [\href{http://xxx.lanl.gov/abs/0804.2907}{{\tt
  arXiv:0804.2907}}].

\bibitem{Estes:2012vm}
J.~Estes, R.~Feldman, and D.~Krym, {\it {Exact Half-BPS Flux Solutions in
  M-theory with $D(2, 1: c': 0)^2$ Symmetry: Local Solutions}},  {\em
  Phys.Rev.} {\bf D87} (2013) 046008,
  [\href{http://xxx.lanl.gov/abs/1209.1845}{{\tt arXiv:1209.1845}}].

\bibitem{Casini:2004bw}
H.~Casini and M.~Huerta, {\it {A Finite Entanglement Entropy and the
  c-theorem}},  {\em Phys.Lett.} {\bf B600} (2004) 142--150,
  [\href{http://xxx.lanl.gov/abs/hep-th/0405111}{{\tt hep-th/0405111}}].

\bibitem{Liu:2012eea}
H.~Liu and M.~Mezei, {\it {A Refinement of Entanglement Entropy and the Number
  of Degrees of Freedom}},  {\em JHEP} {\bf 1304} (2013) 162,
  [\href{http://xxx.lanl.gov/abs/1202.2070}{{\tt arXiv:1202.2070}}].

\bibitem{Klebanov:2012va}
I.~Klebanov, T.~Nishioka, S.~S. Pufu, and B.~Safdi, {\it {Is Renormalized
  Entanglement Entropy Stationary at RG Fixed Points?}},  {\em JHEP} {\bf 1210}
  (2012) 058, [\href{http://xxx.lanl.gov/abs/1207.3360}{{\tt
  arXiv:1207.3360}}].

\bibitem{Freedman:1999gp}
D.~Freedman, S.~Gubser, K.~Pilch, and N.~Warner, {\it {Renormalization Group
  Flows from Holography Supersymmetry and a c-theorem}},  {\em
  Adv.Theor.Math.Phys.} {\bf 3} (1999) 363--417,
  [\href{http://xxx.lanl.gov/abs/hep-th/9904017}{{\tt hep-th/9904017}}].

\bibitem{Chang:2013mca}
H.-C. Chang and A.~Karch, {\it {Entanglement Entropy for Probe Branes}},  {\em
  JHEP} {\bf 1401} (2014) 180, [\href{http://xxx.lanl.gov/abs/1307.5325}{{\tt
  arXiv:1307.5325}}].

\bibitem{Karch:2014ufa}
A.~Karch and C.~Uhlemann, {\it {Generalized Gravitational Entropy of Probe
  Branes: Flavor Entanglement Holographically}},
  \href{http://xxx.lanl.gov/abs/1402.4497}{{\tt arXiv:1402.4497}}.

\bibitem{Anselmi:1997am}
D.~Anselmi, D.~Freedman, M.~Grisaru, and A.~Johansen, {\it {Nonperturbative
  Formulas for Central Functions of Supersymmetric Gauge Theories}},  {\em
  Nucl.Phys.} {\bf B526} (1998) 543--571,
  [\href{http://xxx.lanl.gov/abs/hep-th/9708042}{{\tt hep-th/9708042}}].

\bibitem{Anselmi:1997ys}
D.~Anselmi, J.~Erlich, D.~Freedman, and A.~Johansen, {\it {Positivity
  Constraints on Anomalies in Supersymmetric Gauge Theories}},  {\em Phys.Rev.}
  {\bf D57} (1998) 7570--7588,
  [\href{http://xxx.lanl.gov/abs/hep-th/9711035}{{\tt hep-th/9711035}}].

\bibitem{Grover:2011fa}
T.~Grover, A.~M. Turner, and A.~Vishwanath, {\it {Entanglement Entropy of
  Gapped Phases and Topological Order in Three dimensions}},  {\em Phys.Rev.}
  {\bf B84} (2011) 195120, [\href{http://xxx.lanl.gov/abs/1108.4038}{{\tt
  arXiv:1108.4038}}].

\bibitem{Maciejko:2010tx}
J.~Maciejko, X.-L. Qi, A.~Karch, and S.-C. Zhang, {\it {Fractional Topological
  Insulators in Three Dimensions}},  {\em Phys.Rev.Lett.} {\bf 105} (2010)
  246809, [\href{http://xxx.lanl.gov/abs/1004.3628}{{\tt arXiv:1004.3628}}].

\bibitem{HoyosBadajoz:2010ac}
C.~Hoyos-Badajoz, K.~Jensen, and A.~Karch, {\it {A Holographic Fractional
  Topological Insulator}},  {\em Phys.Rev.} {\bf D82} (2010) 086001,
  [\href{http://xxx.lanl.gov/abs/1007.3253}{{\tt arXiv:1007.3253}}].

\bibitem{Estes:2012nx}
J.~Estes, A.~O'Bannon, E.~Tsatis, and T.~Wrase, {\it {Holographic Wilson Loops,
  Dielectric Interfaces, and Topological Insulators}},  {\em Phys.Rev.} {\bf
  D87} (2013) 106005, [\href{http://xxx.lanl.gov/abs/1210.0534}{{\tt
  arXiv:1210.0534}}].

\bibitem{Nishioka:2013haa}
T.~Nishioka and I.~Yaakov, {\it {Supersymmetric RŽnyi Entropy}},  {\em JHEP}
  {\bf 1310} (2013) 155, [\href{http://xxx.lanl.gov/abs/1306.2958}{{\tt
  arXiv:1306.2958}}].

\bibitem{Sugishita:2013jca}
S.~Sugishita and S.~Terashima, {\it {Exact Results in Supersymmetric Field
  Theories on Manifolds with Boundaries}},  {\em JHEP} {\bf 1311} (2013) 021,
  [\href{http://xxx.lanl.gov/abs/1308.1973}{{\tt arXiv:1308.1973}}].

\bibitem{Honda:2013uca}
D.~Honda and T.~Okuda, {\it {Exact Results for Boundaries and Domain Walls in
  2d Supersymmetric Theories}},  \href{http://xxx.lanl.gov/abs/1308.2217}{{\tt
  arXiv:1308.2217}}.

\bibitem{Hori:2013ika}
K.~Hori and M.~Romo, {\it {Exact Results In Two-Dimensional (2,2)
  Supersymmetric Gauge Theories With Boundary}},
  \href{http://xxx.lanl.gov/abs/1308.2438}{{\tt arXiv:1308.2438}}.

\bibitem{Papadimitriou:2004rz}
I.~Papadimitriou and K.~Skenderis, {\it {Correlation Functions in Holographic
  RG Flows}},  {\em JHEP} {\bf 0410} (2004) 075,
  [\href{http://xxx.lanl.gov/abs/hep-th/0407071}{{\tt hep-th/0407071}}].

\bibitem{Hanany:1996ie}
A.~Hanany and E.~Witten, {\it {Type IIB Superstrings, BPS Monopoles, and Three-
  dimensional Gauge Dynamics}},  {\em Nucl. Phys.} {\bf B492} (1997) 152--190,
  [\href{http://xxx.lanl.gov/abs/hep-th/9611230}{{\tt hep-th/9611230}}].

\bibitem{McAvity:1995zd}
D.~McAvity and H.~Osborn, {\it {Conformal Field Theories Near a Boundary in
  General Dimensions}},  {\em Nucl.Phys.} {\bf B455} (1995) 522--576,
  [\href{http://xxx.lanl.gov/abs/cond-mat/9505127}{{\tt cond-mat/9505127}}].

\bibitem{Assel:2012cp}
B.~Assel, J.~Estes, and M.~Yamazaki, {\it {Large N Free Energy of 3d N=4 SCFTs
  and $AdS_4/CFT_3$}},  {\em JHEP} {\bf 1209} (2012) 074,
  [\href{http://xxx.lanl.gov/abs/1206.2920}{{\tt arXiv:1206.2920}}].

\bibitem{Nishioka:2011dq}
T.~Nishioka, Y.~Tachikawa, and M.~Yamazaki, {\it {3d Partition Function as
  Overlap of Wavefunctions}},  {\em JHEP} {\bf 1108} (2011) 003,
  [\href{http://xxx.lanl.gov/abs/1105.4390}{{\tt arXiv:1105.4390}}].

\end{thebibliography}\endgroup

\end{document}